\documentclass{article}
\usepackage{epsfig}
\newcommand{\be}{\begin{equation}}
\newcommand{\ee}{\end{equation}}
\newcommand{\bea}{\begin{eqnarray}}
\newcommand{\eea}{\end{eqnarray}}
\newcommand{\bt}{\noindent\begin{table}[ht]}
\newcommand{\et}{\end{table}}
\newcommand{\btm}{\noindent}
\newcommand{\etm}{\medskip}
\newcommand{\nn}{\nonumber \\}
\newcommand{\ie}{{\rm i}}
\newcommand{\de}{{\rm d}}
\newcommand{\ex}[1]{{\rm e}^{#1}}
\newcommand{\du}{\,\delta u\,}
\newcommand{\dv}{\,\delta v\,}

\newcommand{\dpr}{^{\prime\prime}}
\newcommand{\cA}{{\cal A}}
\newcommand{\heps}{\hat\epsilon}
\newcommand{\hd}{\hat d}
\newcommand{\nicht}[1]{ }
\newcommand{\cE}[4]{{\cal E}_{#1}\left(#2\left|{#3\atop#4}\right.\right)}
\newcommand{\rb}[1]{\raisebox{-0.5ex}{\epsfig{file=#1,scale=0.4}}}
\newtheorem{fig}{Fig.}
\newtheorem{figm}[fig]{Fig.$^*$}
\newcommand{\Figur}[2]{\begin{fig}\label{#1} #2\end{fig}}
\newcommand{\Figurm}[2]{\begin{figm}\label{#1} #2\end{figm}}

\newfont{\Kapfont}{cmbx10 scaled 1728}

\begin{document}
\vspace*{1cm}

\begin{center}
\Kapfont Floating Bodies of Equilibrium in 2D,\rule[-3mm]{0mm}{5mm}\\
the Tire Track Problem and\rule[-3mm]{0mm}{5mm}\\
Electrons in a Parabolic Magnetic Field
\end{center}
\vspace{1cm}

\begin{center}
\bf Franz Wegner, Institut f\"ur Theoretische Physik \\
Ruprecht-Karls-Universit\"at Heidelberg \\
Philosophenweg 19, D-69120 Heidelberg \\
Email: wegner@tphys.uni-heidelberg.de \\
\nicht{\today \ $<$Float/Fl207a$>$}
\end{center}
\vspace{1cm}

\paragraph*{Abstract} Explicit solutions
of the two-dimensional floating body problem (bodies that can float
in all positions) for relative density $\rho\not=\frac 12$
and of the tire track problem (tire tracks of a bicycle, which do not
allow to determine, which way the bicycle went) are given, which differ
from circles. Starting point is the differential equation given by the
author in \cite{WegnerII,Wegner}. The curves are also trajectories of
charges in a perpendicular parabolic magnetic field.

\nicht{
\begin{center}\epsfig{file=nicefigure.eps,scale=0.5}\\
{\it From figs. \ref{fgp1p3k}, \ref{fgp1p3l}}
\end{center}
\newpage

\tableofcontents

\bigskip } 

\section{Introduction}

In this paper we consider a class of curves, which are solution to three
problems:\\
(i) the two-dimensional version
of the floating body problem asked by Stanislaw Ulam in the Scottish
Book\cite{Scottish} (problem 19): Is a sphere the only solid of uniform density
which will float in water in any position?
Thus we ask for the non-circular cross-section of a long cylindrical log, which
can float in all orientations with the cross-section perpendicular to the
surface of the fluid.\\
(ii) It also deals with the tire
track problem, which asks whether it is possible, to have tire tracks of the
front and the rear wheel of a bicycle, which do not allow to determine which
way the bicycle went.\\
(iii) Finally the boundary of the cross-section describes the trajectory of a
charge moving in a perpendicular parabolic magnetic field.

This paper is an improved and extended version of the solution given in
\cite{WegnerIII}. It is extended insofar as figures for several curves are
given. The discussion of the tire track problem is extended and explicit
results for the cross-sections are given. Moreover the solution is formulated
in terms of the four extreme radii, which makes the solution and symmetries more
transparent.

Two observations\cite{Auerbach,Bracho,Oliveros,Tabachnikov,WegnerI,Wegner} are
on the basis of the floating body problem:\\
(i) Let $h$ be the distance between the centers of gravity of the parts of the
body above and below the water-line. Then the potential energy of the body is a
constant times this height. Thus $h$ has to be constant.\\
(ii) If one rotates the body infinitesimally, then the line between the two
centers of gravity will only remain perpendicular to the water-line, if the
length $2\ell$ of this water-line obeys
\be
\frac 23 \ell^3 \left(\frac 1{\cA_1} + \frac 1{\cA_2}\right) = h, \label{h}
\ee
where $\cA_1$ and $\cA_2$ are the cross-sections above and below the water-line,
which of course are constant. Thus the length $2\ell$ of the chord across the
cross-section along the water-line is constant. Moreover since the areas above
and below the water-line stay constant, the envelope $\gamma$ of the chords are
touched in the middle of the water-lines. Since for relative density (density
of body divided by the density of the fluid) $\rho=1/2$, the water-line is the
same, when the body is rotated by $180^0$, Auerbach\cite{Auerbach} could give
the solution for this case. Then the boundary of the cross-section is obtained
by
starting from a Zindler curve $\gamma$ for the envelope, that is a curve which
closes after the direction of the tangent to the curve has increased by $\pi$
(such curves have typically an odd number of cusps, for examples see red curves
and water-lines in figs. \ref{fgp1p3a} to \ref{fgp1p3c} and \ref{fgp1p5j}). If
one adds tangents of length $\ell$ in both directions to all points of the
curve $\gamma$, then these end points yield the boundary $\Gamma$ of such a
cross-section, provided this boundary is sufficiently convex.
(By {\it sufficiently convex} it is meant, that all water-lines cross the
boundary twice, but not more often).
In the following the case $\rho\not=\frac 12$ will be mainly considered.

Another important property is obtained as follows: Consider two close-by
water-lines $A_1A_2$ and $B_1B_2$ through the cross-section (Fig. \ref{chord1};
The body is kept fixed and the direction
of the gravitational force is rotated.) Put the $x$-axis parallel
to $A_1A_2$. Then the vector $A_1B_1$ is given by $(\de x_1,-\ell_1\de\phi)$
and $A_2B_2$ by $(\de x_2,\ell_2\de\phi)$. Constant length $\de\ell=0$ implies
$\de x_2=\de x_1$. Constant areas $\cA_1$ and $\cA_2$ imply $\de f_1=\frac 12
\ell_1^2\de\phi
=\de f_2=\frac 12 \ell_2^2\de\phi$. Thus $\ell_1=\ell_2=\ell$. This implies,
that the infinitesimal arcs at the perimeter
$\de u_1=\sqrt{(\de x_1)^2+\ell_1^2(\de\phi)^2}$ and
$\de u_2=\sqrt{(\de x_2)^2+\ell_2^2(\de\phi)^2}$
are equal. Thus the part of the perimeter below the water-line is constant. Of
course the same is also true for the part above the water-line.

\bt
\epsfig{file=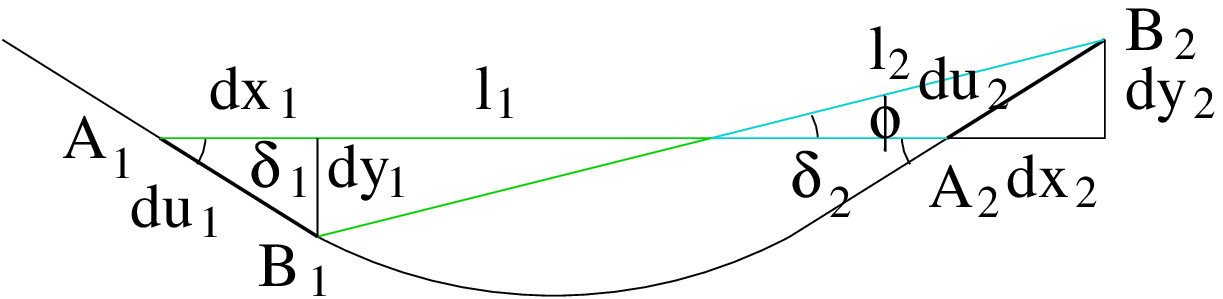,scale=0.5}\hfill
\parbox[b]{3.5cm}{\Figur{chord1}{Boundary and water-lines}}
\et

One can conclude the other way round: If a curve
has the property that if one moves from two fixed points $A_1$ and $A_2$ by
constant arcs $u$ along the
perimeter and the length of the chord remains fixed, then also the
areas separated by the chord stay constant. One argues that since $\ell$ stays
constant one has $\de x_1=\de x_2$. Since $\de u_1=\de u_2$ also 
$\ell_1=\ell_2$ and thus $\de f_1=\de f_2$.
In the main part of the paper this property will be used. By measuring the
distance along the arc
from a fixed point of the boundary one introduces the arc parameter $u$. It will
be shown that for certain differences $2\du$ of the arc parameter the length of
the chord is constant, that is the distance between the point at arc
parameter $u-\du$ and at arc parameter $u+\du$ does not depend on $u$.
I call this the {\it property of constant chord length}.

In the case of $\rho=1/2$ the ends of the water-lines differ by an arc parameter
which is one half of the circumference. Similarly one may look for solutions
where the difference $2\du$ of the arc parameter is a rational fraction of the
circumference. Since however the area bounded by these water-lines and the part
of the circumference is constant
(the total cross-section is constant and the parts below the water also), the
shape of such an equilateral triangle or quadrangle is constant and does not
yield interesting results. An equilateral pentagon, however, of constant area
and constant side length can vary its shape. 
Montejano er al.\cite{Oliveros,Bracho} have used this property to obtain
solutions for these special cases. Unfortunately there was no sufficiently
convex boundary among the solutions.

Another interesting property is that since $\ell_1=\ell_2$, one has
$\de y_1=-\de y_2$,
which implies that the angles $\delta$ between the tangents and the chord are
equal, $\delta_1=\delta_2$.

The floating body problem is related to the tire track problem
(\cite{Tabachnikov,Finn,FinnI} and papers cited therein). It goes back to a
criticism of the discussion between Sherlock Holmes and Watson in {\it The
Adventure of the Priory School}\cite{Doyle} on which way a bicycle went whose
tires' traces are observed. Let the distance between the front and the rear
wheel of the bicycle be $\ell$. The end points of the tangent lines of length
$\ell$ to the trace of the rear wheel in the direction the bicycle went yields
the points of the traces of the front wheel. Thus if the tangent lines in both
directions end at the trace of the front wheel, it is open which way the
bicycle went. Thus curves $\gamma$ for the rear wheel and $\Gamma$ for the
front wheel are solutions for such an ambiguous direction of the bicycle.
The tire track problem consists in finding such curves $\Gamma$ and $\gamma$
different from circles and straight lines. It is equivalent to the
two-dimensional floating body
problem with the exception that the tire tracks need not close or may wind
several times around some point.

Starting from $r=r_0(1+\epsilon\cos(n\psi))+O(\epsilon^2)$ for the boundary of
the floating body
and carrying the expansion in $\epsilon$ through to seventh order the
author conjectured that one obtains solutions for $n-2$ different densities
$\rho$ for the same boundary\cite{WegnerII,Wegner}. This means that one and the
same boundary $\Gamma$ has the above mentioned property for several different
chord lengths $2\ell$. Assuming now that this is also true, if the curve does
not close as it is allowed in the tire track problem, one may assume that it
also holds if the curve closes apart from an infinitesimal small angle
difference $\delta\chi$ with chords of infinitesimal length $2\ell$. Then a
differential equation of third order for this curve can be derived.
After one integration this differential equation yields

\be
\kappa(r):=\frac{r^2+2r^{\prime 2}-rr\dpr}{(r^2+r^{\prime 2})^{3/2}} = 4ar^2+2b,
\label{kappa1}
\ee
in polar coordinates $(r,\psi)$ with $r$ function of $\psi$ and the prime
indicating differentiation with respect to $\psi$. $\kappa$ is the curvature of
the curve. $a$ is given by the limit $\frac{16}3a=\lim\ell^3/\delta\chi$ for vanishing $l$ and $\delta\chi$ (defined by $\chi-\hat\chi$ of sect. \ref{propdist}).
$b$ is the integration constant for this first integral. Thus these curves
describe the trajectories of particles of mass $m$, charge $e$ and velocity $w$
in a perpendicular magnetic field, which depends quadratically on $r$,
\be
B(r) = -\frac{mw}e (4ar^2+2b). \label{Br}
\ee
Another integral of the third-order differential equation is
\be
\frac{r+r\dpr}{(r^2+r^{\prime 2})^{3/2}} = -2ar + \frac{2c}{r^3}.
\label{invcurve}
\ee
If we introduce the inverse radius $r=1/\bar r$, then this equation can be
rewritten
\be
\bar{\kappa}(\bar r) = \frac{\bar r^2+2\bar r^{\prime 2}-\bar r\bar r\dpr}
{(\bar r^2+\bar r^{\prime 2})^{3/2}} = -\frac{2a}{\bar r^4} + 2c.
\ee
Thus $\bar r=1/r$ describes the tractory of a charge in a perpendicular magnetic field
\be
\bar B(\bar r) = \frac{mw}e \left(\frac{2a}{\bar r^4} - 2c \right).
\ee
Elimination of $r\dpr$ from eqs. (\ref{kappa1}, \ref{invcurve}) yields our basic
differential equation
\be
\frac 1{\sqrt{r^2+r^{\prime 2}}} = ar^2+b+cr^{-2}. \label{diffc}
\ee

It will be shown that the resulting curves solve the floating body problem,
provided they are sufficiently convex and closed. Indeed there are such
solutions. For the tire track problem the restrictions are less rigorous. But
among them there are solutions, which require artistic mastery of the bicyclist,
since he/she has to move back and forth. In more difficult cases the bicycle
has to allow the steerer to be rotated by more than 180$^0$ degrees.

Quite generally the following remarkable property of these curves will be shown:
Consider two copies of these curves (not necessarily closed). Choose an
arbitrary point on each curve. Then there exists always an angle by which the
two curves can be rotated against each other, so that the distance between the
points on the two curves stays constant, if one moves from the given points on
both curves by the same arc distance. I call this the {\it property of constant
distance}. This remarkable property made it possible
to obtain the above-mentioned differential equation by choosing the two points
infinitesimally close.

In \cite{WegnerII,Wegner} the following necessary condition was shown:
If for some chord the angles $\delta$ between the chord and the tangents on the
curve obey $\delta_1=\delta_2$ at $r_1\not=r_2$ then also the
derivative $\frac{\de(\delta_1-\delta_2)}{\de u}$ vanishes. This is a necessary
condition, but it is not sufficient. I am indebted to Serge Tabachnikov for
pointing this out to me. Here the differential equation (\ref{diffc}) will be
solved explicitly as a function of the arc parameter $u$, and it will be shown
that the curve has the desired property.

The outline of the paper is as follows: In sect. \ref{basic} an appropriate
parametrization is introduced. Depending on the parameters $a$, $b$, and $c$
one may have no or
one or two (real) branches. Two branches means that the curves are not
congruent. The equations are given in terms of the arc parameter.
In sect. \ref{radius} the radius $r$ as a function of the arc parameter $u$ is
determined.
It is given in terms of Weierstrass $\wp$-functions. The invariants $g_2$ and
$g_3$ are real.
In the case of one branch the discrimant is negative, otherwise positive.
This is discussed in sect. \ref{onetwo}.
In the following section the polar angle and thus the whole curve is determined
as function of $u$. Starting from the differential equation (\ref{diffc}) the
solution can be written
\be
z(u):=x+\ie y= C\ex{\ie\chi} \frac{\sigma(u-3v)}{\sigma(u+v)} \ex{2\zeta(2v)u},
\label{zu1}
\ee
where $u$ is the (real) arc parameter. The purely imaginary $v$ and the real
invariants $g_2$ and $g_3$ of the Weierstrass integrals $\zeta$ and $\sigma$,
and the constant $C$ depend on the coefficients $a$, $b$, and $c$. $\chi$ is an
arbitrary angle which defines the orientation of the curve around the origin
and can be chosen arbitrarily.

Instead of solving (\ref{diffc}) one could immediately start with the curve
defined by (\ref{zu1}) and first show that the arc-parameter $u$ indeed
measures the distance along the curve, which is done in subsect. \ref{more}.

Secondly the above mentioned property of constant distance is shown in sect.
\ref{chord}: Given a
difference $\du$, there exists a difference of angles, by which two copies of
the curve described by $z$ and $\hat z$ have to be rotated against each other,
so that the distance $2\ell=|z(u+\du)-\hat z(u-\du)|$ does not depend on $u$
(eq. \ref{chdu}). If the difference of the angles is such that the two curves fall onto each other, then one obtains the
desired property of constant chord length for floating bodies and tire tracks.
The property of constant distance holds even between the two different
branches in the case of positive discriminant. A remark on the expression of
Finn for the curvatures of the curves is added in subsection \ref{RemFinn}.
Finally the area above and below the water-line is determined and it is shown
that the centers of gravity of the parts of the cross-section above and below
the water-line lie on circles (compare Auerbach\cite{Auerbach}).

In the following section the periodicity of the curves is discussed and figures
of several curves are shown. In addition a number of these examples are shown as
animations on the internet\cite{movie}. These examples are
indicated by an asterisk at the figure caption.

The limit case, where the discriminant vanishes, is considered in section
\ref{calD}. In this case the curves can be expressed in terms of trigonometric
and exponential functions. Moreover there is a simple construction principle of
these curves explained in subsection \ref{elem}.
In section \ref{linear} the limit is considered, in which the curve oscillates
around a straight line. (This does not constitute a solution of the floating
body problem, but of the tire-track problem.) There one finds an infinite set
of $\du$s, for which the length of the chord between the points of arc
parameter $u-\du$ and $u+\du$ is independent of $u$ and where $2\du$ is not a
period of the curve. The curves found in this
section are also trajectories of electrons in a linearly varying magnetic field
as considered by Evers et al\cite{Evers}.

In the following section the {\it carousels} by Bracho, Montejano and Oliveros
\cite{Oliveros,Bracho} are discussed, where five chords or line segments close to
an equilateral pentagon. The solution given here
allows to reconstruct these curves. Some of them will be shown and discussed.

I cannot claim that the solutions given here are the only solutions to the
floating body problem, but I am not aware of any other solutions. There remains
the question whether starting from the known solutions small deformations can
yield new solutions similarly as a class of solutions were found from deforming
the circle \cite{WegnerI, Wegner}. The solutions of the tire track problem are
not restricted to closed and not terminating curves. In view of the paper by
Finn \cite{Finn,FinnI}
who describes a procedure to continue from an initial back-tire track segment
one expects a large manifold of solutions for this latter problem.

\section{Basic Equations\label{basic}}

\subsection{A first discussion}

The curves described by the differential equation (\ref{diffc}) will now be
considered quite generally. For this purpose the angle $\phi$ is introduced
indicating the direction of the tangent to the curve with respect to the radial
vector by
\be
\frac{\de r}{\de\psi}=r\tan\phi
\ee
\bt
\epsfig{file=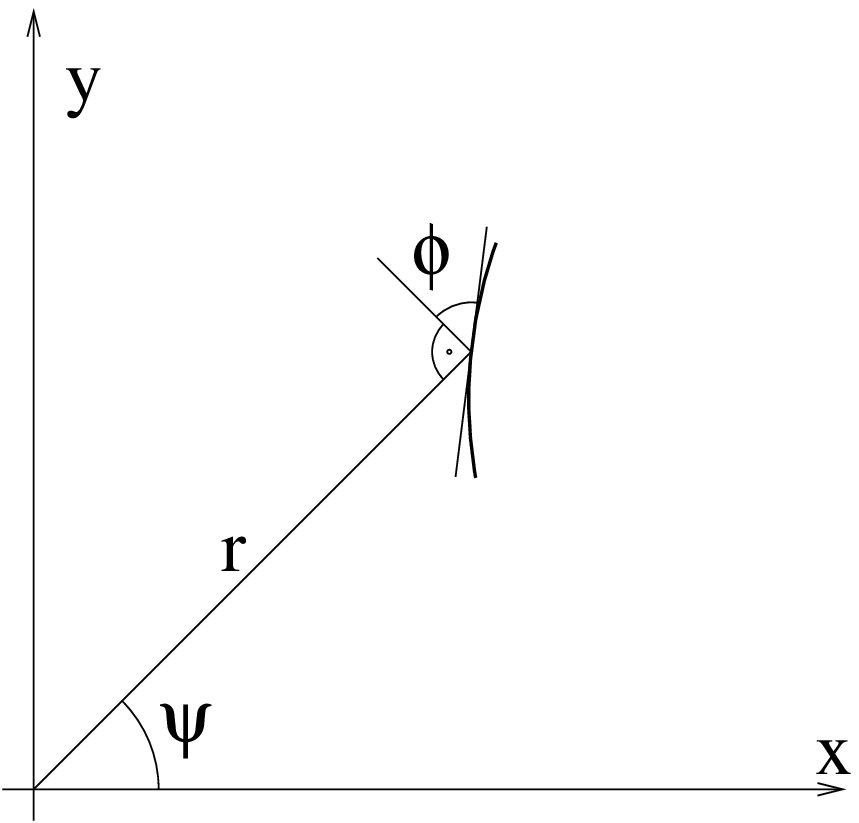,scale=0.4} \hspace{1cm}
\parbox[b]{6cm}{
\Figur{polar}{Figure indicating the angle $\phi$,
part of the curve (heavy line) and the tangent to the curve.}}
\et
as shown in fig. \ref{polar}. Then eq. (\ref{diffc}) can be written
\be
aq^2+bq+c=\sqrt{q} \cos\phi,
\ee
where the square $q=r^2$ of the radius is introduced.
Thus the curves obey
\be
-\sqrt q \le aq^2+bq+c \le \sqrt q. \label{qeq}
\ee
Whenever
\be
aq^2+bq+c=\pm\sqrt{q}
\ee
is reached, the curve has found an extreme distance from the origin.

Depending on
the constants $a$, $b$, and $c$ the inequality (\ref{qeq}) allows for zero, one
or two curves. This can be seen by plotting the parabola \rb{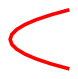} depicting
$\pm\sqrt q$
and the parabola \rb{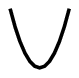} showing $aq^2+bq+c$ versus $q$. To the extend
segments of the parabola \rb{pars.eps}
lie inside the parabola \rb{parw.eps} one obtains curves. These segments are
indicated by thick lines. Various possible cases are shown in fig. \ref{cases}.
\bt
\begin{tabular}{ccc}
\epsfig{file=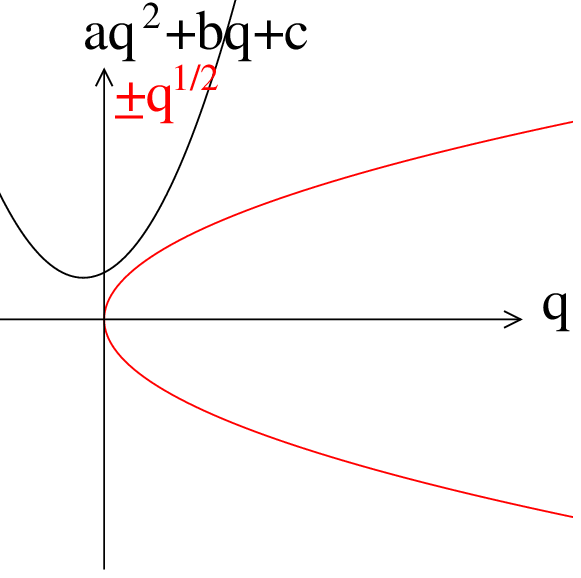,scale=0.5,angle=270} &
\epsfig{file=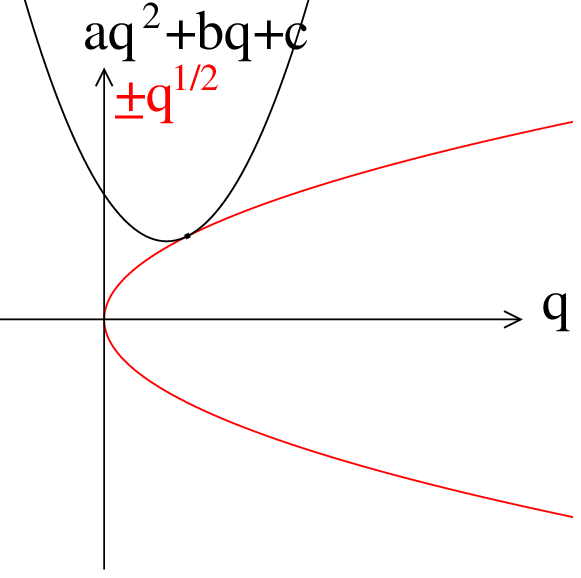,scale=0.5,angle=270} &
\epsfig{file=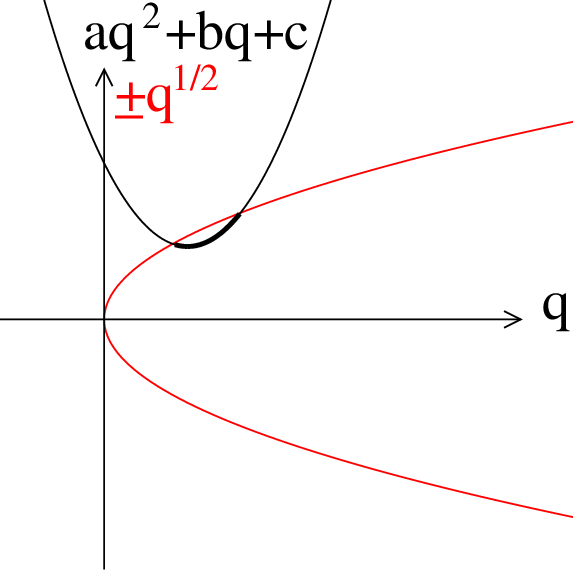,scale=0.5,angle=270} \\
\it Fig. \ref{cases}a & \it Fig. \ref{cases}b & \it Fig. \ref{cases}c
\end{tabular}

\begin{tabular}{ccc}
\epsfig{file=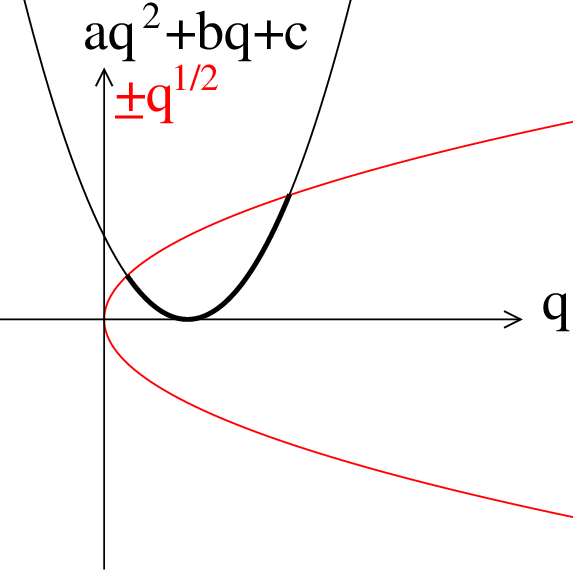,scale=0.5,angle=270} &
\epsfig{file=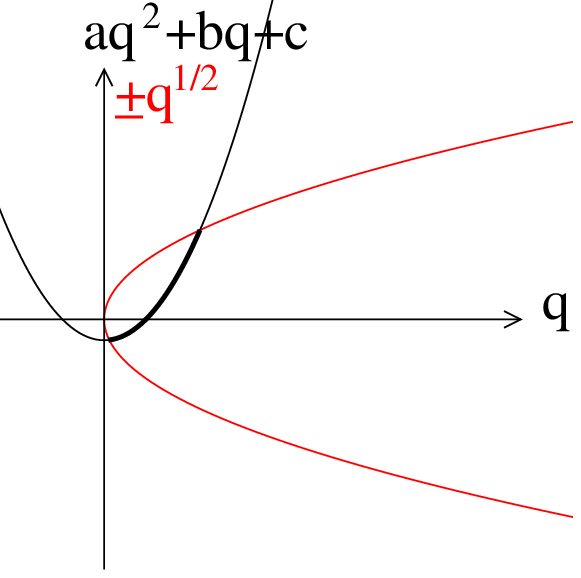,scale=0.5,angle=270} &
\epsfig{file=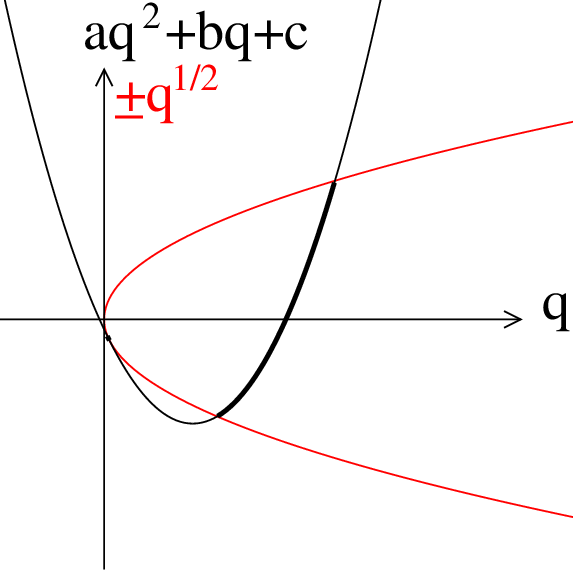,scale=0.5,angle=270} \\
\it Fig. \ref{cases}d & \it Fig. \ref{cases}e & \it Fig. \ref{cases}f
\end{tabular}

\begin{tabular}{ccc}
\epsfig{file=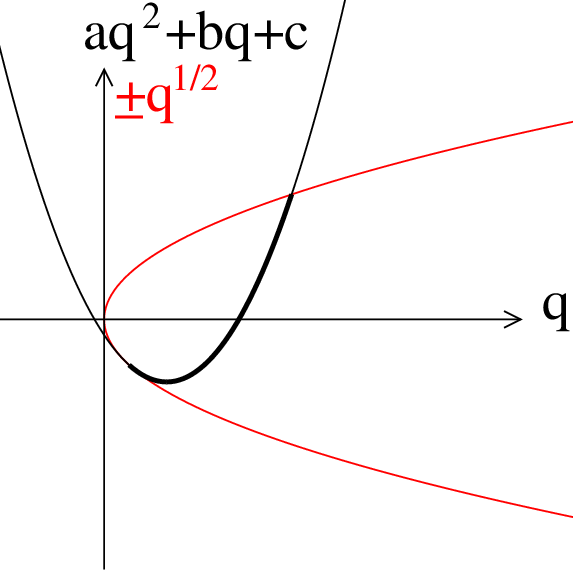,scale=0.5,angle=270} &
\epsfig{file=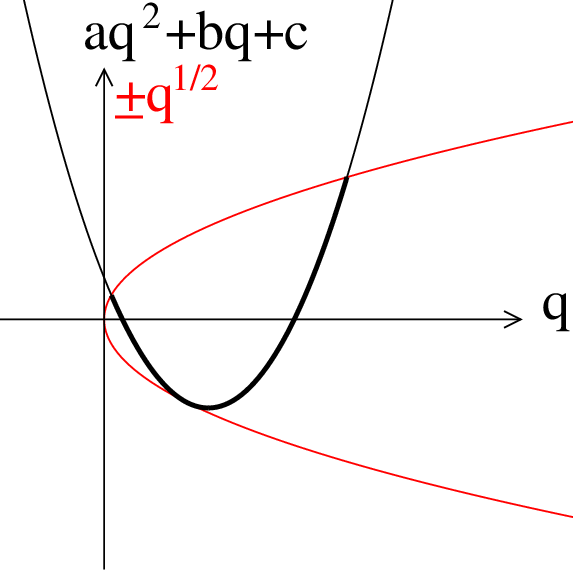,scale=0.5,angle=270} &
\epsfig{file=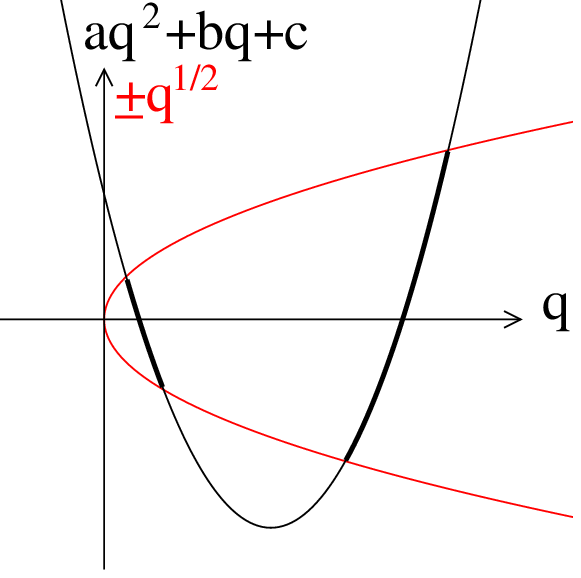,scale=0.5,angle=270} \\
\it Fig. \ref{cases}g & \it Fig. \ref{cases}h & \it Fig. \ref{cases}i
\end{tabular}

\Figur{cases}{Various cases of the relative position of the two parabola}
\et
Figure \ref{cases}a does not allow for any real curves, since the parabola
\rb{pars.eps} lies completely outside \rb{parw.eps}. In fig. \ref{cases}b the
parabola \rb{pars.eps} touches \rb{parw.eps}. Only one radius is allowed, which
corresponds to a circle. Fig. \ref{cases}c allows
for some variation of the radius. It yields a curve
oscillating between the two extreme radii. From this case one obtains good
solutions for the floating body and the tire track problem.
In fig. \ref{cases}d the parabola \rb{pars.eps} touches
the $q$-axis. Thus the curve shows a tangent in radial direction. In fig.
\ref{cases}e the
curve goes backward in $\psi$-direction close to the smallest radius. Fig.
\ref{cases}f allows for a small circle
plus a curve. Fig. \ref{cases}g is the limit case where the curve approaches
asymptotically the circle. Fig. \ref{cases}h allows for two curves both of
which approach
asymptotically a common circle, one from outside, the other from inside. Fig.
\ref{cases}i is the generic case of two curves.

The extreme distances from the origin are reached, when $\cos\phi=\pm 1$. This
yields
\be
ar_i^4+br_i^2+c-r_i=0 \label{ricond}
\ee
where for real $r_i$ the $|r_i|$ are the extreme radii of the curves.
$r_i$ is chosen positive, if $\cos\phi=+1$, and negative, if
$\cos\phi=-1$.
The polynomial (\ref{ricond}) can be expressed in terms of these extreme radii
$r_i$,
\be
ar^4+br^2+c-r = a \prod_{i=1}^4 (r-r_i).
\ee
Comparing the coefficients of the polynomial one obtains
\bea
0 &=& \sum_i r_i, \label{sumr} \\
b &=& a \sum_{i<j} r_i r_j = \frac a2 (\sum_i r_i)^2 - \frac a2 \sum_i r_i^2
= - \frac a2 \sum_i q_i, \label{b} \\
-1 &=& -a (r_1r_2r_3+r_1r_2r_4+r_1r_3r_4+r_2r_3r_4) = -a P_r, \\
c &=& a r_1r_2r_3r_4 = a \hat P, \label{c}
\eea
which allows the coefficients $a$, $b$, $c$ to be expressed by the $r_i$.
Thus the sum of the four $r_i$ vanishes.
Here
\be
q_i=r_i^2
\ee
is introduced and use is made of the polynomials $P_r$ and $\hat P$ defined in
appendix \ref{symm} in eqs. (\ref{Pr}) and (\ref{hatP}) and of eq. (\ref{Pra}).
In particular one obtains
\be
a=\frac 1{P_r}. \label{a}
\ee
In the following $a$ is assumed to be positive. Reversing simultaneously the
signs of $a$, $b$, and $c$ corresponds to reversing all $r_i$, which yields the
same curves, but a different orientation. If $a=0$, then according to
(\ref{kappa1}) the curve is a circle ($b\not=0$) or a straight line ($b=0$).

\subsection{Parametrization according to the shape}

Multiplication of all $r_i$ by the same factor yields similar curves, but no new
shape. Since the four $r_i$ obey the restriction $\sum_i r_i=0$, only two
parameters are left, which determine the shape of the curves. Thus the
parametrization
\be
r_{4,3}=r_0 (1\pm\epsilon), \quad r_{1,2}= r_0 (-1\pm\heps)
\ee
is introduced, where $r_0$ sets the scale of the curve, whereas $\epsilon$ and
$\heps$ determine its shape.
$\epsilon$ and $\heps$ can always be chosen either real or purely imaginary. A
sign change of $\epsilon$ and/or $\heps$ corresponds to a permutation of the
$r_i$. Therefore the $\epsilon^2,\heps^2$ plane is plotted in figure
\ref{param}. Both $\epsilon^2$ and $\heps^2$ are real. $r_0$ is assumed to be
real.
\bt
\epsfig{file=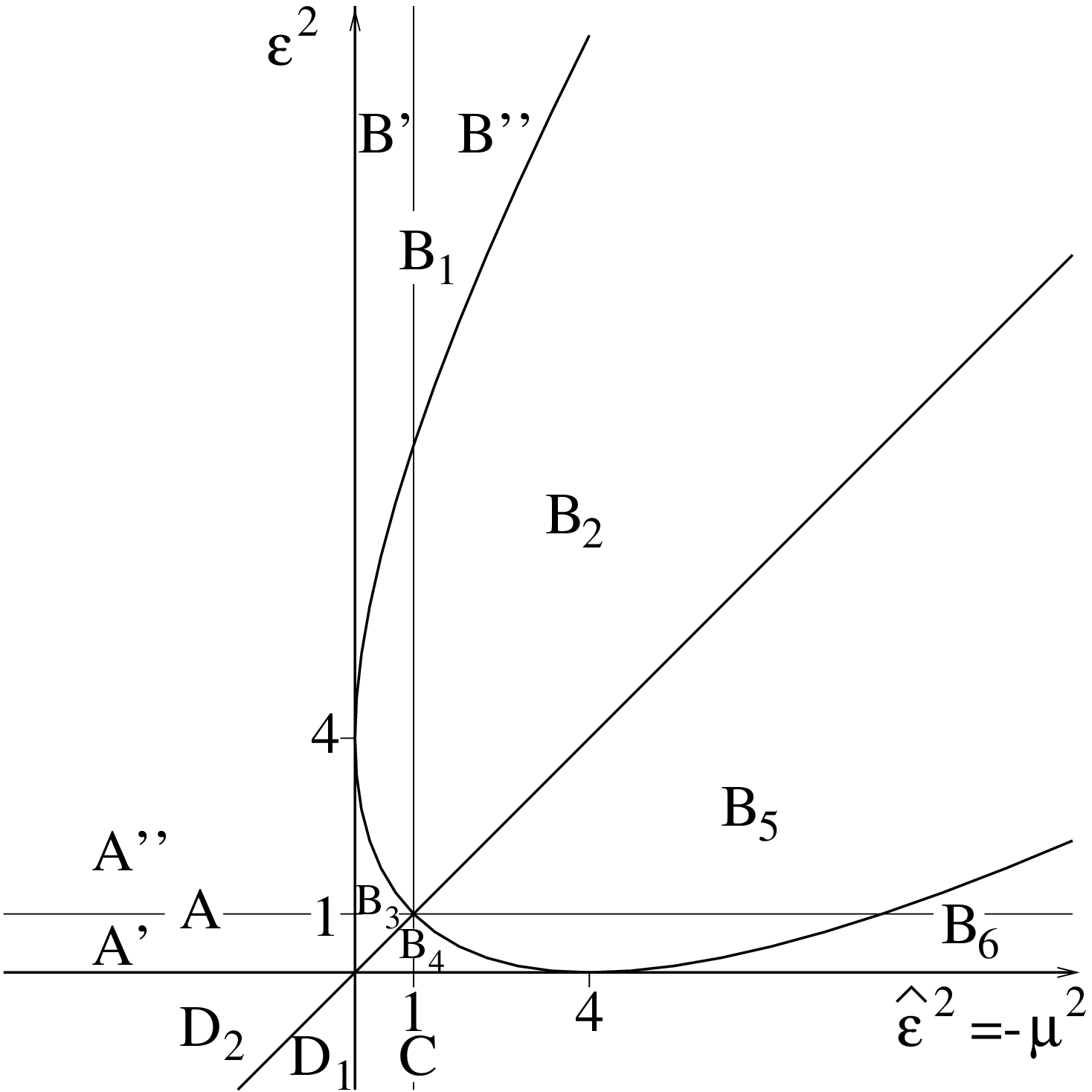,scale=0.5} \hfill
\parbox[b]{3cm}{\Figur{param}{Parameter space for $\epsilon^2$ and $\heps^2$.}}
\et

Consider the four quadrants in fig. \ref{param}. $\epsilon$ is
real and $\heps=-\ie\mu$ is imaginary in the left upper quadrant A. Here one has
one real curve. In the right
upper quadrant B both $\epsilon$ and $\heps$ are real. There one obtains two
branches.
In the right lower quadrant C $\epsilon$ is imaginary and $\heps$ is real. By
exchanging $\epsilon$ and $\heps$ one comes back to quadrant A. Finally
in the lower left quadrant D both $\epsilon$ and $\heps$ are imaginary. This
does not correspond to any real curve.
Thus it is sufficient to consider the quadrants A and B.

Without restriction of the general case
\be
\epsilon-2>\heps>0.
\ee
can be required in region B.
This corresponds to the ordering $q_4>q_3>q_2>q_1$ and is covered by the region
B$_1$. Fig. \ref{casei} repeats fig. \ref{cases}i with the assignments
of the $r_i$.
\bt
\epsfig{file=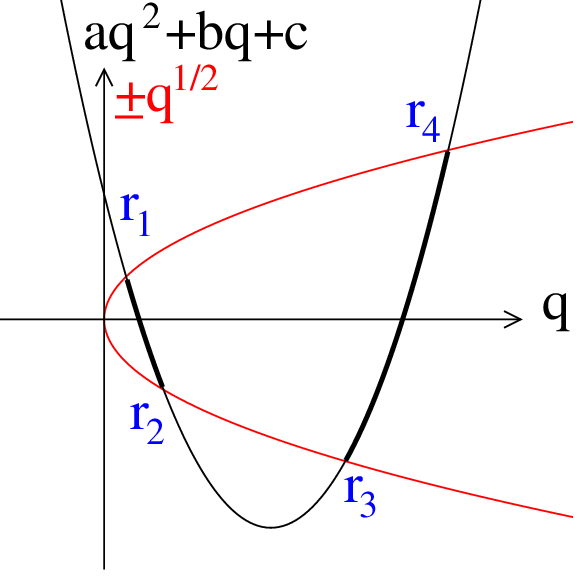,angle=270,scale=0.5} \hspace{1cm}
\parbox[t]{4cm}{\Figur{casei}{The $r_i$ in Fig. \ref{cases}i.}}
\et

The parobola in figure \ref{param}
\be
[(\epsilon-2)^2-\heps^2][(\epsilon+2)^2-\heps^2]
= (\epsilon^2-\heps^2)^2 -8(\epsilon^2+\heps^2-2) = 0
\ee
and the line $\heps^2=\epsilon^2$
decomposes the whole quadrant B $\heps^2>0$, $\epsilon^2>0$ into six regions. If
one allows an arbitrary assignment of the $r_i$ to $r_0(1\pm\epsilon)$ and
$r_0(-1\pm\heps)$ then one obtains 24 equivalent permutations. Since
only the squares of $\epsilon$ and $\heps$ are plotted, this reduces to six sets
of $(\epsilon^2,\heps^2)$, which are represented by the six regions B$_1$ to
B$_6$.

\subsection{Equations to be solved}

From equation (\ref{diffc}) one obtains
\be
\left(\frac{\de r}{\de\psi}\right)^2 = \frac {r^4}{(ar^4+br^2+c)^2}-r^2
\ee
and thus
\be
\frac{\de\psi}{\de q} = \frac{aq^2+bq+c}{2q\sqrt{q-(aq^2+bq+c)^2}} \label{psi}
\ee
with $q=r^2$. If one denotes the distance on the curve from some fixed point
(arc-parameter) by $u$ then
one obtains
\be
\frac{\de u}{\de\psi} = \sqrt{r^2+r^{\prime 2}}
=\frac 1{ar^2+b+cr^{-2}} = \frac 1{aq+b+cq^{-1}} \label{diffupsi}
\ee
and
\be
\frac{\de u}{\de q} = \frac 1{2\sqrt{q-(aq^2+bq+c)^2}}
\ee
First (sect. \ref{radius}) $q(u)$ will be calculated from
\be
\left(\frac{\de q}{\de u}\right)^2 = -4a^2 \prod_{i=1}^4 (q-q_i), \label{diffq}
\ee
\be
u = \int \frac{\de q}{2a\sqrt{-\prod_{i=1}^4(q-q_i)}},
\ee
since
\bea
(aq^2+bq+c)^2-q &=& (ar^4+br^2+c-r)(ar^4+br^2+c+r) \nn
&=& a^2 \prod_i(r-r_i)\prod_i(r+r_i) = a^2 \prod_i(q-q_i),
\eea
Secondly $\psi(u)$ (sect. \ref{angle}) is determined from
\be
\psi(u) = \int\de u (aq+b+cq^{-1}). \label{ipsiu}
\ee

Using eqs. (\ref{kappa1}, \ref{b}) the curvature is given by
\be
\kappa = 4ar^2+2b = a(4r^2-\sum_i r_i^2). \label{kappa}
\ee
One concludes that a convex curve oscillating
between $r_0(1+\epsilon)$ and $r_0(1-\epsilon)$ has to fulfill
\be
4\epsilon < \epsilon^2-\heps^2. \label{convex}
\ee
This condition guarantees for positive $\epsilon$ that the curvature does not
change sign.

\section{Radius as Function of Arc Parameter\label{radius}}

The solution of the differential equation (\ref{diffq}) reads
\be
q=q_4 \frac{\wp(u)-p_3}{\wp(u)-p_1} \label{qu}
\ee
with the Weierstrass $\wp$-function. To show this, one calculates the derivative
\bea
\frac{\de q}{\de u} &=& -q_4 (p_1-p_3) \frac{\wp'(u)}{(\wp(u)-p_1)^2} \\
&=& -\frac{2q_4(p_1-p_3)}{(\wp(u)-p_1)^2}
\sqrt{(\wp(u)-e_1)(\wp(u)-e_2)(\wp(u)-e_3)}.
\eea
From eq. (\ref{qu}) one obtains
\be
\frac 1{\wp(u)-p_1} = \frac 1{p_1-p_3}\left(\frac q{q_4}-1\right), \quad
\frac{\wp(u)-e_i}{\wp(u)-p_1} = \frac{p_1-e_i}{p_1-p_3}
\left(\frac q{q_4} - \frac{p_3-e_i}{p_1-e_i}\right).
\ee
Eq. (\ref{diffq}) is fulfilled provided the relations
\be
-4a^2 = \frac{4\prod_{i=1}^3(p_1-e_i)}{r_4^4(p_1-p_3)^2}, \quad
q_i = q_4 \frac{p_3-e_i}{p_1-e_i}, \quad i\not=4 \label{piei}
\ee
hold. The last eqs. and $\sum_{i=1}^3 e_i=0$ yield four linear homogeneous
eqs. for $p_1$, $p_3$ and the $e_i$. Together with the first equation one
obtains the solutions
\bea
p_1 &=& a^2 q_4 (-q_4+q_1+q_2+q_3) - \frac{a^2}3 P_{\rm m}, \label{p1} \\
p_3 &=& a^2 ( -\frac{q_1q_2q_3}{q_4} +q_1q_2+q_1q_3+q_2q_3)
- \frac{a^2}3 P_{\rm m}, \label{p3} \\
e_1 &=& a^2 (q_1q_4+q_2q_3)- \frac{a^2}3 P_{\rm m}, \label{e1} \\
e_2 &=& a^2 (q_2q_4+q_1q_3)- \frac{a^2}3 P_{\rm m}, \label{e2} \\
e_3 &=& a^2 (q_3q_4+q_1q_2)- \frac{a^2}3 P_{\rm m} \label{e3}
\eea
with the polynomial $P_{\rm m}$ as defined in eq. (\ref{Pm}).
Permutations of the $r_i$ yield permutations of the $e_i$.
Therefore the invariants $g_2=-4(e_1e_2+e_1e_3+e_2e_3)$ and $g_3=4e_1e_2e_3$ and
also $\wp(u)$ are invariant under all permutations of the $r_i$.

For future purposes the parameter $v$ is introduced by
\be
p_1=\wp(v),
\ee
which yields
\bea
\wp'(v) &=& \mp 2\ie a^3 (q_4-q_1)(q_4-q_2)(q_4-q_3), \label{wpvs}\\
\wp^{\dpr}(v) &=& 2a^4(q_4-q_1)(q_4-q_2)(q_4-q_3)(3q_4-q_1-q_2-q_3).
\eea

From these expressions one concludes
\bea
\wp(2v) &=& -\frac{a^2}4 \big(q_1^2+q_2^2+q_3^2+q_4^2\big)
+\frac{a^2}6 P_{\rm m}, \label{wp2v} \\
\wp'(2v) &=& \mp 2\ie a^3 P_q = \mp \frac{2\ie}{P_r} = \mp2\ie a\label{wp2vs}
\eea
with the polynomial $P_q$ defined in (\ref{Pq}) and by use of eqs. (\ref{a},
\ref{Prq}). These expressions are invariant under all permutations of the
$r_i$.

Using the condition $\sum_i r_i = 0$ one obtains
\be
p_3=\wp(3v). \label{wp3va}
\ee
To show this one starts from
\bea
\wp(3v) &=& -\wp(v)-\wp(2v) + \frac{(\wp'(v)-\wp'(2v))^2}{4(\wp(v)-\wp(2v))^2},
\\
\wp(v) &=& -\wp(v)-\wp(2v) + \frac{(\wp'(v)+\wp'(2v))^2}{4(\wp(v)-\wp(2v))^2},
\eea
from which one concludes
\be
\wp(3v) = \wp(v) - \frac{\wp'(v)\wp'(2v)}{(\wp(v)-\wp(2v))^2}. \label{wp3v}
\ee
Comparison shows
\be
\wp(v)-\wp(2v) = a^2 \big(-2\hat P + q_4(q_1+q_2+q_3-q_4)\big).
\ee
with the polynomial $\hat P$ defined in eq. (\ref{hatP}). 
By means of eqs. (\ref{r4Pr}, \ref{Prr}) one finds
\be
\wp(v)-\wp(2v) = -2 a^2 r_4 P_r
\ee
and
\be
(\wp(v)-\wp(2v))^2 = 4a^4 q_4 P_q.
\ee
Insertion into eq. (\ref{wp3v}) yields eq. (\ref{wp3va}).
One obtains the derivative $\wp'(3v)$ by means of (\ref{det})
\be
\wp'(3v) = -\frac{r_1r_2r_3}{r_4^3} \wp'(v), \label{wp3vs}
\ee
which by use of eq. (\ref{Prr}) can be written
\be
\wp'(3v) = -\frac{\hat P}{q_4^2} \wp'(v).
\ee

The solution (\ref{qu}) can be written
\be
q(u) = q_4 \cE 0 u {3v,-3v}{v,-v}
=q_4 \frac{\sigma(u+3v)\sigma(u-3v)}{\sigma(u+v)\sigma(u-v)}
\frac{\sigma^2(v)}{\sigma^2(3v)},
\ee
where the arc parameter $u$ is measured from the extreme radius $r=r_4$.
The notation $\cE n u{\{u_i\}}{\{v_i\}}$ for double-periodic functions of $u$ is
explained in appendix \ref{cE}. Apart from the origin it has zeroes at $u_i$
and poles at $v_i$ and behaves like $u^n$ at the origin.
Correspondingly the radius reads
\be
r(u) = C
\left(\frac{\sigma(u+3v)\sigma(3v-u)}{\sigma(u+v)\sigma(v-u)}\right)^{1/2},
\quad
C= r_4 \frac{\sigma(v)}{\sigma(3v)}
\ee
The prefactor $C$ is rewritten by means of the identity
\be
\frac{\sigma(3v)}{\sigma(v)\sigma^2(2v)} = \wp(v)-\wp(2v),
\ee
which can be seen by realizing that both the left and right hand-side of this
equation is double-periodic and can be written
\be
\frac 34 \cE{-2}v{\frac{2\omega_1}3,\frac{-2\omega_1}3,
\frac{2\omega_2}3,\frac{-2\omega_2}3,
\frac{2\omega_1+2\omega_2}3,\frac{-2\omega_1-2\omega_2}3,
\frac{2\omega_1-2\omega_2}3,\frac{2\omega_2-2\omega_1}3}
{\omega_1,-\omega_1,\omega_2,-\omega_2,\omega_1+\omega_2,-\omega_1-\omega_2}.
\ee
One obtains
\be
C = \frac{r_4}{\sigma^2(2v)(\wp(v)-\wp(2v))}
= \frac{-P_r}{2 \sigma^2(2v)}
= \frac{\ie}{\wp'(2v)\sigma^2(2v)}
= \frac{-\ie\sigma^2(2v)}{\sigma(4v)}, \label{CC}
\ee
where eqs. (\ref{wp2vs}, \ref{wpssigma}) have been used. Thus the factor $r_4$
drops out and the prefactor $C$ is invariant against any permutations of the
$r_i$.

\section{Single Branches and Pairs of Branches\label{onetwo}}

In the preceding section the radius square $q$ has been obtained as a function
of the arc parameter $u$ in eq. (\ref{qu}). At $u=0$ the function $\wp(u)$
diverges and $q$ approaches $q_4=r_4^2$. As $u$ is increased up to $\omega_3$
one obtains $\wp(\omega_3)=e_3$ and $q$ approaches its minimum
$q=q_4(e_3-p_3)/(e_3-p_1)=q_3$ along the curve. Then $q$ increases until at
$u=2\omega_3$ it returns to $q=q_4$. Instead starting at $u=0$ with
$q=q_4$ one can start at $u=\omega_3$ with $q=q_3$. Therefore consider
\be
q(u+\omega_i) = q_4 \frac{\wp(u+\omega_i)-\wp(3v)}{\wp(u+\omega_i)-\wp(v)},
\label{qplom}
\ee
with half-periods $\omega_i$, $i=1,2,3$. $\omega$ is called a
half-period, if it is not a period of $\wp$, but $2\omega$
is a period. The expression (\ref{qplom}) can be rewritten
\bea
q(u+\omega_i) &=& q_4 \cE 0{u+\omega_i}{3v,-3v}{v,-v} \nn
&=& q_4 \frac{\wp(\omega_i)-\wp(3v)}{\wp(\omega_i)-\wp(v)}
\cE 0u{3v-\omega_i,-3v-\omega_i}{v-\omega_i,-v-\omega_i} \nn
&=& q_4 \frac{e_i-p_3}{e_i-p_1}
\frac{\wp(u)-\wp(3v-\omega_i)}{\wp(u)-\wp(v-\omega_i)} \nn
&=& q_i \frac{\wp(u)-p_3^{(i)}}{\wp(u)-p_1^{(i)}}
\eea
with
\be
p_1^{(i)} = \wp(v^{(i)}), \quad p_3^{(i)} = \wp(3v^{(i)}), \quad
v^{(i)} = v-\omega_i.
\ee
Explicit calculation by means of
\be
\wp(v-\omega_i) = \frac{e_i^2+e_je_k+e_i\wp(v)}{\wp(v)-e_i},
\ee
where $j$ and $k$ are the elements out of (1,2,3) different from $i$,
shows that $p_1^{(i)}$ and $p_3^{(i)}$ are obtained from $p_1$ and $p_3$ by
exchanging $q_i$ and $q_4$ in the expressions (\ref{p1}) and (\ref{p3}).

The expression (\ref{wp2v}) for $\wp(2v)$ is invariant under permutations of the
$q_i$. Solving for $v$ from this equation yields besides $v$ from (\ref{p1})
also the $v^{(i)}$, $i=1,2,3$. In the following $v$ will be chosen, so that $\Im
\wp'(v) <0$. Then the upper sign in eqs. (\ref{wpvs}, \ref{wp2vs}) apply. Since
from (\ref{det}) one obtains with $u=-\omega_i$
\be
\wp'(v^{(i)}) = - \frac{\wp(v-\omega_i)-e_i}{\wp(v)-e_i} \wp'(v),
\ee
one finds that $\wp'(v^{(i)})$ is obtained from $\wp'(v)$ by exchanging $q_i$
and $q_4$ and one has also to take the upper sign.

\subsection{Single branch}

If two of the $r_i$ are complex and two are real, then one (real) curve is
obtained. The choice of complex $r_2=r_1^*$ and of real $r_4$ and $r_3$ yields
\be
q_2=q_1^*, \quad q_4>q_3.
\ee
Then $e_3$ is real, whereas $e_1$ and $e_2$ constitute a conjugate complex pair
\be
e_2=e_1^*, \quad e_3=e_3^*.
\ee
Consequently the discriminant
\be
{\cal D} = 16(e_3-e_1)^2 (e_3-e_2)^2 (e_2-e_1)^2 <0 \label{discr}
\ee
is negative. For negative discriminant $\cal D$ the periods and half-periods are
located in the complex plane as shown in fig. \ref{weierdn}
\bt
\epsfig{file=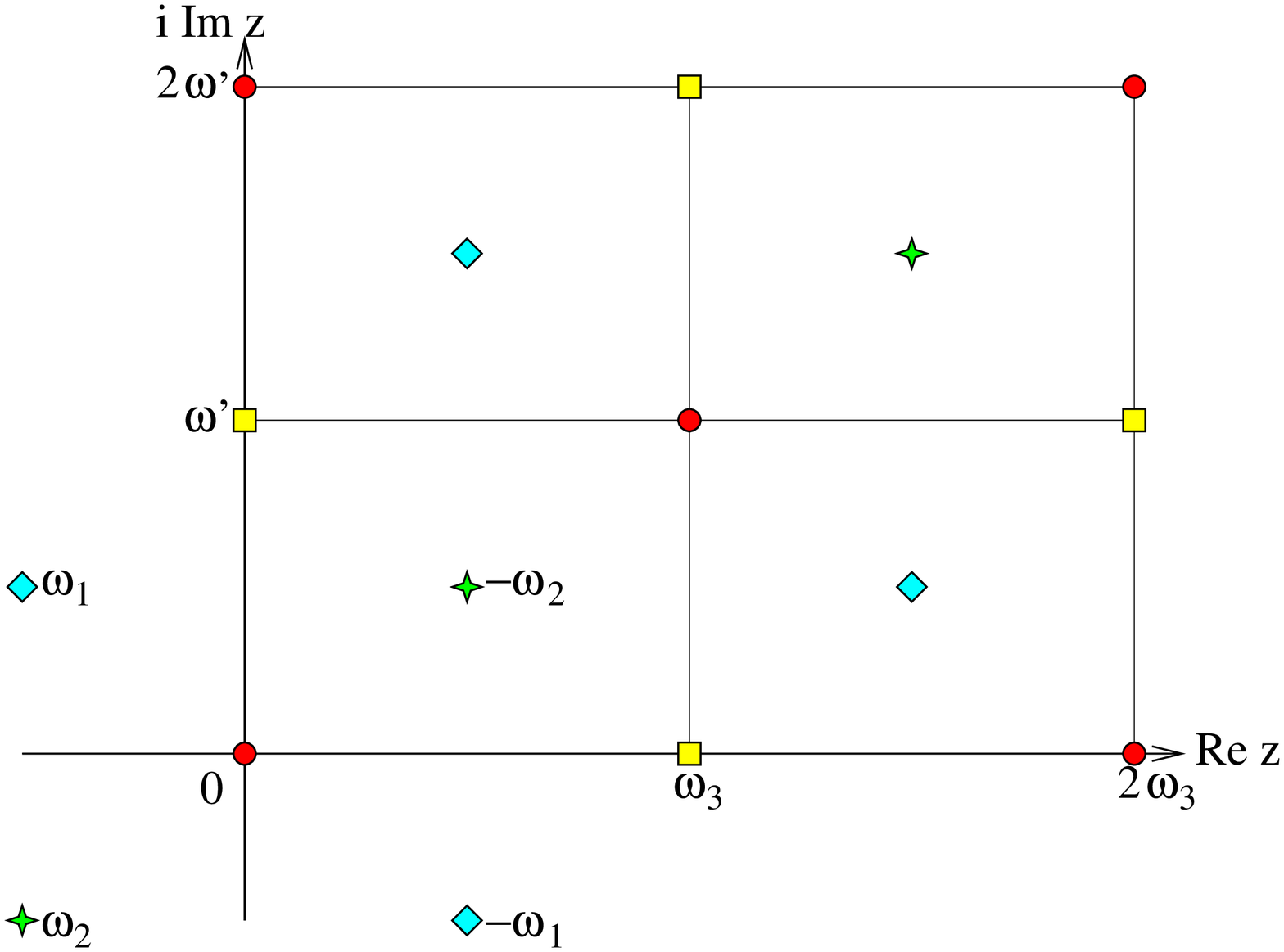,scale=0.4}\\
\Figur{weierdn}{Two ele\-men\-ta\-ry cells of
pe\-ri\-o\-di\-ci\-ty for negative discriminant
inside the rectangle.}
\et

The rectangle $0\le \Re z < 2\omega_3$,
$0\le \Im z < 2\omega'/\ie$ covers two elementary cells for the
double-periodic function $\wp(z)$. The
function is real along the straight lines drawn. For constant $\Im z$
(horizontal lines) it varies between $+\infty$ and $e_3$.
For constant $\Re z$ (vertical lines) it varies between $-\infty$ and $e_3$.
The singularities are at the points indicated by $\rb{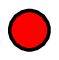}$,
$\wp(\rb{weierkr.eps})=\infty$.
The derivative of the function $\wp(z)$ vanishes at half-periods $\omega$.
There are three different half-periods which do
not differ by periods. They are indicated by $\rb{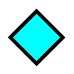}$,
$\rb{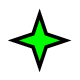}$, and $\rb{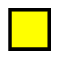}$. Thus one has $\wp'(\rb{weierrq.eps})
= \wp'(\rb{weierst.eps}) = \wp'(\rb{weierqu.eps})=0$. At these points
the function approaches $\wp(\omega_i)=e_i$. Examples are
$\wp(\omega_1)=\wp(-\omega_1)=\wp(\rb{weierrq.eps})=e_1$ with
$2\omega_1=\omega'-\omega_3$,
$\wp(\omega_2)=\wp(-\omega_2)=\wp(\rb{weierst.eps})=e_2$ with
$2\omega_2=-\omega'-\omega_3$,
$\wp(\omega_3)=\wp(\omega')=\wp(\rb{weierqu.eps})=e_3$.

Next consider the location of $p_1$. In section \ref{radius} the $p_1$ related
to $q_4$ was considered. To indicate this connection one may write
$p_1=p_1^{(4)}$ and generally introduce
\be
p_1^{(i)} = a^2 q_i (-2q_i+\sum_{j=1}^4 q_j) -\frac{a^2}3 P_{\rm m}. \label{p1i}
\ee
Here one is interested in the two real $p_1^{(4)}$ and $p_1^{(3)}$. One obtains
\bea
p_1^{(4)} -e_3 &=& -a^2(q_4-q_1)(q_4-q_2) <0, \\
p_1^{(3)} -e_3 &=& -a^2(q_3-q_1)(q_3-q_2) <0.
\eea
Thus one has
\be
p_1^{(4)}, p_1^{(3)} < e_3.
\ee
The real part of $v^{(i)}$ given by
\be
\wp(v^{(i)}) = p_1^{(i)} \label{vipi}
\ee
lies on the vertical lines in fig. \ref{weierdn}
\be
\Re v^{(4)}, \Re v^{(3)} = n\omega_3
\ee
with integer $n$. In the following the main choice for $v^{(4)}=v$ will be the
one with
\be
\Re v = 0, \quad 0<\Im v<\omega',
\ee
so that
\be
\Re \wp'(v)=0, \quad \Im\wp'(v)<0. \label{sign1}
\ee
in agreement with the requirement introduced before.
As mentioned before adding $\omega_3$ to $u$ is equivalent to exchanging $q_4$
and $q_3$. Both choices yield the same curve. To exchange $q_4$ with $q_1$ or
$q_2$ does not yield a positive $q$ and thus a real curve.

The $\omega_i$ are obtained from
\bea
\omega_3 = \frac{{\bf K}(\frac 12-\frac{3e_3}{4H_3})}{\sqrt{H_3}}, &&
\omega' = \ie\frac{{\bf K}(\frac 12+\frac{3e_3}{4H_3})}{\sqrt{H_3}}, \\
H_3 &=& \sqrt{(e_3-e_1)(e_3-e_2)},
\eea
where for the elliptic integral ${\bf K}(m)$ the notation of Abramowitz
and Stegun\cite{Abramowitz} is used. The reader is refered to chapter 17 of this
handbook for elliptic integrals and chapter 18 for Weierstrass elliptic
functions
(watch permutations of indices $i$ of $e_i$ and $\omega_i$ in
comparison to the use here), and to appendix \ref{Wei} of the present paper.

\subsection{Pair of branches}

If all four $r_i$ are real, then one obtains two branches. The $r_i$ should be
ordered according to
\be
q_4 > q_3 > q_2 > q_1.
\ee
Then the differences of the $e_1$, $e_2$, $e_3$ obey
\bea
e_3-e_2 &=& a^2(q_3-q_2)(q_4-q_1) > 0, \\
e_2-e_1 &=& a^2(q_4-q_3)(q_2-q_1) > 0
\eea
and thus
\be
e_3 > e_2 > e_1.
\ee

In this case, where all three $e_i$ are real, the discriminant $\cal D$, eq.
(\ref{discr}), is positive. Then the periods and half-periods are located in the
complex plane as shown in fig. \ref{weierdp}.

\bt
\epsfig{file=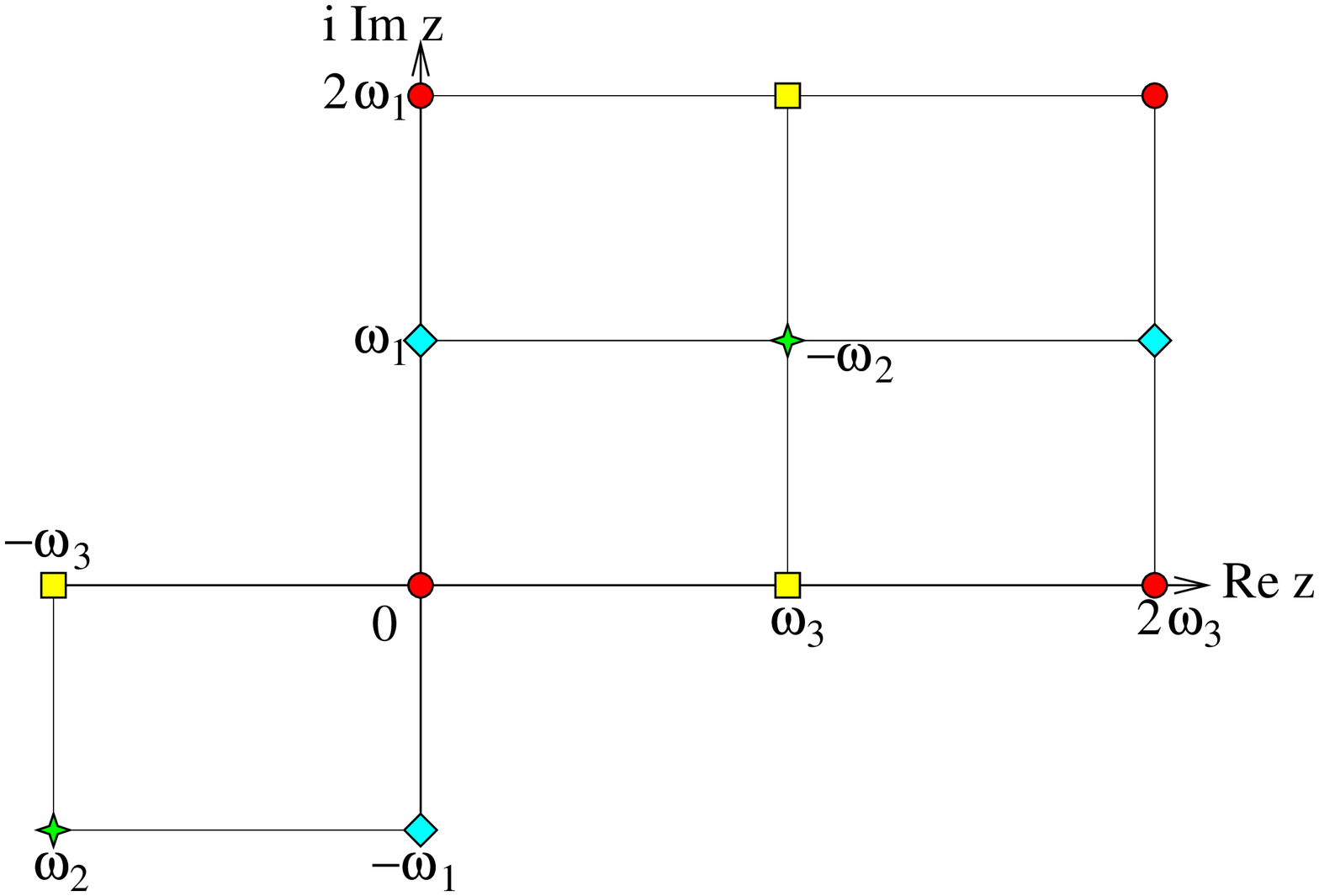,scale=0.5}\\
\Figur{weierdp}{One ele\-men\-ta\-ry cell of
pe\-ri\-o\-di\-ci\-ty inside the upper right rectangle for positive
discriminant.}
\et

One elementary cell of periodicity is given by the rectangle
$0\le\Re z<2\omega_3$, $0\le\Im z<2\omega_1/\ie$.
Again the function is real along the straight lines drawn. The function
approaches infinity at $\wp(\rb{weierkr.eps})=\infty$. The half-periods are
indicated by \rb{weierqu.eps}, \rb{weierrq.eps} and \rb{weierst.eps}. The
derivatives vanish at these half-periods,
$\wp'(\rb{weierqu.eps})=\wp'(\rb{weierrq.eps})=\wp'(\rb{weierst.eps})=0$. The
values of $\wp$ at the half-periods are denoted by $\wp(\omega_i)=e_i$,
$\wp(\omega_1)=\wp(\rb{weierrq.eps})=e_1$,
$\wp(\omega_2)=\wp(\rb{weierst.eps})=e_2$,
$\wp(\omega_3)=\wp(\rb{weierqu.eps})=e_3$.

One obtains the inequalities for the $p_1^{(i)}$ defined in (\ref{p1i})
\bea
p_1^{(4)} - e_1 &=& -a^2(q_4-q_2)(q_4-q_3) < 0, \\
p_1^{(1)} - e_1 &=& -a^2(q_2-q_1)(q_3-q_1) < 0, \\
p_1^{(3)} - e_2 &=& a^2(q_3-q_2)(q_4-q_3) >0, \\
p_1^{(3)} - e_3 &=& -a^2(q_3-q_1)(q_3-q_2) <0, \\
p_1^{(2)} - e_2 &=& a^2(q_2-q_1)(q_3-q_1) >0, \\
p_1^{(2)} - e_3 &=& -a^2(q_3-q_2)(q_4-q_2) <0.
\eea
Thus one has
\be
p_1^{(4)}, p_1^{(1)} < e_1 < e_2 < p_1^{(3)}, p_1^{(2)} < e_3.
\ee
Consequently one finds for the real part of $v^{(i)}$ defined in (\ref{vipi})
\be
\Re v^{(4)}, \Re v^{(1)} = 2n \omega_3, \quad
\Re v^{(3)}, \Re v^{(2)} = (2n+1) \omega_3
\ee
with integer $n$.
In the following the main choice for $v^{(4)}=v$ will be the one with
\be
\Re v = 0, \quad 0<\Im v<\omega_1,
\ee
so that as in the case of one curve eq. (\ref{sign1}) and the upper signs in
eqs. (\ref{wpvs}, \ref{wp2vs}) apply.
Adding $\omega_3$ to $u$ again yields the same curve with $u$ running through
\rb{weierkr.eps} and \rb{weierqu.eps} and the radius oscillating between
$\sqrt{q_4}$ and $\sqrt{q_3}$. However, adding $\omega_1$
or $\omega_2$ yields a different variation of the radius between $\sqrt{q_1}$
and $\sqrt{q_2}$. It corresponds to $u$ running along a horizontal line in fig.
\ref{weierdp} through \rb{weierrq.eps} and \rb{weierst.eps}, where $\wp$ is
again real.
This is equivalent to subtracting $\omega_1$ or $\omega_2$ from $v$. Thus in
this case one obtains two branches.

The $\omega_i$ are given by
\be
\omega_3 = \frac{{\bf K}(\frac{e_2-e_1}{e_3-e_1})}{\sqrt{e_3-e_1}}, \quad
\omega_1 = \ie\frac{{\bf K}(\frac{e_3-e_2}{e_3-e_1})}{\sqrt{e_3-e_1}}.
\ee

There are the special cases where two of the radii $r_i$ coincide. In these
cases the discriminant vanishes. In fig. \ref{cases}f one has $r_1=r_2$. This
case yields a small circle plus a curve with varying distance from the
origin. In this case the imaginary period goes to infinity. In fig.
\ref{cases}h, where $r_2=r_3$ there are two curves asymptotically approaching
the same circle. There the real period approaches infinity. The limit case is
shown in fig. \ref{cases}g, where $r_1=r_2=r_3$. In this case both periods
approach infinity. These limit cases are considered in section \ref{calD}.

\section{Angle and Curve as Function of the Arc Parameter\label{angle}}

\subsection{The angle}

The angle $\psi$ as function of $u$ is determined from eq. (\ref{ipsiu})
\be
\psi(u) = \int\de u (aq+b+cq^{-1}) 
= \int\de u \left(aq_4 \frac{\wp(u)-p_3}{\wp(u)-p_1} + b
+ \frac c{q_4} \frac{\wp(u)-p_1}{\wp(u)-p_3}\right).
\ee
With
\be
\int\de u \frac 1{\wp(u)-\wp(v)}
= \frac 1{\wp'(v)} \left( 2u\zeta(v) +
\ln\left(\frac{\sigma(u-v)}{\sigma(u+v)}\right)\right)
\ee
one obtains
\be
\psi(u) = \chi + c_0 u + c_1 \ln\left(\frac{\sigma(v-u)}{\sigma(v+u)}\right)
+ c_3 \ln\left(\frac{\sigma(3v-u)}{\sigma(3v+u)}\right)
\ee
with the real integration constant $\chi$ and
\bea
c_0 &=& aq_4+b+\frac c{q_4} + 2c_1\zeta(v) + 2c_3\zeta(3v), \\
c_1 &=& \frac{aq_4(p_1-p_3)}{\wp'(v)}, \\
c_3 &=& \frac{c(p_3-p_1)}{q_4\wp'(3v)}.
\eea
Evaluation of $c_1$ and $c_3$ yields
\be
c_1=c_3= - \frac{\ie}2,
\ee
where the sign convention (\ref{sign1}) is applied.
Further one obtains from (\ref{ricond})
\be
aq_4+b+\frac c{q_4} = \frac 1{r_4}.
\ee
This yields
\be
\psi(u) = \chi + c_0 u - \frac{\ie}2
\ln\left(\frac{\sigma(v-u)}{\sigma(v+u)}\right)
- \frac{\ie}2 \ln\left(\frac{\sigma(3v-u)}{\sigma(3v+u)}\right)
\ee
with
\be
c_0 = \frac 1{r_4} -\ie(\zeta(v)+\zeta(3v)).
\ee
Using
\be
\zeta(2v\pm v) = \zeta(2v) \pm\zeta(v)
+ \frac{\wp'(2v)\mp\wp'(v)}{2(\wp(2v)-\wp(v))}
\ee
one obtains
\be
\zeta(3v)+\zeta(v) = 2\zeta(2v) + \frac{\wp'(2v)}{\wp(2v)-\wp(v)}
=2\zeta(2v) -\frac{\ie 2a^3 P_q}{2a^2 P_r r_4}
=2\zeta(2v) -\frac{\ie}{r_4}
\ee
and thus
\be
c_0 = - 2\ie\zeta(2v),
\ee
which yields
\be
\ex{\ie\psi(u)} = \ex{\ie\chi} \left(\frac{\sigma(v-u)\sigma(3v-u)}
{\sigma(v+u)\sigma(3v+u)}\right)^{1/2} \ex{2u\zeta(2v)}. \label{expsi}
\ee

\subsection{The curve}

Combining the result for the radius and the angle one obtains the curve as a
function of the arc parameter $u$
\be
z(u) := x+\ie y = r(u) \ex{\ie\psi(u)} = \ex{\ie\chi} \frac{P_r}{2\sigma^2(2v)}
\frac{\sigma(u-3v)}{\sigma(u+v)} \ex{2u\zeta(2v)} \label{zu}
\ee
as claimed in (\ref{zu1}).

Changing simultaneously the sign of $u$ and $v$ does not change $z$,
\be
z(u,-v) = z(-u,v).
\ee
Thus changing the sign of $v$ is equivalent to changing the orientation of
measuring the arc-parameter.

Taking the complex conjugate of $z$ yields
\be
(\ex{-\ie\chi}z(u,v))^* = \ex{-\ie\chi}z(u,-v) = \ex{-\ie\chi}z(-u,v).
\ee
Thus the curve has a mirror axis which runs through the origin and
$\ex{\ie\chi}$.

By means of eqs. (\ref{zetaper}, \ref{sigmaper}) one finds that adding a period
$2\omega$ to $v$ does not alter $z$,
\be
z(u,v+2\omega) = z(u,v).
\ee

One obtains
\bea
\ex{\ie\psi(u+2\omega_3)} &=& \ex{\ie\psi(u)} \ex{\ie\psi_{\rm per}},
\label{exppsi1} \\
\ex{\ie\psi_{\rm per}} &=& \ex{-8v\zeta(\omega_3) + 4\omega_3 \zeta(2v)}
\label{exppsi}
\eea
Thus the curve is periodic under rotations by $\psi_{\rm per}$. This equation
determines the angle $\psi_{\rm per}$ only modulo $2\pi$.
Define
\be
\psi_{\rm c} := 4\pi + 8v\ie\zeta(\omega_3) - 4\ie\omega_3 \zeta(2v).
\label{psic}
\ee
Then $\ex{\ie\psi_{\rm per}}=\ex{\ie\psi_{\rm c}}$.
In sect. \ref{period} it will be shown that this angle equals $\psi_{\rm per}$
in
the region A' of fig. \ref{param}. There it will be discussed, how 
$\psi_{\rm per}$ and $\psi_{\rm c}$ are related in the other regions.

Adding a half period to $v$ yields
\be
\ex{\ie\psi_{\rm per}(v+\omega)}
= \ex{\ie\psi_{\rm per}(v)}
\ex{-8\ie\omega\zeta(\omega_3)+8\ie\omega_3\zeta(\omega)}
= \ex{\ie\psi_{\rm per}(v)},
\ee
where use has been made of the Legendre-relation (\ref{Legendre}).
Thus two curves with $v$ differing by a half-period have the same
$\ex{\ie\psi_{\rm per}}$.

\subsection{More on the curve\label{more}}

Instead of verifying that (\ref{zu}) obeys the differential equation
(\ref{diffc}) one may immediately start with the expression
\be
z=x+\ie y = C \frac{\sigma(u+v')}{\sigma(u+v)} \ex{\ie c_0 u}
\ee
and demonstrate that it is solution to the problem by first showing that $u$
is really the arc parameter provided one has the appropriate relations between
$v$, $v'$, and $c_0$, which requires
\be
\left|\frac{\de z}{\de u}\right|=1 \label{dzdu}
\ee
and secondly show that it is possible to find differences $2\du$ of the arc
parameter, so that the distance $|z(u+\du)-z(u-\du)|$ does not depend on $u$.
Here it will be shown that (\ref{dzdu}) is fulfilled with appropriate choices of
$v'$, $c_0$, and $C$ for given real $g_2$, $g_3$ and imaginary $v$. The
investigation of the property of constant distance is postponed to sect.
\ref{chord}. One obtains
\bea
\frac{\de z}{\de u} = z \Phi(u), &&
\Phi(u) = \zeta(u+v')-\zeta(u+v)+\ie c_0, \\
\frac{\de z^*}{\de u} = z^* \Phi^*(u), &&
\Phi^*(u) = \zeta(u+v'^*)-\zeta(u+v^*)+\ie c^*_0
\eea
for real $u$. Thus
\be
\frac{\de z}{\de u}\frac{\de z^*}{\de u}
= \Phi(u) \Phi^*(u) zz^* = 1
\ee
has to hold for real $u$ with
\be
zz^*=CC^*
\frac{\sigma(u+v')\sigma(u+v^{\prime *})}{\sigma(u+v)\sigma(u+v^*)}
\ex{\ie(c_0-c^*_0)u}.
\ee
Since in this representation $\Phi(u) \Phi^*(u) zz^*$ is holomorphic in
$u$, it has to be constant for complex $u$.
Since $\Phi(u)$ and $\Phi^*(u)$ are periodic functions
in $u$, also $zz^*$ has to have this property, which yields
\bea
\Phi(u) =c_u \cE 0 u {u_1,u_2}{-v,-v'}, &&
\Phi^*(u) =c^*_u \cE 0 u {u^*_1,u^*_2}{-v^*,-v^{\prime *}}, \nn
zz^*  &=& c_z \cE 0 u {-v',-v^{\prime *}}{-v,-v^*}, \label{cc}
\eea
where $u_1$, $u_2$ depend on the choice of $c_0$, and $c_u$ and $c_z$ are
constant prefactors.
Further $v'+v^{\prime *}$ can differ from $v+v^*$ only by an integer multiple of
$2\omega_3$.
The product of the three functions in (\ref{cc}) yields
\be
\cE 0 u{u_1,u_2,u^*_1,u^*_2}{-v,-v,-v^*,-v^*}.
\ee
In order that this function is constant one chooses modulo periods $2\omega$
\be
u_1=u_2=-v^*.
\ee
(The choice $u_1=-v$ or $u_2=-v$ would yield infinite $c_u$.) Thus
\be
\Phi(u) \propto \cE 0 u{-v^*,-v^*}{-v,-v'},
\ee
which requires
\be
v'=-v+2v^*+2\omega.
\ee
Thus one concludes
\be
\Re v' = \Re v + 2\Re\omega, \quad
\Im v' = -3\Im v +2\Im\omega.
\ee
The real part of $v$ can be absorbed into $u$. Thus one may choose $v$ purely
imaginary. Since the functions $\cal E$ are invariant against addition of a
period to any of its arguments, one may choose $\omega=0$, which yields
$v'=-3v$.
Since $\Phi(u)$ has to vanish for $u=-v^*$, one obtains
\be
\ie c_0 = 2\zeta(2v).
\ee
$\Phi(u)$ can be written
\be
\Phi(u) = \zeta(u-3v)-\zeta(u+v)+2\zeta(2v) =
-\frac{\wp'(2v)}{\wp(2v)-\wp(u-v)}, \label{Phiu}
\ee
where eq. (\ref{zetaadd2}) has been used. The prefactors in eqs. (\ref{cc}) are
\be
c_u = -\zeta(3v)+2\zeta(2v)-\zeta(v) =
\frac{-\sigma(v)\sigma(4v)}{\sigma^2(2v)\sigma(3v)}, \quad
c_z = C^2 \frac{\sigma^2(3v)}{\sigma^2(v)}.
\ee
Thus one has the condition
\be
c_u c^*_u c_z = -C^2 \frac{\sigma^2(4v)}{\sigma^4(2v)} = 1
\ee
in agreement with eq. (\ref{CC}).

\section{Line Segments of Constant Length\label{chord}}

\subsection{Property of constant distance\label{propdist}}

In this section the distance $2\ell$ between points at $u+\du$ and $u-\du$ on
the same or on different curves of equal or different branch is considered
\be
(2\ell)^2 = (z^*(u+\du)-\hat z^*(u-\du))(z(u+\du)-\hat z(u-\du)) \label{ell}
\ee
in order to find line segments of constant length $2\ell$ independent of $u$,
\bea
z(u) &=& C \ex{\ie\chi} \frac{\sigma(u-3v)}{\sigma(u+v)} \ex{2u\zeta(2v)},\\
\hat z(u) &=& \hat C \ex{\ie\hat\chi} \frac{\sigma(u-3\hat v)} {\sigma(u+\hat
v)} \ex{2u\zeta(2\hat v)}.
\eea
For simplicities sake assume that $v$ and $\hat v$ are purely imaginary.
Then the conjugate complex of $z$ and $\hat z$ is obtained simply by reversing
the
sign of $v$, $\hat v$, $\chi$ and $\hat\chi$.
Since $u$ and $\du$ are real, they are left unchanged upon determining the
conjugate complex of $z$ and $\hat z$ and only after this conjugation they are
extended into the complex plane. Thus one has
\bea
z^*(u) &=& C \ex{-\ie\chi} \frac{\sigma(u+3v)}{\sigma(u-v)} \ex{-2u\zeta(2v)},\\
\hat z^*(u) &=& \hat C \ex{-\ie\hat\chi} \frac{\sigma(u+3\hat v)} {\sigma(u-\hat
v)} \ex{-2u\zeta(2\hat v)}.
\eea
The expression $(2\ell)^2$ becomes an analytic function in $u$ and $\du$.
Choosing
\be
\hat v = v + \nu\omega_1
\ee
allows not only to consider line segments between copies of one and the same
curve
(even $\nu$), but also to connect two curves of different branches (odd $\nu$).

Similar to eq. (\ref{exppsi}) addition of $2\omega$ to the argument of
$u$ multiplies both $z$ and $\hat z$ by the same factor
$\ex{8v\zeta(\omega)-4\omega\zeta(2v)}$
and divides both $z^*$ and $\hat z^*$ by this factor. Thus $4\ell^2$ is a
periodic function in $u$.

Apparently $z^*(u+\du)$ has a pole at $u=v-\du$ and $\hat z^*(u-\du)$ a pole at
$u=\hat v+\du$. In order that $(2\ell)^2$ is constant, the other factor
$z(u+\du)-\hat z(u-\du)$
has to vanish for these values of $u$,
\bea
&& z(v)-\hat z(v-2\du) = \nn
&& - C \ex{\ie\chi} \ex{2v\zeta(2v)}
- \hat C \ex{\ie\hat\chi} \frac{\sigma(-2\du+v-3\hat v)}{\sigma(-2\du+v+\hat v)}
\ex{-4\du\zeta(2\hat v)+2v\zeta(2\hat v)} = 0, \\
&& z(\hat v+2\du) - \hat z(\hat v) = \nn
&& C \ex{\ie\chi} \frac{\sigma(2\du+\hat v-3v)}{\sigma(2\du+v+\hat v)}
\ex{4\du\zeta(2v)+2\hat v\zeta(2v)}
+ \hat C \ex{\ie\hat\chi} \ex{2\hat v\zeta(2\hat v)}= 0.
\eea
Both eqs. yield
\be
\ex{\ie(\chi-\hat\chi)} = (-1)^{1-\nu}
\ex{-2\du(\zeta(2v)+\zeta(2\hat v))}
\frac{\sigma(2\du+v+\hat v)}{\sigma(2\du-v-\hat v)}. \label{chdu}
\ee
Similarly $z(u+\du)$ has a pole at $u=-v-\du$ and $\hat z(u-\du)$ has a pole at
$u=-\hat v+\du$. It turns out that $z^*(u+\du)-z^*(u-\du)$ also vanish, if
(\ref{chdu}) is fulfilled.
Thus if condition (\ref{chdu}) between $\chi-\hat\chi$ and $\du$ is fulfilled,
then the distance $2\ell$ does not depend on the arc parameter $u$. This proves
the property of constant distance.

Next the number of solutions will be determined. First one realizes that
$\chi-\hat\chi$ is a monotonic increasing function of $\du$, which can be
seen as follows.
The derivative
\be
\frac{\de(\chi-\hat\chi)}{\de\du} = 2\ie \Big(\zeta(2v)+\zeta(2\hat v)
-\zeta(2\du+v+\hat v)+\zeta(2\du-v-\hat v)\Big)
\ee
can be rewritten
\be
\frac{\de(\chi-\hat\chi)}{\de\du} = \left\{\begin{array}{cc}
\frac{2\ie\wp'(2v)}{\wp(2\du)-\wp(2v)} & \mbox{for } \hat v=v, \\
\frac{-2\ie\wp'(2v+\omega_1)}{e_1-\wp(2v+\omega_1)}
\frac{\wp(2\du)-e_1}{\wp(2\du)-\wp(2v+\omega_1)} & \mbox{for } \hat
v=v+\omega_1.
\end{array}\right.
\ee
One convinces oneself that all numerators and denominators are positive. Thus
$\chi-\hat\chi$ is a monotonic increasing function of $\du$.
Next consider the change of $\chi-\hat\chi$ during one period, that is while
increasing $\du$ by $\omega_3$,
\be
\Delta\chi=\Delta\arg\sigma(2\du+v+\hat v)-\Delta\arg\sigma(2\du-v-\hat v)
+2\ie\omega_3(\zeta(2v)+\zeta(2\hat v)).
\ee
Since $0<\Im(2v)<\omega'/2$ and $0<\Im(2v)<\omega_1$ one obtains from
eqs. (\ref{argsigma}, \ref{argsigma1})
\be
\Delta\chi= -2\pi -4\ie(v+\hat v)\zeta(\omega_3)
+2\ie\omega_3(\zeta(2v)+\zeta(2\hat v))
=2\pi-\frac{\psi_{\rm c}(v)+\psi_{\rm c}(\hat v)}2.
\ee
If curves are considered, which close after $n$ periods, that is for which
\be
\psi_{\rm c}(v) = 2\pi\frac mn \label{psicmn}
\ee
with $m,n$ coprime holds, then $\chi-\hat\chi$ increases by
\be
n\Delta\chi = 2\pi \times \left\{ \begin{array}{cc}
n-m & \mbox{for } \hat v=v, \\
2n-m & \mbox{for } \hat v=v+\omega_1,
\end{array} \right.
\ee
since from eqs. (\ref{psic}) and (\ref{Legendre}) one obtains
\be
\psi_{\rm c}(v+\omega_1) = \psi_{\rm c}(v) -4\pi.
\ee
The numbers $n-m$ and $2n-m$ are the number of solutions $\du$ for given
$\chi-\hat\chi$ after they repeat, since with $\du$ also $\du+n\omega_3$ is a
solution, which yields line segments between the same points of curves. If
chords on one and the same curve are considered ($\hat v=v$, $\hat\chi=\chi$),
which
constitutes the original problem, then one solution is the trivial solution
$\du=0$. Thus one is left with only $n-m-1$ non-trivial solutions, which
moreover appear pairwise, since with $\du$ also $-\du$ is a solution. For
$m/n=1/2$, for example one has only the trivial solution. If however two copies
of the curve are rotated against each other, then a set of line segments of
constant length is obtained.
Examples are shown in figures \ref{fgp1p2b} and \ref{fgp1p2c}.
The asterisk at the figure number indicates that there is a corresponding
animation on the internet\cite{movie}, where one can see the chord or segment of constant length moving.

\bt
\parbox[b]{5cm}{\epsfig{file=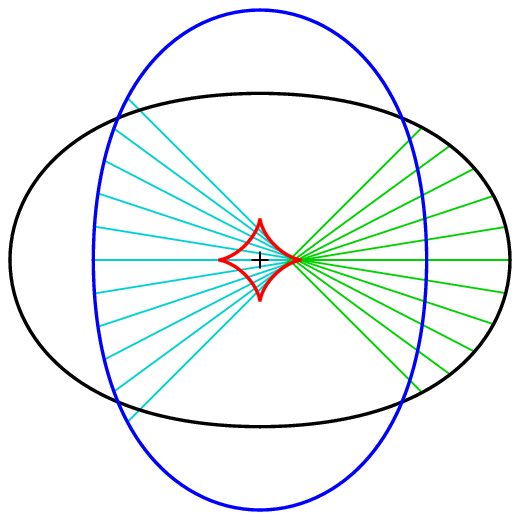,scale=0.5}\\
\Figurm{fgp1p2b}{$m/n=1/2$, $\epsilon=0.2$}}
\hfill
\parbox[b]{5cm}{\epsfig{file=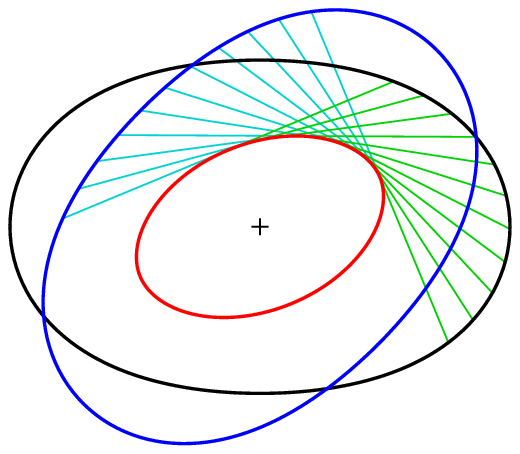,scale=0.5}\\
\Figurm{fgp1p2c}{$m/n=1/2$, $\epsilon=0.2$}}
\et

Here and in the following the curves $\Gamma$ are shown in black and blue, the
chords and line-segments are shown in green and cyan. The color switches in the
middle, where the lines touch the red envelopes $\gamma$.

\subsection{Length of the line segments}

Next a simple expression for the length of the line segments is derived assuming
that $\du$ has been determined. Since $4\ell^2$ is independent of $u$, one
considers the expression in the vicinity of diverging $z^*(u+\du)$, that is
around
$u=-\du+v$. There the residuum is given by
\be
{\rm res} = C \ex{-\ie\chi} \sigma(4v) \ex{-2v\zeta(2v)}.
\ee
This divergence is compensated by the zero of $z(u+\du)-\hat z(u-\du)$. In order
to obtain $4\ell^2$ the residuum has to be multiplied by the derivate
\bea
&& \left.\frac{\de z(u+\du)}{\de u}\right|_{u=-\du+v}
-\left.\frac{\de\hat z(u-\du)}{\de u}\right|_{u=-\du+v} \\
&=& \Phi(v)z(v) - \hat\Phi(-2\du+v)\hat z(-2\du+v)
\eea
Using (\ref{Phiu}) one obtains
\be
\Phi(v) = 0, \quad
\hat\Phi(-2\du+v) = \frac{-\wp'(2\hat v)}{\wp(2\hat v)-\wp(-2\du+v-\hat v)}.
\ee
Thus the residuum has to be multiplied by $-\hat\Phi(-2\du+v)\hat z(-2\du+v)$,
and since $\hat z(-2\du+v)=z(v)$, one obtains
\be
4\ell^2 = {\rm res} \cdot (-z(v) \hat\Phi(-2\du+v))
= -C^2 \sigma(4v) \frac{\wp'(2\hat v)}{\wp(2\hat v)-\wp(-2\du+v-\hat v)}.
\ee
Making use of eq. (\ref{CC}) and that $2v$ and $2\hat v$ differ by a
period $2\omega$, one obtains finally
\be
4\ell^2 = \frac 1{\wp(2\du+v-\hat v)-\wp(2v)}. \label{ell0}
\ee
This expression yields bounds on the length of the line segment. Within one
and the same curve one uses
\be
e_3\le\wp(2\du)\le\infty
\ee
For $\wp(2\du)=\infty$ one has $4\ell^2=0$. In the other limit case one finds
\be
e_3-\wp(2v) = \frac{a^2}4 (q_4+q_3-q_2-q_1)^2
= \frac 1{(r_1+r_2)^2} = \frac 1{(r_3+r_4)^2},
\ee
where eqs. (\ref{a}), (\ref{Pr}), and (\ref{qq}) have been used. This yields the
bounds
\be
0 \le 2\ell \le |r_1+r_2| = r_3+r_4. \label{bound1}
\ee
In the case of two branches being connected by a line segment one has
\be
e_1\le\wp(2\du+\omega_1)\le e_2.
\ee
Then the bounds read
\be
|r_1+r_3| = r_2+r_4 \le 2\ell \le |r_2+r_3| = r_1+r_4. \label{bound2}
\ee

There are obvious geometric bounds. For line segments between points on the same
curve
one has the upper bounds $r_4+|r_3|$ and if there is a second branch also
$|r_2|+|r_1|$. Lower bound is 0. For line segments between different branches
one has
upper bounds $r_4+|r_1|$ and $|r_3|+|r_2|$ and lower bounds $r_4-|r_2|$ and
$|r_3|-|r_1|$. It turns out that the limits given by eq. (\ref{ell0}) are in
some cases more restrictive. They are given in the tables (\ref{tbl1}) and
(\ref{tbl2}).

\bt
\be
\begin{array}{c|cccc|cc}
\mbox{Region} & \multicolumn{4}{c}{\mbox{Sign of}} &
\multicolumn{2}{c}{\mbox{Same curve}} \\
 & r_4 & r_3 & r_2 & r_1 & 2\ell_{\rm min} & 2\ell_{\rm max} \\
 &&&&& 0 & 2r_0 \\ \hline
 A' & + & + && & 0 & r_4+r_3 \\
 A\dpr & + & - && & 0 & r_4-|r_3| \\
 B' & + & - & - & - & 0 & r_4-|r_3|=|r_1|+|r_2| \\
 B\dpr & + & - & - & + & 0 & r_4-|r_3|=|r_2|-r_1  
\end{array} \label{tbl1}
\ee
\be
\begin{array}{c|cc}
\mbox{Region} & \multicolumn{2}{c}{\mbox{Different curves}} \\
 & 2\ell_{\rm min} & 2\ell_{\rm max} \\
 & r_0(\epsilon-\heps) & r_0(\epsilon+\heps) \\ \hline
 B' & r_4-|r_2| = |r_1|+|r_3| & r_4-|r_1| = |r_3|+|r_2| \\
 B\dpr & r_4-|r_2|=|r_3|-r_1 & r_4+r_1 = |r_3|+|r_2| 
\end{array} \label{tbl2}
\ee
\et

\subsection{Remark on the approach by Finn\label{RemFinn}}

David Finn considers the bicycle problem in \cite{Finn,FinnI}. Denoting the
curvature of trace of the rear wheel by $\kappa_{\beta}$ he obtains for the
curvature of the trace of the front wheel
\be
\kappa(u\pm\du) = \frac{\kappa_{\beta}}{(1+(\ell\kappa_{\beta})^2)^{1/2}}
\pm\ell\frac{\frac{\de\kappa_{\beta}}{\de u}}{1+(\ell\kappa_{\beta})^2},
\ee
where the two signs apply to the bicycle riding in both directions. He gives
$\kappa_{\beta}$ for an intervall $2\du$ and continues so that this equation is
fulfilled. If one eliminates $\kappa_{\beta}$ then one obtains
\bea
\frac{\de(\kappa(u+\du)+\kappa(u-\du))}{\de u}
&=& (\kappa(u+\du)-\kappa(u-\du))\nn
&\times& \sqrt{\frac 1{\ell^2} -\frac 14 (\kappa(u+\du)+\kappa(u-\du))^2}.
\eea
Performing some algebra one sees that this equation is fulfilled by our
solution. To do this one may express $\kappa$ and $\ell$ in terms of $\wp(u)$,
$\wp(\du)$ and $\wp(v)$.

\subsection{The area}

Next the area $\cA_2$ below the water-line is determined, which is the area of
the
sector $\cA_{\rm s}$ from the origin to the end points of the water-line with
arc parameters $u-\du$ and $u+\du$
minus the area $\cA_{\rm t}$ of the triangle with corners at the origin and at
the end-points of the water-line,
\be
\cA_2 = \cA_{\rm s} - \cA_{\rm t}.
\ee
The area of the sector is obtained by integrating
\be
\de\cA_{\rm s} = \frac 12 (x\de y-y\de x) = \frac 1{4\ie}(z^*\de z-z\de z^*).
\ee
Use of $\de z=z\Phi(u)\de u$, $\de z^*=z^*\Phi^*(u)\de u$,
$zz^*\Phi(u)\Phi^*(u)=1$ from section \ref{more} yields
\be
\de\cA_{\rm s} = \frac 1{4\ie} zz^*(\Phi(u)-\Phi^*(u))\de u
=\frac 1{4\ie} \left(\frac 1{\Phi^*(u)}-\frac 1{\Phi(u)}\right)\de u.
\ee
Insertion of eq. (\ref{Phiu}) and use of eq. (\ref{wp2vs}) yields
\bea
\cA_{\rm s} &=& \int_{u-\du}^{u+\du} \frac{\de u}{8a}
(2\wp(2v)-\wp(u+v)-\wp(u-v)) \\
&=& \frac 1{8a} \big(4\du\wp(2v) +\zeta(u+\du+v) +\zeta(u+\du-v) \nn
&& -\zeta(u-\du+v)-\zeta(u-\du-v)\big) \\
&=&  \frac 1{8a} \left( 4\du\wp(2v)
+2\zeta(\du+v)+ 2\zeta(\du-v) \right.\nn
&& \left. + \frac{\wp'(\du+v)}{\wp(\du+v)-\wp(u)}
+\frac{\wp'(\du-v)}{\wp(\du-v)-\wp(u)} \right).
\eea
The area $\cA_{\rm per}$ of the sector by increasing $u$ by $2\omega_3$, that is
$\du=\omega_3$ is given by
\be
\cA_{\rm per} = \frac 1{2a}(\omega_3\wp(2v)+\zeta(\omega_3)).
\ee
Then the density $\rho$ for a cross-section with $\psi_{\rm per}=2\pi/n$ is
given by
\be
\rho=\frac{\cA_2}{n\cA_{\rm per}}.
\ee
The area of the triangle spanned by the origin and the points on the curve at
arc parameters $u+\du$ and $u-\du$ is given by
\bea
\cA_{\rm t} &=& \frac 12 \big(x(u-\du) y(u+\du) - x(u+\du) y(u-\du) \big) \\
&=& \frac 1{4\pi\ie} \big( z^*(u-\du) z(u+\du) - z^*(u+\du) z(u-\du) \big) \\
&=& \frac{C^2}{4\ie}
\left(\frac{\sigma(u-\du+3v)\sigma(u+\du-3v)}{\sigma(u-\du-v)\sigma(u+\du+v)}
\ex{4\du\zeta(2v)} \right. \nn
&& \left.
-\frac{\sigma(u+\du+3v)\sigma(u-\du-3v)}{\sigma(u+\du-v)\sigma(u-\du+v)}
\ex{-4\du\zeta(2v)}\right) \\
&=& \frac{C^2}{4\ie} \left( \frac{\sigma^2(\du-3v)}{\sigma^2(\du+v)}
\frac{\wp(u)-\wp(\du-3v)}{\wp(u)-\wp(\du+v)} \ex{4\du\zeta(2v)} \right. \nn
&& \left. -\frac{\sigma^2(\du+3v)}{\sigma^2(\du-v)}
\frac{\wp(u)-\wp(\du+3v)}{\wp(u)-\wp(\du-v)} \ex{-4\du\zeta(2v)}\right).
\eea
In the last expression one rewrites
\be
\frac{\sigma^2(\du\mp 3v)}{\sigma^2(\du\pm v)}
= \frac{\sigma(2\du\mp 2v)}{\sigma(2\du\pm 2v)}
\frac{\wp'(\du\pm v)\sigma(\mp 4v)}{\wp(\du\mp 3v)-\wp(\du\pm v)},
\ee
where numerator and denominator on the left hand side have been multiplied by
$\sigma^2(\du\pm v)$ and eqs. (\ref{sigmaadd}, \ref{wpssigma}) have been used.
Use of eqs. (\ref{wp2vs}, \ref{CC}) yields
\be
\cA_2 = \cA_0 + \frac{\wp'(\du+v)}{8a(\wp(\du+v)-\wp(u))}(1+f_+)
+ \frac{\wp'(\du-v)}{8a(\wp(\du-v)-\wp(u))} (1+f_-)
\ee
with
\bea
\cA_0 &=& \frac 1{8a}
\left(4\du\wp(2v)+2\zeta(\du+v)+2\zeta(\du-v)
\right. \\
&& \left. +f_+ \frac{\wp'(\du+v)}{\wp(\du-3v)-\wp(\du+v)}
+f_- \frac{\wp'(\du-v)}{\wp(\du+3v)-\wp(\du-v)} \right), \nn
f_{\pm} &=& \frac{\sigma(2\du\mp 2v)}{\sigma(2\du\pm 2v)}\ex{\pm 4\du\zeta(2v)}.
\eea
In order that the area $\cA_2$ does not depend on $u$, the factors $1+f_{\pm}$
have to vanish. This yields exactly the condition (\ref{chdu}) with 
$\chi=\hat\chi$, $\nu=0$, $\hat v=v$ for the chord of
constant length. This was expected. The constant area below
the water-line is then given by $\cA_2=\cA_0$ with $f_{\pm}=-1$.
The expression for the area $\cA_0$ can be simplified. One obtains
\be
\frac{\wp'(\du\pm v)}{\wp(\du\mp 3v)-\wp(\du\pm v)}
= 2\zeta(\du\pm v) - \zeta(2\du\mp 2v) \mp \zeta(4v)
\ee
by using theorem I of appendix \ref{cE}. The functions on both sides of the
equation are periodic in $\du$, their singularities within an elementary
cell are simple poles at $\du=\mp v$ with residua 2, and at $\du=\pm v$ and
$\du=\pm v+\omega_i$ with residua $-1/2$. The constant $\mp\zeta(4v)$ is
obtained by evaluating both sides at $\du=\mp v+\omega_i$. Thus one obtains
\be
\cA_2=\cA_0 = \frac 1{8a}
\left(4\du\wp(2v)+\zeta(2\du-2v)+\zeta(2\du+2v)\right).
\ee

\subsection{Centers of gravity}

Let $z_{\rm g2}$ denote the center of gravity of the area below the water-line
and $\phi$ denote the angle of $z(u+\du)-z(u-\du)$ against the real axis in the
complex $z$-plane. Thus
\be
z(u+\du)-z(u-du) = 2\ell \ex{\ie\phi}. \label{zpm}
\ee
An infinitesimal change of $\phi$ yields
\be
\cA_2 \de z_{\rm g2} = \frac 23 (z(u+\du)-z(u-du)) \de f,
\ee
which with $\de f=\ell^2/2$ and eq. (\ref{zpm}) yields
\be
\cA_2 \de z_{\rm g2} = \frac 23 \ell^3 \ex{\ie\phi} \de \phi
\ee
and thus
\be
z_{\rm g2} = \frac{2\ell^3}{3\cA_2} \left( -\ie \ex{\ie\phi} +{\rm const}_2
\right).
\ee
Similarly one obtains for the center of gravity $z_{\rm g1}$ of the area above
the water-line
\be
z_{\rm g1} = \frac{2\ell^3}{3\cA_1} \left( \ie \ex{\ie\phi} +{\rm const}_1
\right).
\ee
Thus both centers of gravity move on circles. This is a necessary and sufficient condition for the solution (see Auerbach\cite{Auerbach} and references therein).
Increasing $u$ by $2\omega_3$ will
increase $\phi$ by $\psi_{\rm per}$, and the center of gravity is rotated by
$\psi_{\rm per}$. If it is not an integer multiple of $2\pi$, then both
constants vanish. If it is such a multiple, then continuity normally shows that
it will vanish, too. We realize that then the line between the centers of
gravity is perpendicular to the water-line and the distance $h$ between both
centers agrees with the expression (\ref{h}). This derivation did not use the
solution of (\ref{diffc}).

\section{Periodicity and Figures\label{period}}

\subsection{Limit of small $\epsilon$}

In the limit of small $\epsilon$ one obtains (in this subsection we put $r_0=1$)
\be
e_{1,2} = -\frac{4+\mu^2}{12\mu^2} \pm \frac{2\ie\epsilon}{\mu^3}
+O(\epsilon^2), \quad
e_3 =\frac{4+\mu^2}{6\mu^2}+O(\epsilon^2).
\ee
Thus for $\epsilon=0$ the limiting case of vanishing discriminant is approached
with $\omega'$ tending to $\ie\infty$ and
\be
\omega_3 = \frac{\pi\mu}{\sqrt{4+\mu^2}}
\ee
from
\be
e_3 = \left(\frac{\pi}{2\omega_3}\right)^2\frac 23.
\ee
In order to determine $\omega'$ one may use the representation of the
Weierstrass functions in appendix \ref{Weier}. It is sufficient to use the
terms up to linear order in $q$. With
\be
q=\ex{2\pi\ie\omega_1/\omega_3} = -\ex{\pi\ie\omega'/\omega_3}
=-\hat q^2, \quad \hat q=\ex{\pi\ie\omega'/2\omega_3}
\ee
one obtains
\be
\wp(z) = \left(\frac{\pi}{2\omega_3}\right)^2
\left(\frac 2{1-\cos u} +8q(1-\cos u) -\frac 13\right), \quad
u=\frac{\pi z}{\omega_3}.
\ee
For argument $\frac{\omega'}2+\delta z$ and small imaginary part of
$\delta z$ one obtains
\be
\wp(\frac{\omega'}2+\delta z) = \left(\frac{\pi}{2\omega_3}\right)^2
\left(-\frac 13-8\ie\hat q\sin\frac{\pi\delta z}{\omega_3}\right).
\ee
One concludes from $e_1=\wp(\frac{\omega'-\omega_3}2)$
\be
\hat q = \frac{\epsilon}{\mu(4+\mu^2)}.
\ee
Next $v=\frac{\omega'}2+\delta v$ is determined from
\be
p_1 = \wp(v) = -\frac{\mu^2+4}{12\mu^2} - \frac{\epsilon}{\mu^2}
\ee
and yields
\be
\sin(\frac{\pi\delta v}{\omega_3})= -\frac{\ie\mu}2.
\ee
Now $\psi_{\rm c}$, eq. (\ref{psic}) can be evaluated with
$\omega'=\omega_3+2\omega_1$
\bea
\psi_{\rm c} &=& 4\pi +8v\ie \zeta(\omega_3) -4\ie \omega_3\zeta(2v) \nn
&=& 4\pi +\ie(4\omega'+8\delta v)\zeta(\omega_3)
-4\ie\omega_3\zeta(\omega'+2\delta v) \nn
&=& 4\pi +\ie(4\omega_3+8\omega_1+8\delta v)\zeta(\omega_3)
-4\ie\omega_3\zeta(\omega_3+2\omega_1+2\delta v) \\
&=& 4\pi +8\ie(\omega_1\zeta(\omega_3)-\omega_3\zeta(\omega_1))
+\ie(4\omega_3+8\delta v)\zeta(\omega_3)
-4\ie\omega_3\zeta(\omega_3+2\delta v). \nonumber
\eea
The term $8\ie(\omega_1\zeta(\omega_3)-\omega_3\zeta(\omega_1))$
in the last line yields $-4\pi$ due to Legendre's relation
(\ref{Legendre}), the second one can be evaluated by means of eq. (\ref{repz}),
where $q$ is now negligible and yields
\be
\psi_{\rm c}
=-2\pi\ie \cot\left(\frac{\pi}2+\frac{\pi\delta v}{\omega_3}\right)
=\pi\ie \tan\left(\frac{\pi\delta v}{\omega_3}\right)
= 2\pi\frac{\mu}{\sqrt{4+\mu^2}}.
\ee
Thus one obtains
\be
\psi_{\rm per} = 2\omega_3=2\pi\mu/\sqrt{4+\mu^2} = \psi_{\rm c}
\ee
in the limit of small $\epsilon$.

Two dimensional bodies which can float in all directions are given by
$\psi_{\rm per}=2\pi/n$, thus for $m=1$ and sufficiently small $\epsilon$.
In this limit the $\du$ can be determined from eq. (\ref{chdu}) with
\bea
\zeta(\omega'+\delta z) &=& \frac{n^2}{12} (\omega'+\delta z)
-n\ie -\frac n2 \tan(\frac{n\delta z}2) + O(q), \\
\sigma(\omega'+\delta z) &=& \frac{2i}{n\hat q} \ex{-\ie n\delta z}
\cos(\frac{n\delta z}2) \ex{n^2(\omega'+\delta z)^2/24} + O(\hat q),
\eea
where eqs. (\ref{repz}) and (\ref{reps}) and $\frac{\pi}{\omega_3}=n$,
$\frac{\eta_3}{\omega_3}=\frac{n^2}{12}$ have been used. Then eq. (\ref{chdu})
yields
\be
\tan(n\du) = n \tan(\du),
\ee
in agreement with the results obtained in refs. \cite{WegnerI,WegnerII,Wegner},
where $\du$ corresponds to $\frac{\pi}2-\delta_0$ and in ref.
\cite{Tabachnikov}, where $\du$ corresponds to $\pi\rho$.

A few cross-sections of the bodies are
shown in figs. \ref{fgp1p3a} to \ref{fgp1p7b}. For odd $n$ the innermost
envelope corresponds to density $\rho=1/2$.
\medskip

\btm
\parbox[b]{3.2cm}{\epsfig{file=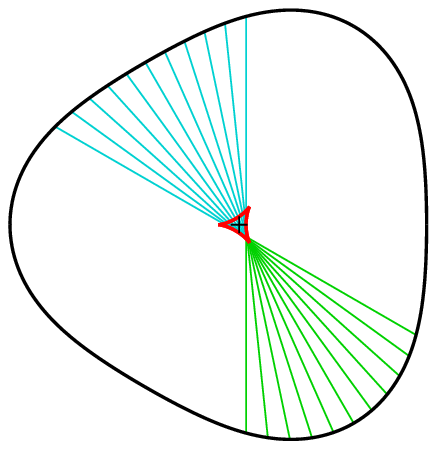,scale=0.5,angle=90}\\
\Figur{fgp1p3a}{$m/n=1/3$, $\epsilon=0.1$}}
\hfill
\parbox[b]{3.2cm}{\epsfig{file=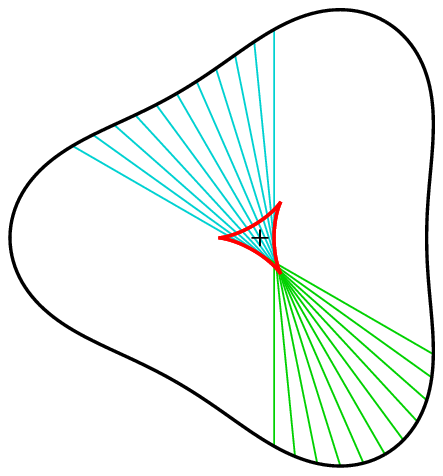,scale=0.5,angle=90}\\
\Figur{fgp1p3b}{$m/n=1/3$, $\epsilon=0.2$}}
\hfill
\parbox[b]{3.2cm}{\epsfig{file=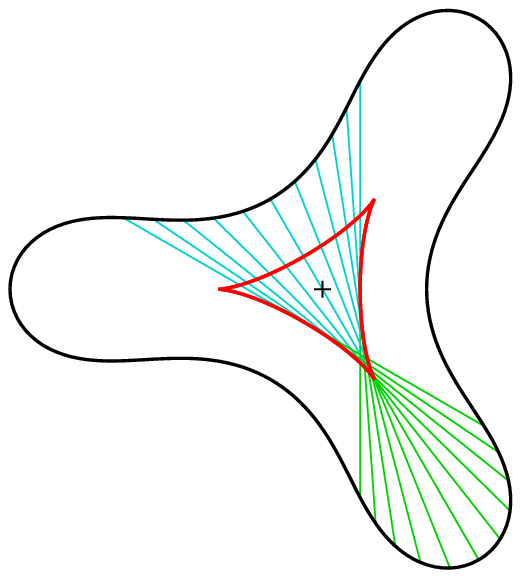,scale=0.5,angle=90}\\
\Figur{fgp1p3c}{$m/n=1/3$, $\epsilon=0.5$}}
\etm

\btm
\parbox[b]{3.2cm}{\epsfig{file=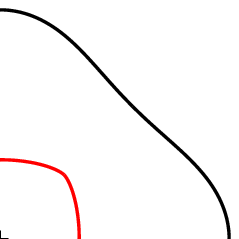,scale=0.5}\\
\Figur{fgp1p4a}{$m/n=1/4$, $\epsilon=0.1$}}
\hfill
\parbox[b]{3.2cm}{\epsfig{file=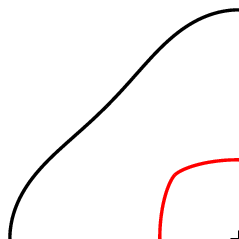,scale=0.5,angle=90}\\
\Figurm{fgp1p4b}{$m/n=1/4$, $\epsilon=0.1$}}
\hfill
\parbox[b]{3.2cm}{\epsfig{file=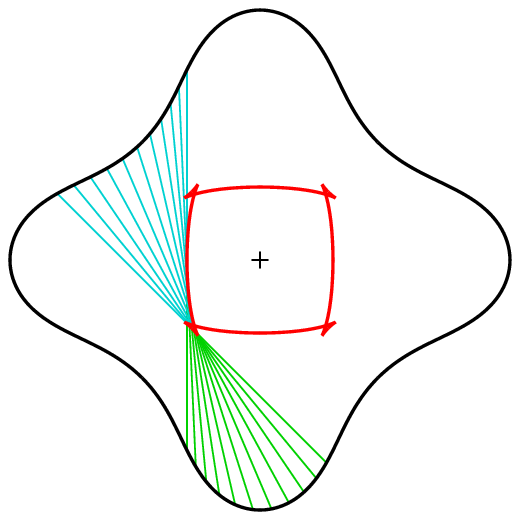,scale=0.5,angle=90}\\
\Figur{fgp1p4c}{$m/n=1/4$, $\epsilon=0.2$}}
\etm

\btm
\parbox[b]{3.2cm}{\epsfig{file=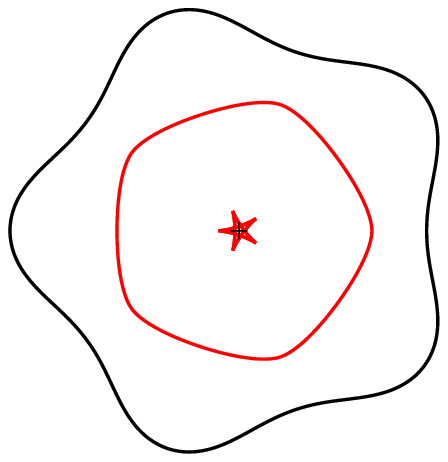,scale=0.5,angle=90}\\
\Figur{fgp1p5a}{$m/n=1/5$, $\epsilon=0.1$}}
\hfill
\parbox[b]{3.2cm}{\epsfig{file=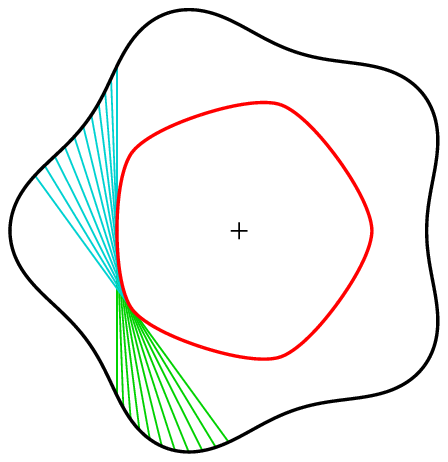,scale=0.5,angle=90}\\
\Figurm{fgp1p5b}{$m/n=1/5$, $\epsilon=0.1$}}
\hfill
\parbox[b]{3.2cm}{\epsfig{file=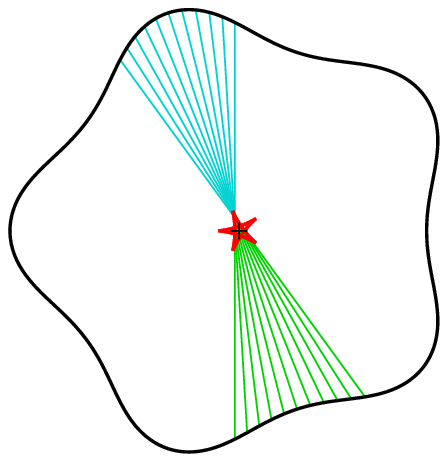,scale=0.5,angle=90}\\
\Figurm{fgp1p5j}{$m/n=1/5$, $\epsilon=0.1$}}
\etm

\btm
\parbox[b]{3.2cm}{\epsfig{file=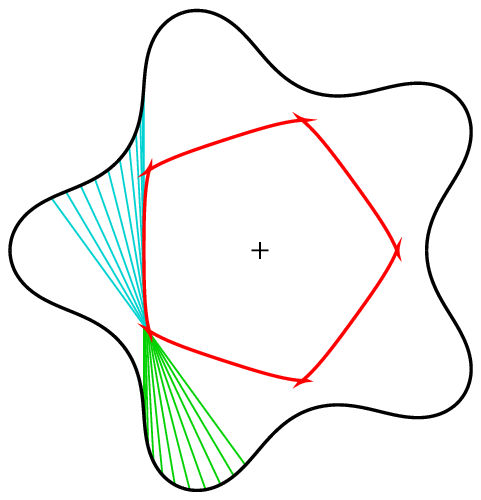,scale=0.5,angle=90}\\
\Figurm{fgp1p5c}{$m/n=1/5$, $\epsilon=0.2$}}
\hfill
\parbox[b]{3.2cm}{\epsfig{file=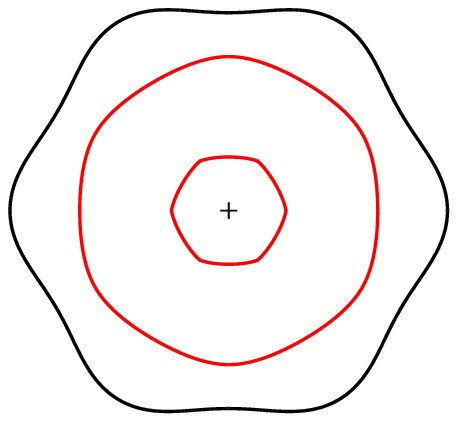,scale=0.5,angle=90}\\
\Figur{fgp1p6a}{$m/n=1/6$, $\epsilon=0.05$}}
\hfill
\parbox[b]{3.2cm}{\epsfig{file=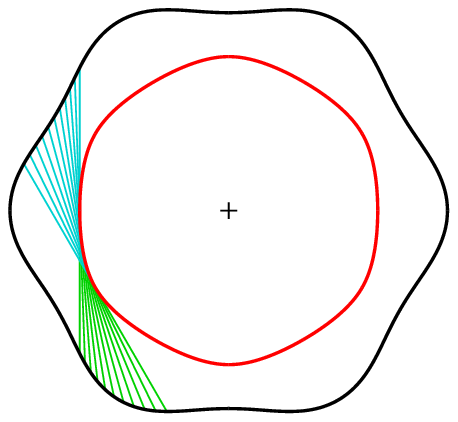,scale=0.5,angle=90}\\
\Figurm{fgp1p6b}{$m/n=1/6$, $\epsilon=0.05$}}
\etm

\btm
\parbox[b]{3.2cm}{\epsfig{file=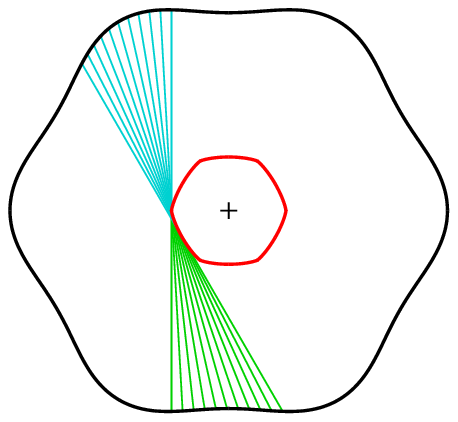,scale=0.5,angle=90}\\
\Figurm{fgp1p6c}{$m/n=1/6$, $\epsilon=0.05$}}
\hspace{1cm}
\parbox[b]{3.2cm}{\epsfig{file=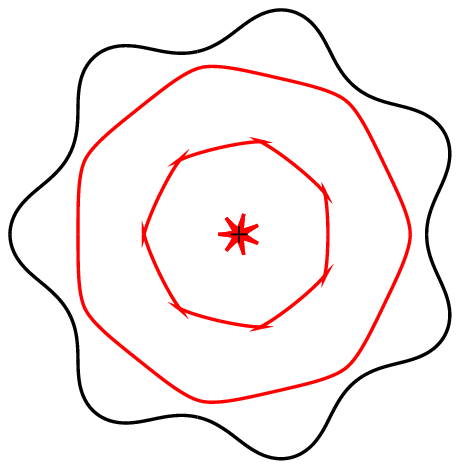,scale=0.5}\\
\Figur{fgp1p7b}{$m/n=1/7$, $\epsilon=0.1$}}
\etm

\nicht{A few cross-sections of the bodies are
shown in figs. \ref{fgp1p3a} to \ref{fgp1p7b}. For odd $n$ the innermost
envelope corresponds to density $\rho=1/2$.}

\subsection{Periodicity}

In eq. (\ref{psic}) an angle of periodicity $\psi_{\rm c}$ has been defined.
Here the periodicity is discussed for several regions in fig. \ref{param}. The
angle of periodicity $\psi_{\rm per}$ is defined as the change of the angle
$\psi$, as one moves from a point of extremal radius $r_i$ along the curve
until a point of this extremal radius is reached again. Its sign is defined by
the requirement that watching from the origin one starts moving
counterclockwise. This yields
\bea
\psi_{\rm per} &=& \frac{\Delta\psi}{{\rm sign}
\left.(\frac{\de\psi}{\de u})\right|_{r=r_i}}, \\
\Delta\psi &=& \psi(u+2\omega_3)-\psi(u)
\eea
Since due to eqs. (\ref{diffupsi}, \ref{ricond})
\be
\left.\frac{\de\psi}{\de u}\right|_{r=r_i} = aq_i +b +\frac c{q_i} = \frac
1{r_i}
\ee
one obtains
\be
\psi_{\rm per} = {\rm sign\,r_i}\, \Delta\psi.
\ee
$\Delta\psi$ is obtained from eq. (\ref{zu}) by the increase of
\be
{\rm arg\,}\sigma(u-3v) -{\rm arg\,}\sigma(u+v) +2u \Im\zeta(2v)
\ee
as $u$ is increased by $2\omega_3$. As shown in (\ref{argsigma},
\ref{argsigma1}) this depends on the imaginary part of the argument of the
functions $\sigma$.
Thus in addition to
\be
\Delta\psi_0 = 8v\ie\zeta(\omega_3) -4\ie\omega_3\zeta(2v)
\ee
one obtains extra multiple of $2\pi$. In region $A'$ one obtains $3\pi$ from
$\sigma(u-3v)$ and $\pi$ from $\sigma(u+v)$. This changes as one crosses to
$A\dpr$, since there $\Im v<\omega'/(3\ie)$. This can be seen from the sign
change of $\wp'(3v)$, eq. (\ref{wp3vs}) due to the sign change of $r_i$.
One observes, that now the curve passes at its minimum radius on the other side
of the origin.
In the table (\ref{table3}) the ranges of $v$ are listed in the various regions
of
fig. \ref{param} and the corresponding $\psi_{\rm per}$ are given:

\be
\begin{array}{|lc|cccc|}
\hline
\mbox{region} & \mbox{inequality} & \psi^{4}_{\rm per} &
\psi^{3}_{\rm per} & \psi^{2}_{\rm per} & \psi^{1}_{\rm per} \\ \hline
A' & \frac{\omega'}{3\ie} < \Im v < \frac{\omega'}{2\ie} & \psi_{\rm c}
& \psi_{\rm c} & & \\
A\dpr & 0 < \Im v < \frac{\omega'}{3\ie} & \psi_{\rm c}-2\pi
& 2\pi-\psi_{\rm c} & & \\
B' & 0 < \Im v < \frac{\omega_1}{3\ie} & \psi_{\rm c}-2\pi
& 2\pi-\psi_{\rm c} & 4\pi-\psi_{\rm c} & 4\pi-\psi_{\rm c} \\
B\dpr & \frac{\omega_1}{3\ie} < \Im v < \frac{\omega_1}{2\ie} & \psi_{\rm
c}-2\pi
& 2\pi-\psi_{\rm c} & 2\pi-\psi_{\rm c} & \psi_{\rm c}-2\pi \\ \hline
\end{array} \label{table3}
\ee

As examples figures out of the various regions are shown for $\psi_{\rm per}=0$.
One of them looks like an eight.
\medskip

\btm
\parbox[b]{3.2cm}{\epsfig{file=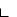,scale=0.5}\\
\Figur{fg0p1a}{$m/n=0/1$, $\epsilon=0.5$,\\ region $A'$}}
\hspace{1cm}
\parbox[b]{3.2cm}{\epsfig{file=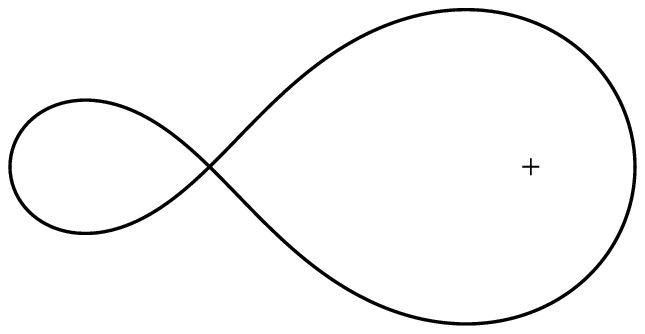,scale=0.5}\\
\Figur{fg0p1b}{$m/n=0/1$, $\epsilon=1.5$,\\ region $A\dpr$}}
\etm

\btm
\epsfig{file=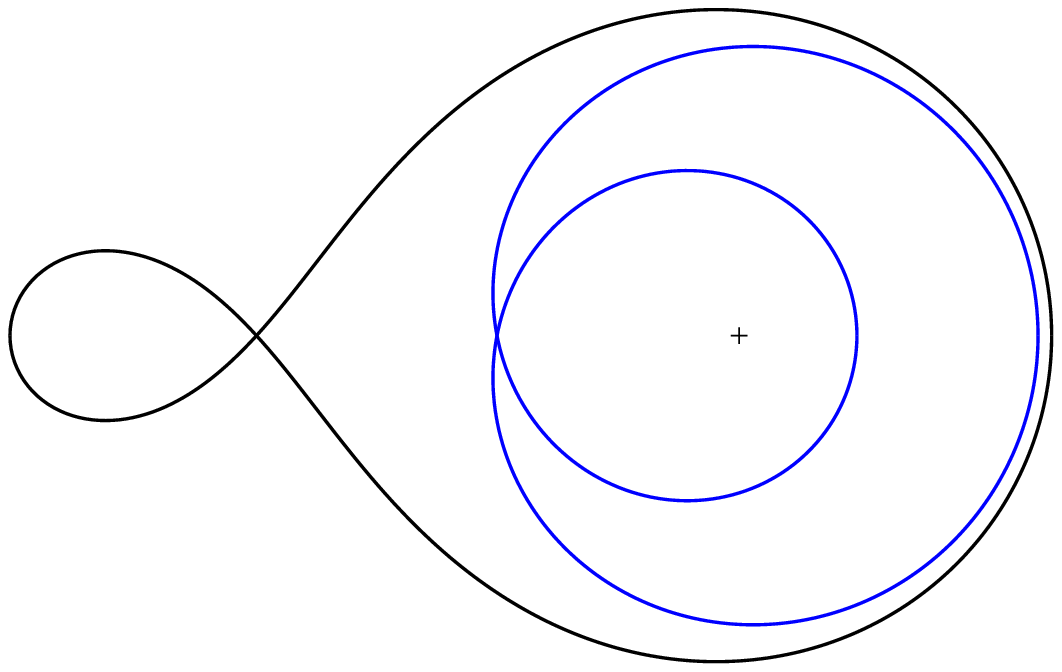,scale=0.4} \hspace{1cm}
\parbox[b]{5cm}{\Figur{fg0p1c}{$m/n=0/1$, $\epsilon=2.5$,\\ region $B'$}}
\etm

\bt
\epsfig{file=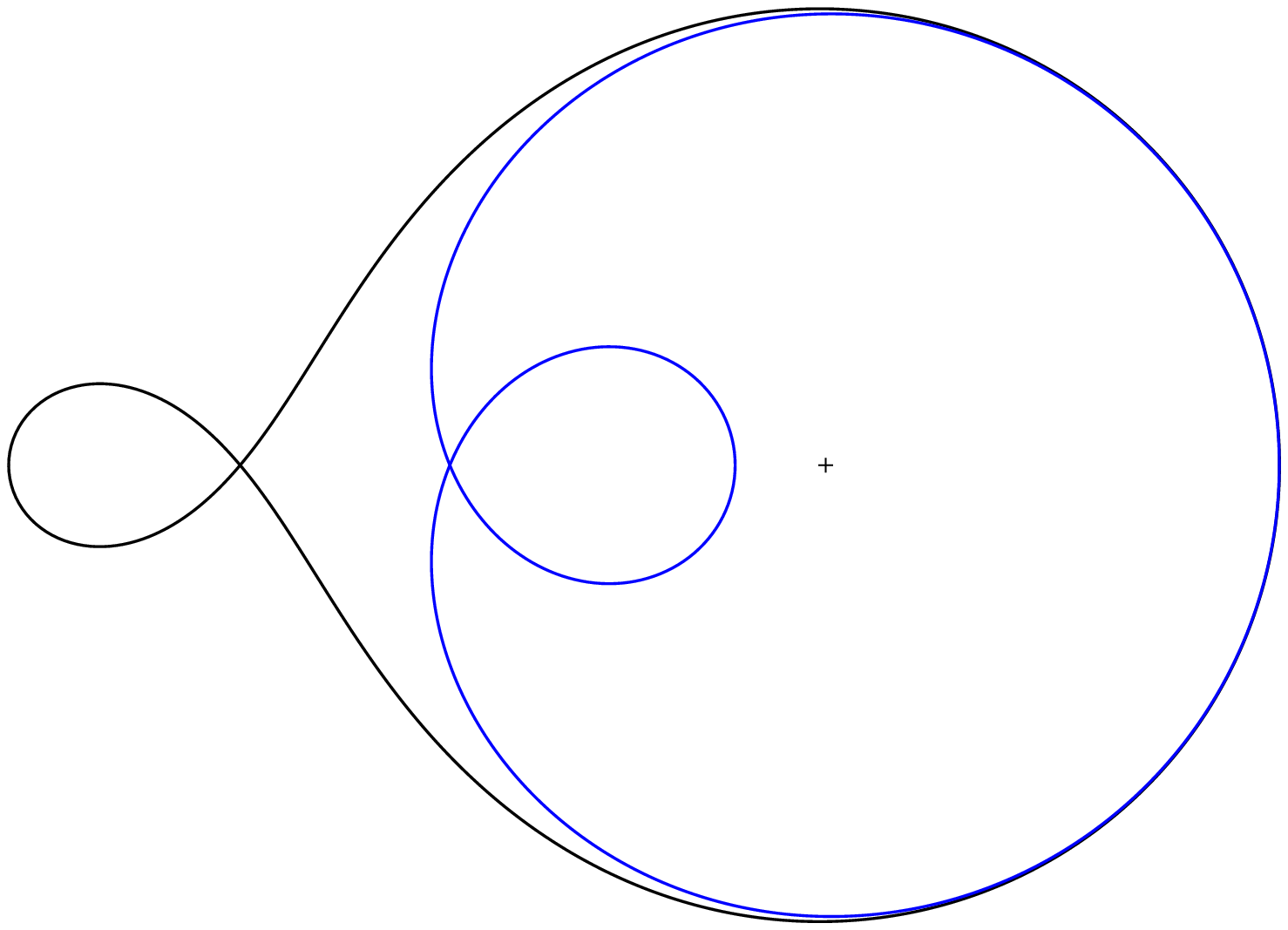,scale=0.4}\hspace{1cm}
\parbox[b]{3.2cm}{\Figur{fg0p1d}{$m/n=0/1$, $\epsilon=3.5$, region $B\dpr$}}
\et

Here and in the following the parametrization (\ref{psicmn})
$\psi_{\rm c} = 2\pi \frac mn$ is used.
It should be noted, that $v$, $\wp(v)$, and $\psi_{\rm c}$ vary continuously
across all the borders between the regions $A'$, $A\dpr$, $B'$, and $B\dpr$.
$\psi_{\rm c}$ approaches for fixed $\epsilon$ the limits
\be
\lim_{\mu\rightarrow\infty} \psi_{\rm c} = 2\pi, \quad
\lim_{\mu\rightarrow 0,\epsilon\le 2} \psi_{\rm c} = -\infty, \quad
\lim_{\heps\rightarrow\epsilon-2,\epsilon\ge 2} \psi_{\rm c} = -\infty.
\ee
Probably (I did not really check this) $\psi_{\rm c}$ is a
monotonic decreasing function of $\heps^2$ for fixed $\epsilon$.

\subsection{More figures}

In the following more curves are shown. Now larger values of $\epsilon$ are
chosen as already done for the case $m/n=0/1$. The next examples are for
$m/n=1/3,1/4,1/5$, and $1/6$. One observes that one obtains two branches for
$\epsilon>5/2$ and $m/n=1/3$ (figs. \ref{fgp1p3o} - \ref{fgp1p3p}).

\btm
\parbox[b]{5cm}{\epsfig{file=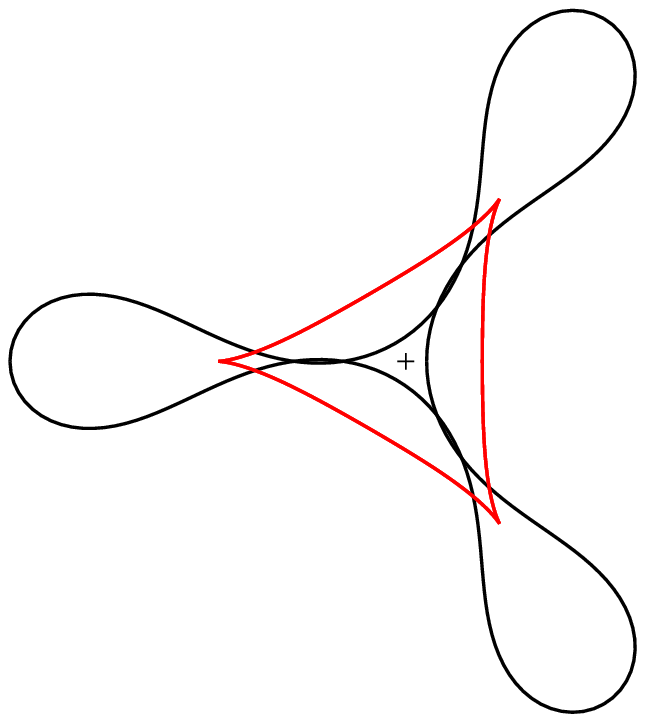,scale=0.5}\\
\Figur{fgp1p3d}{$m/n=1/3$, $\epsilon=0.9$}}
\hspace{1cm}
\parbox[b]{5cm}{\epsfig{file=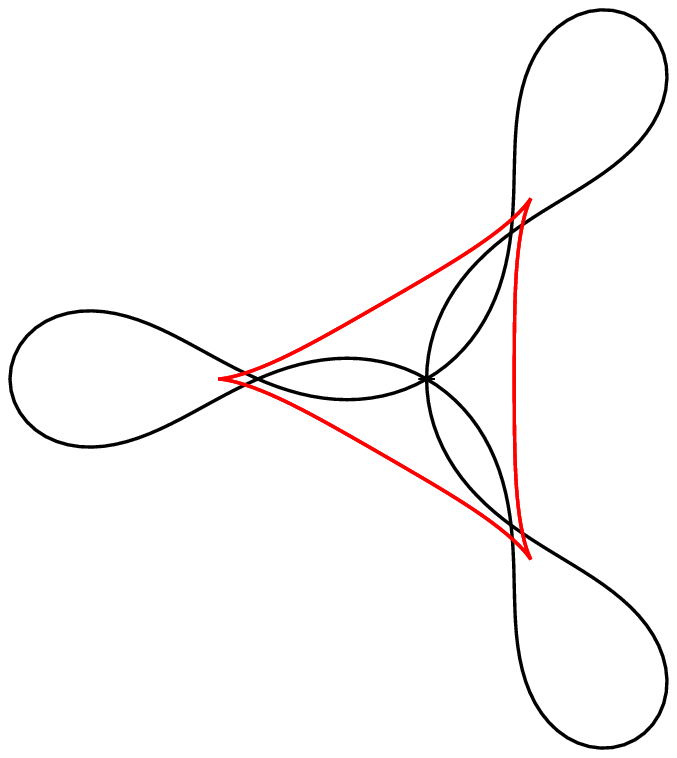,scale=0.5}\\
\Figur{fgp1p3e}{$m/n=1/3$, $\epsilon=1.0$}}
\etm

\btm
\parbox[b]{5cm}{\epsfig{file=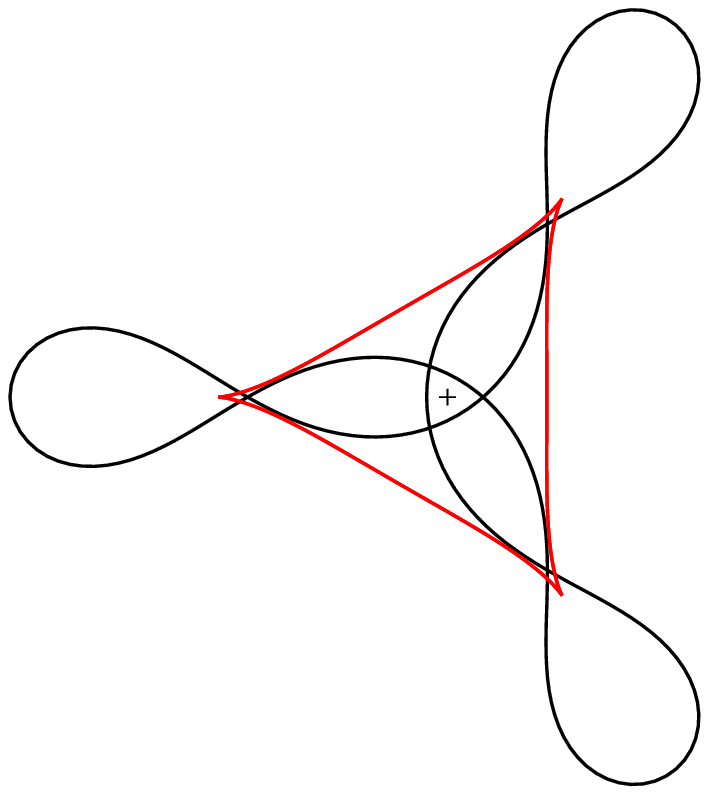,scale=0.5}\\
\Figur{fgp1p3f}{$m/n=1/3$, $\epsilon=1.1$}}
\hspace{1cm}
\parbox[b]{5cm}{\epsfig{file=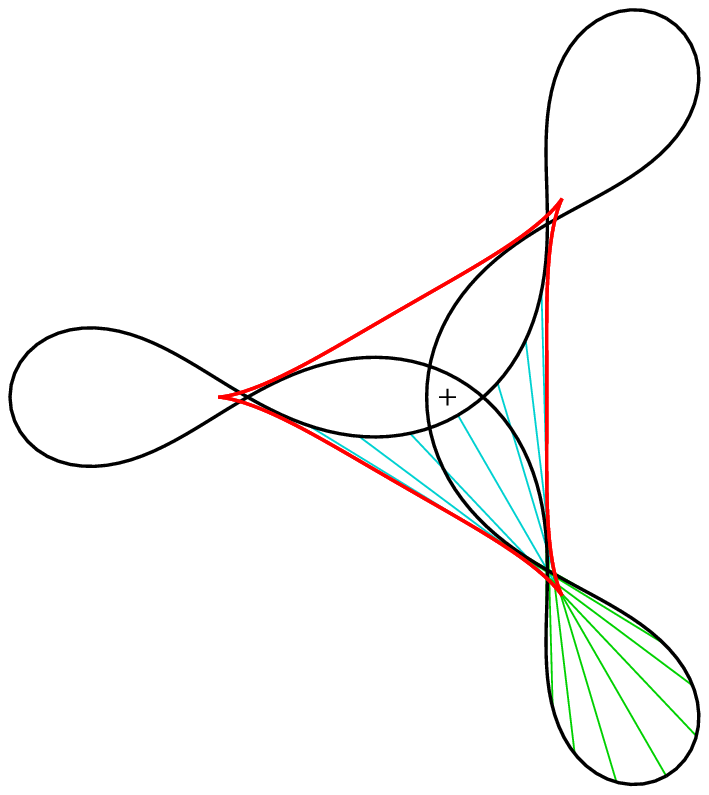,scale=0.5}\\
\Figur{fgp1p3g}{$m/n=1/3$, $\epsilon=1.1$}}
\etm

\btm
\parbox[b]{5cm}{\epsfig{file=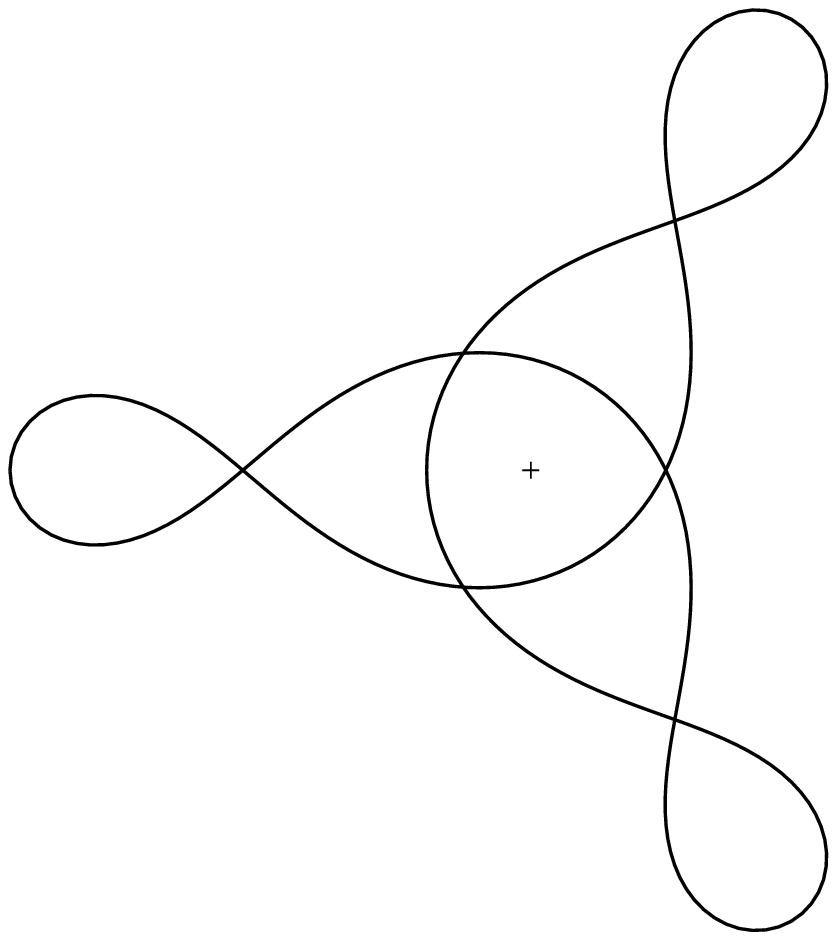,scale=0.3}\\
\Figur{fgp1p3m}{$m/n=1/3$, $\epsilon=1.5$}}
\hspace{1cm}
\parbox[b]{5cm}{\epsfig{file=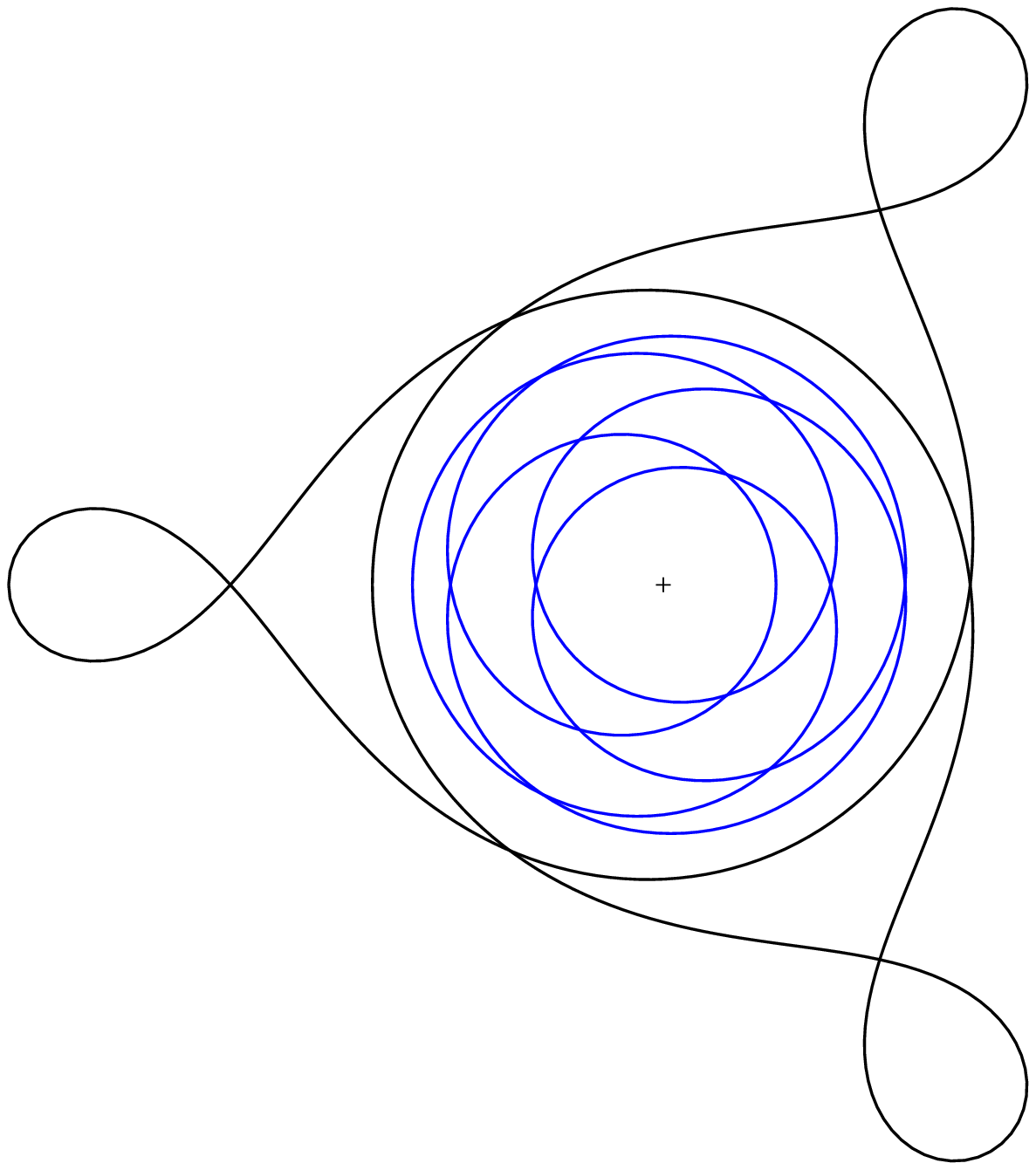,scale=0.3}\\
\Figur{fgp1p3o}{$m/n=1/3$, $\epsilon=2.6$}}
\etm

\btm
\parbox[b]{5cm}{\epsfig{file=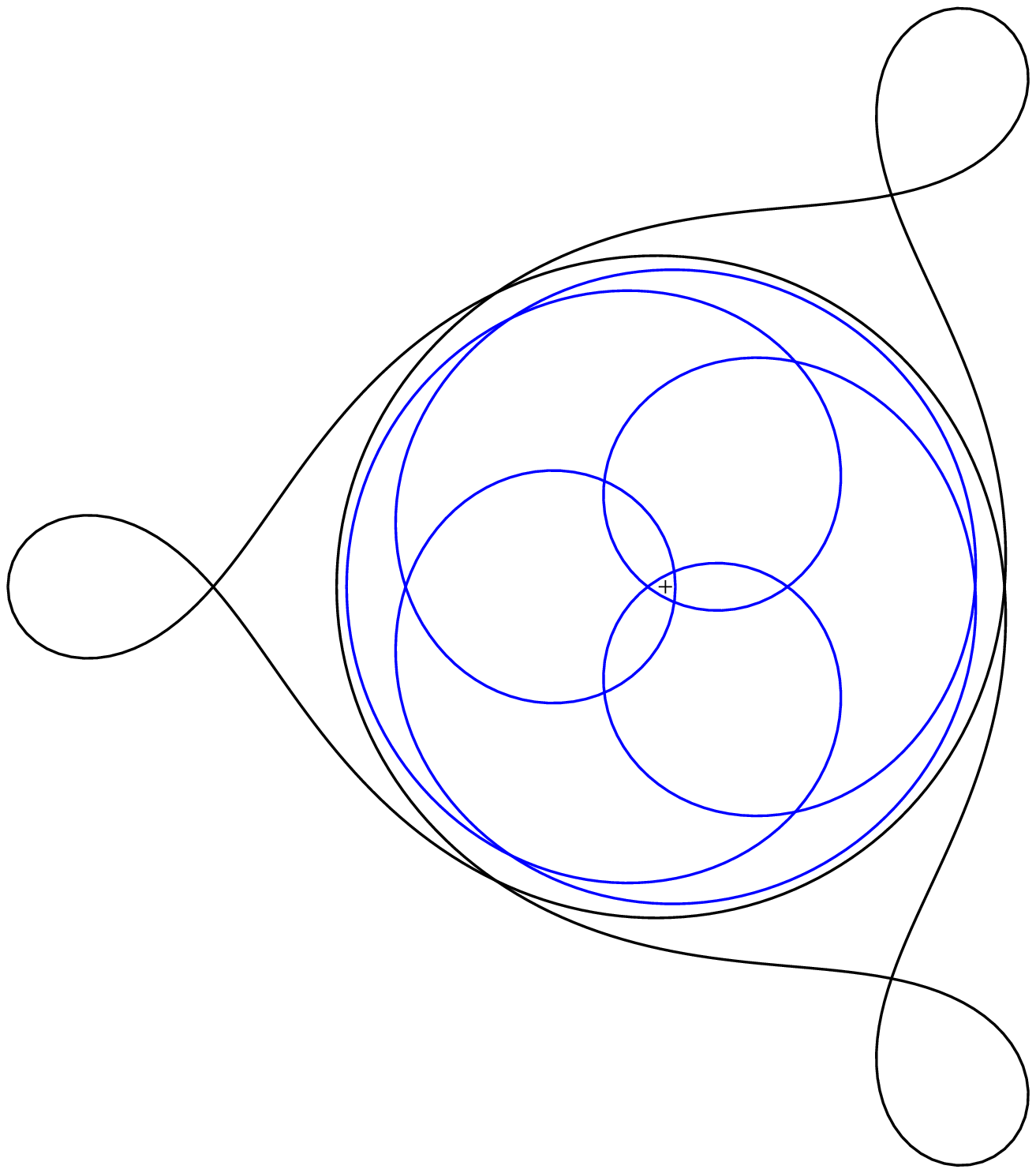,scale=0.3}\\
\Figur{fgp1p3h}{$m/n=1/3$, $\epsilon=3.0$}}
\hspace{1cm}
\parbox[b]{5cm}{\epsfig{file=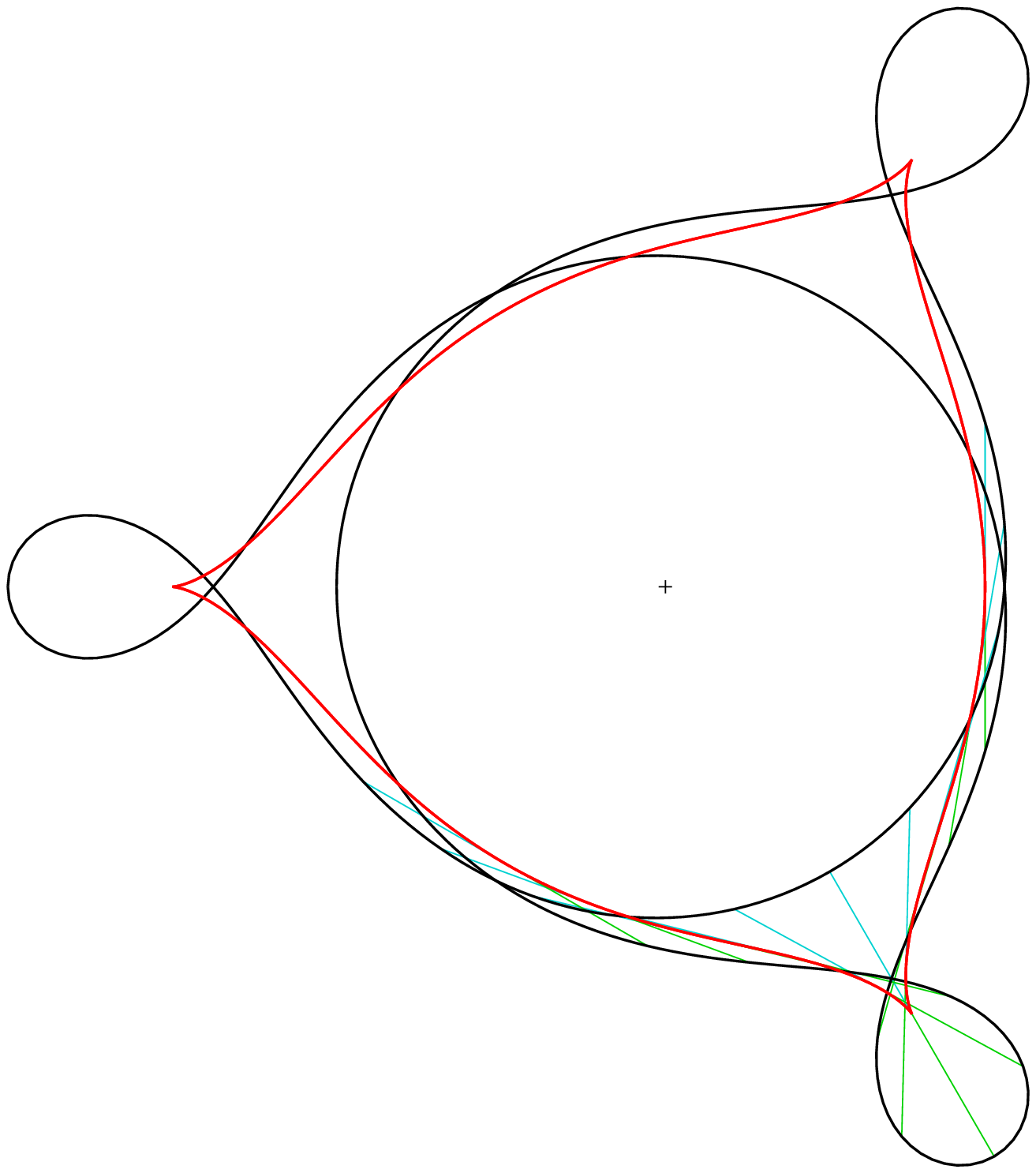,scale=0.3}\\
\Figurm{fgp1p3i}{$m/n=1/3$, $\epsilon=3.0$}}
\etm

\btm
\parbox[b]{5cm}{\epsfig{file=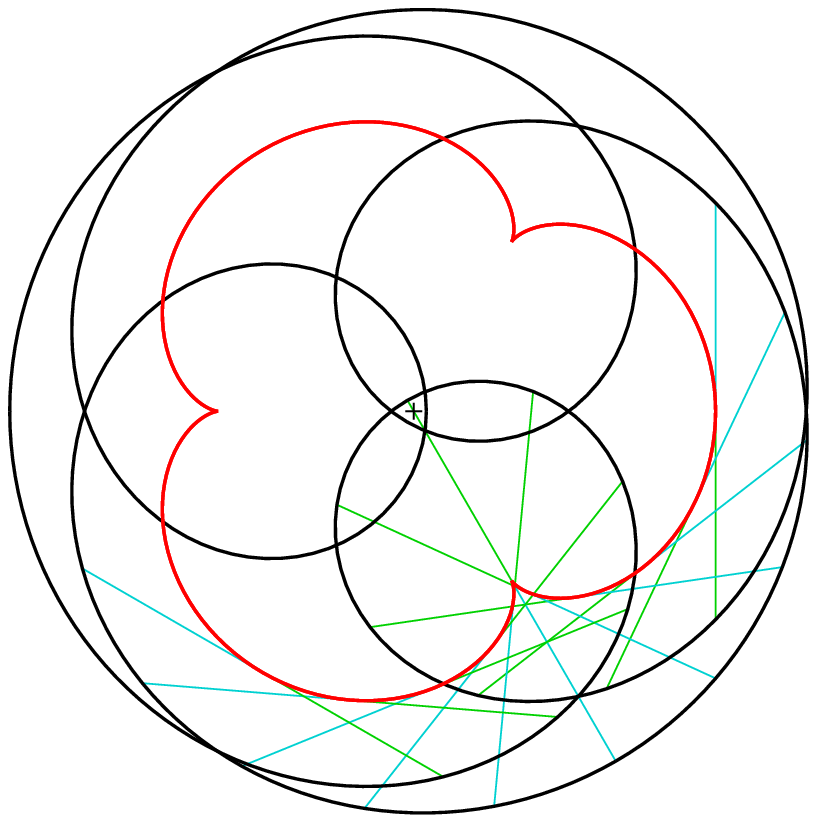,scale=0.3}\\
\Figurm{fgp1p3j}{$m/n=1/3$, $\epsilon=3.0$}}
\hspace{1cm}
\parbox[b]{5cm}{\epsfig{file=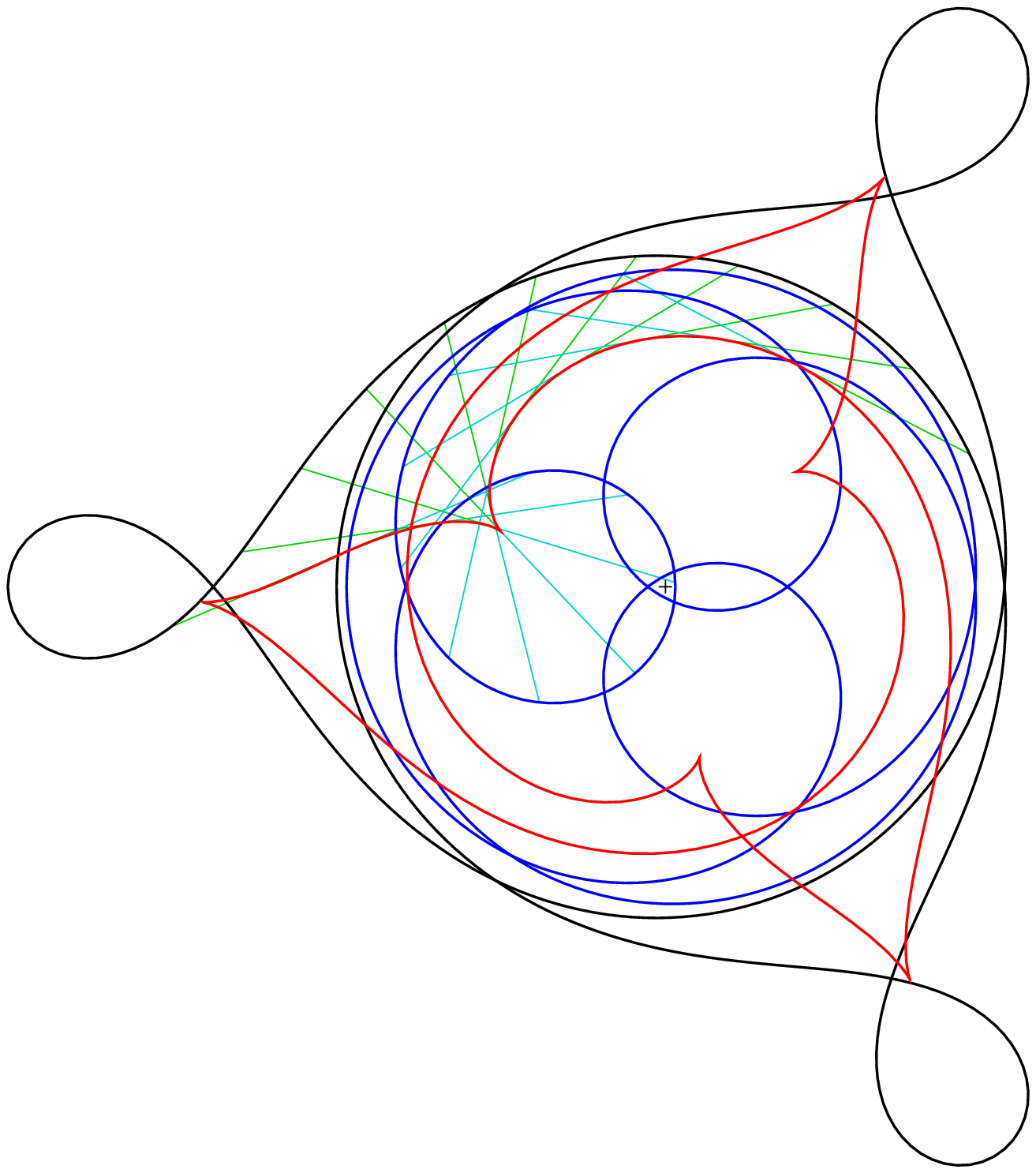,scale=0.3}\\
\Figurm{fgp1p3k}{$m/n=1/3$, $\epsilon=3.0$}}
\etm

\btm
\parbox[b]{5cm}{\epsfig{file=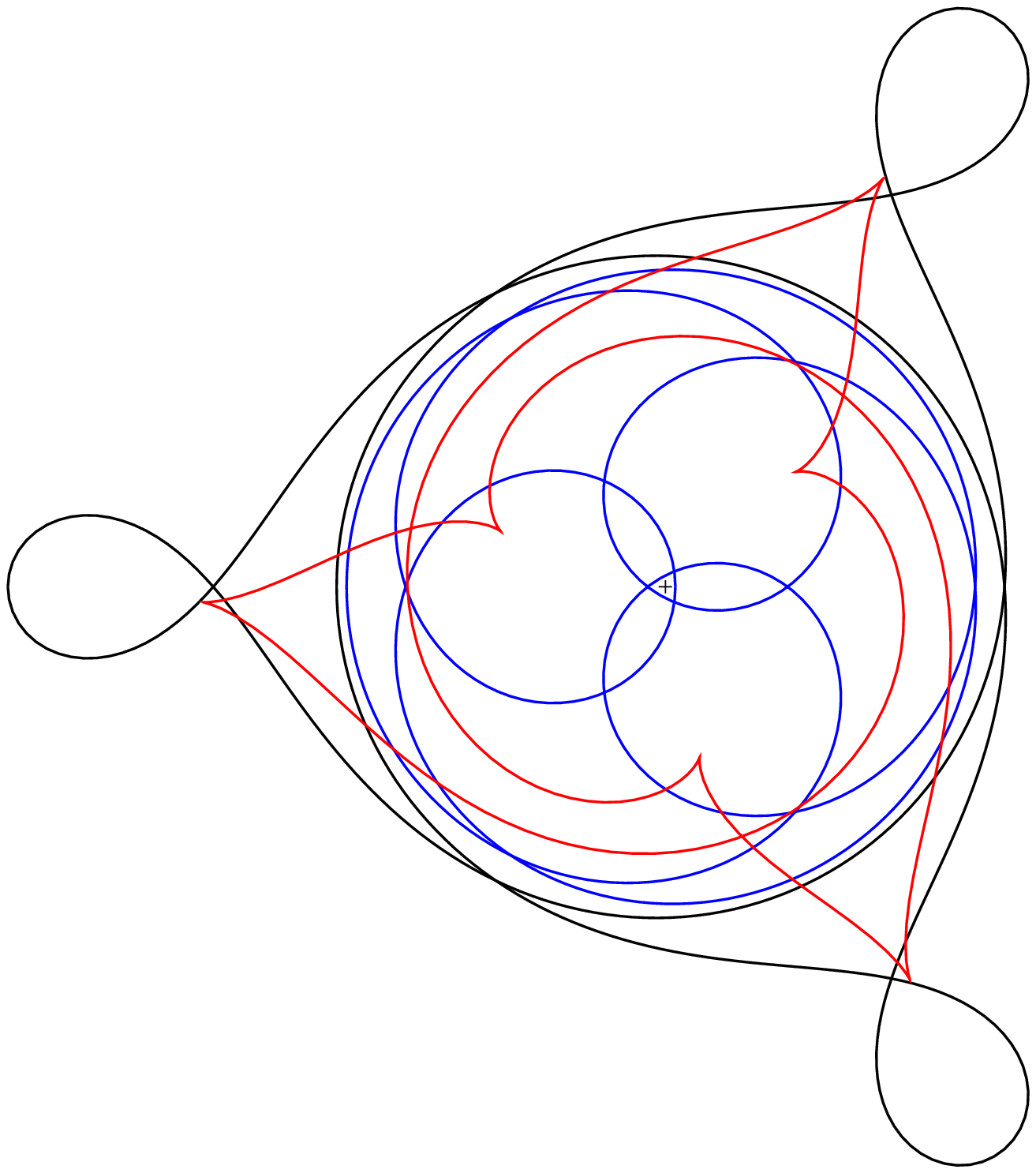,scale=0.3}\\
\Figur{fgp1p3l}{$m/n=1/3$, $\epsilon=3.0$}}
\hspace{1cm}
\parbox[b]{5cm}{\epsfig{file=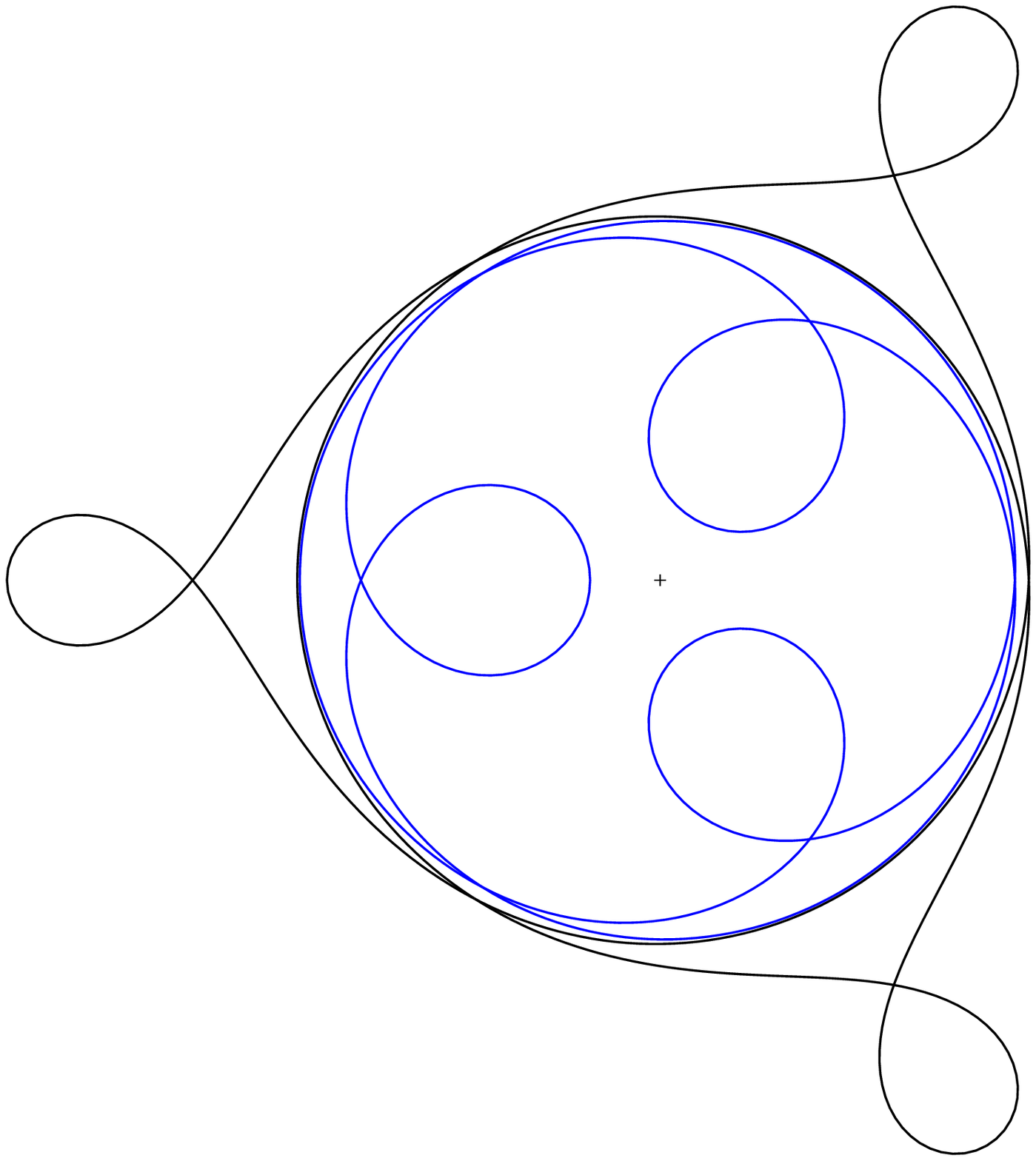,scale=0.3}\\
\Figur{fgp1p3p}{$m/n=1/3$, $\epsilon=3.5$}}
\etm

\btm
\parbox[b]{5cm}{\epsfig{file=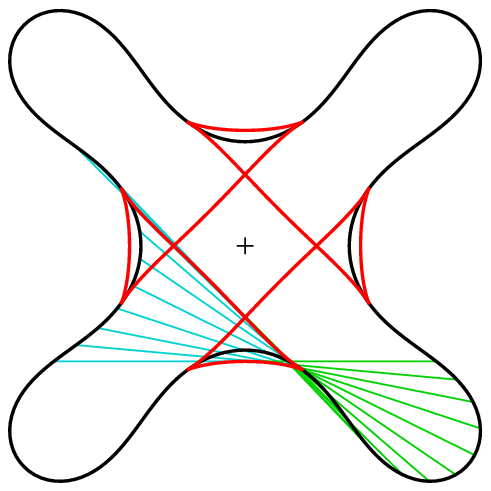,scale=0.5}\\
\Figur{fgp1p4d}{$m/n=1/4$, $\epsilon=0.5$}}
\hspace{1cm}
\parbox[b]{5cm}{\epsfig{file=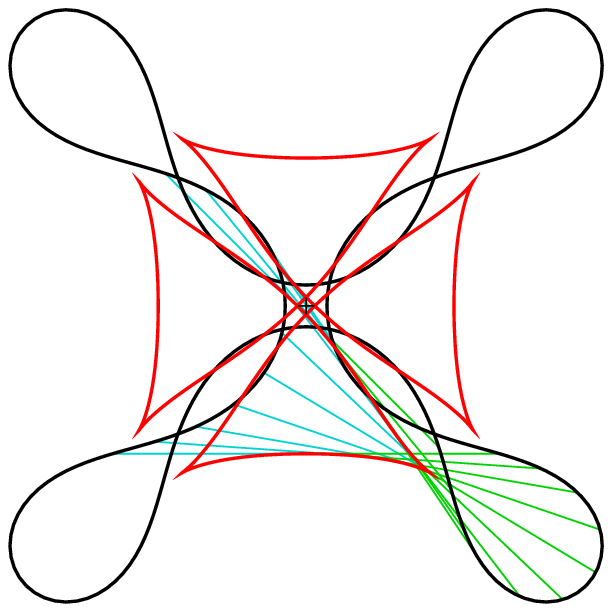,scale=0.5}\\
\Figur{fgp1p4e}{$m/n=1/4$, $\epsilon=0.9$}}
\etm

\btm
\parbox[b]{5cm}{\epsfig{file=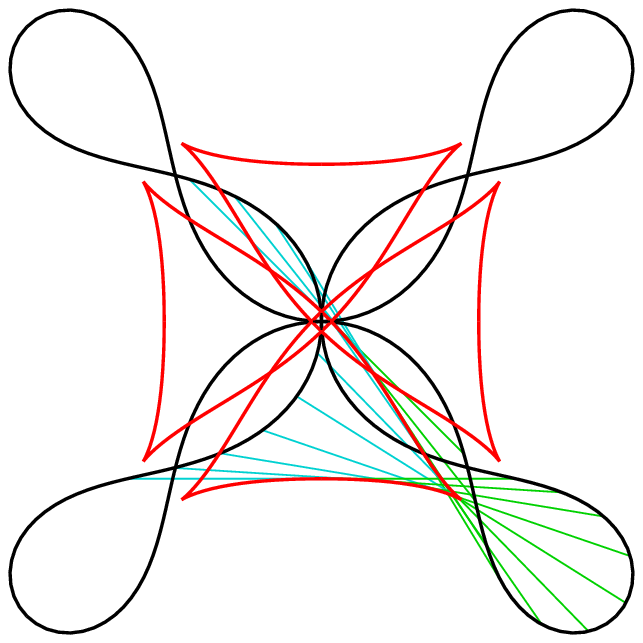,scale=0.5}\\
\Figur{fgp1p4f}{$m/n=1/4$, $\epsilon=1.0$}}
\hspace{1cm}
\parbox[b]{5cm}{\epsfig{file=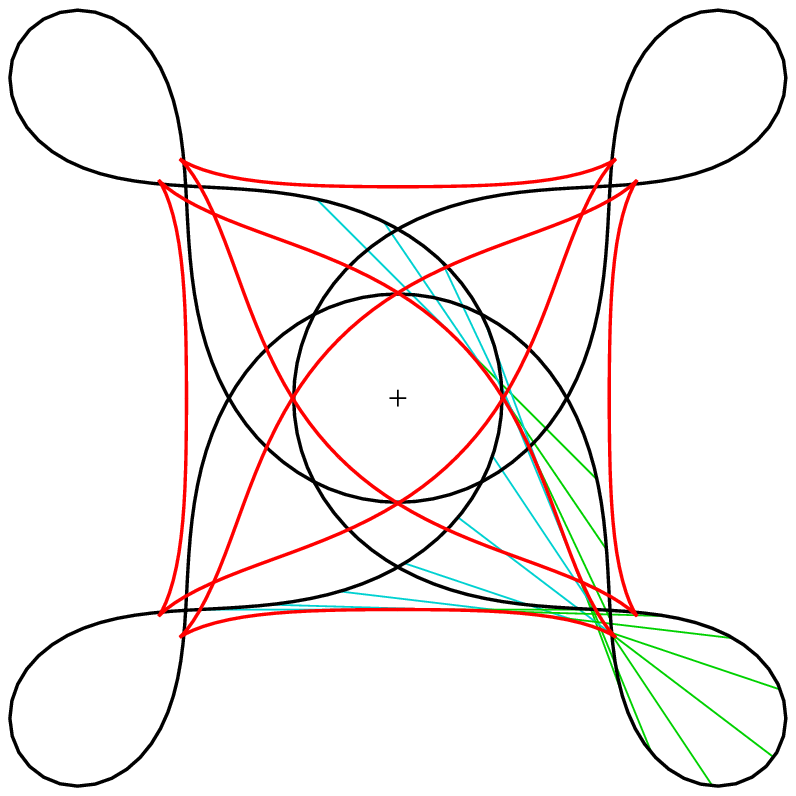,scale=0.5}\\
\Figur{fgp1p4g}{$m/n=1/4$, $\epsilon=1.5$}}
\etm

\btm
\parbox[b]{5cm}{\epsfig{file=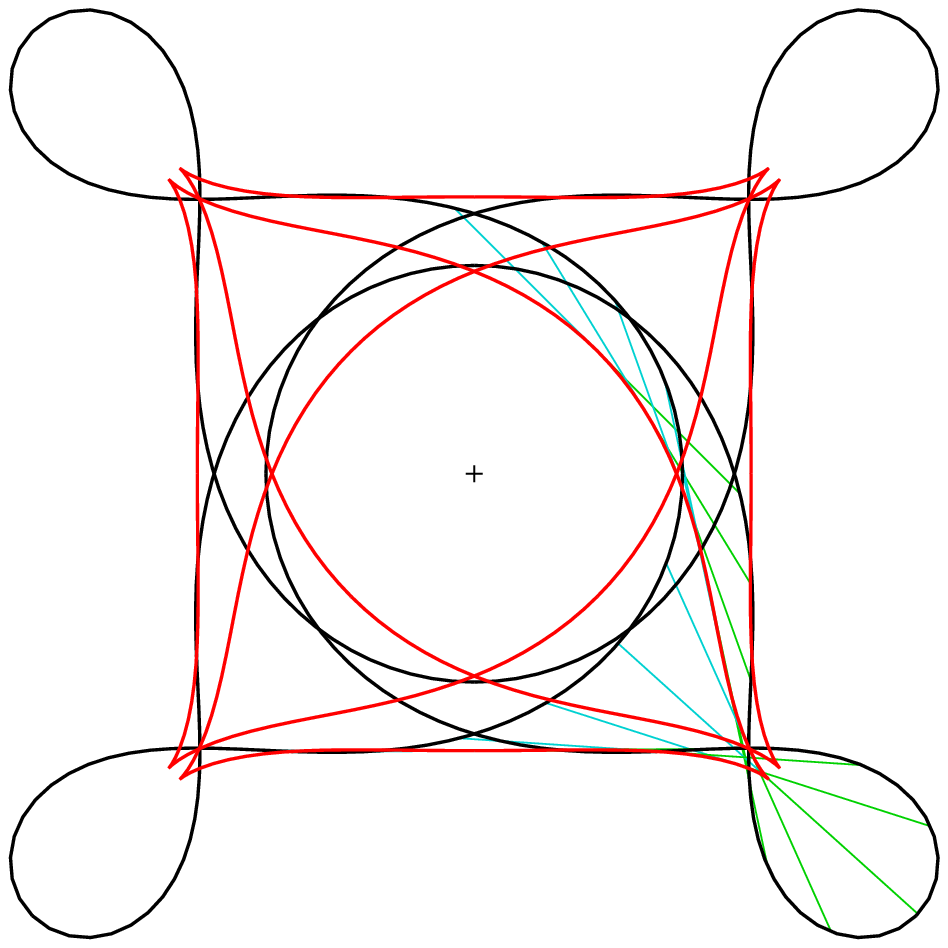,scale=0.4}\\
\Figur{fgp1p4h}{$m/n=1/4$, $\epsilon=2.0$}}
\hspace{1cm}
\parbox[b]{5cm}{\epsfig{file=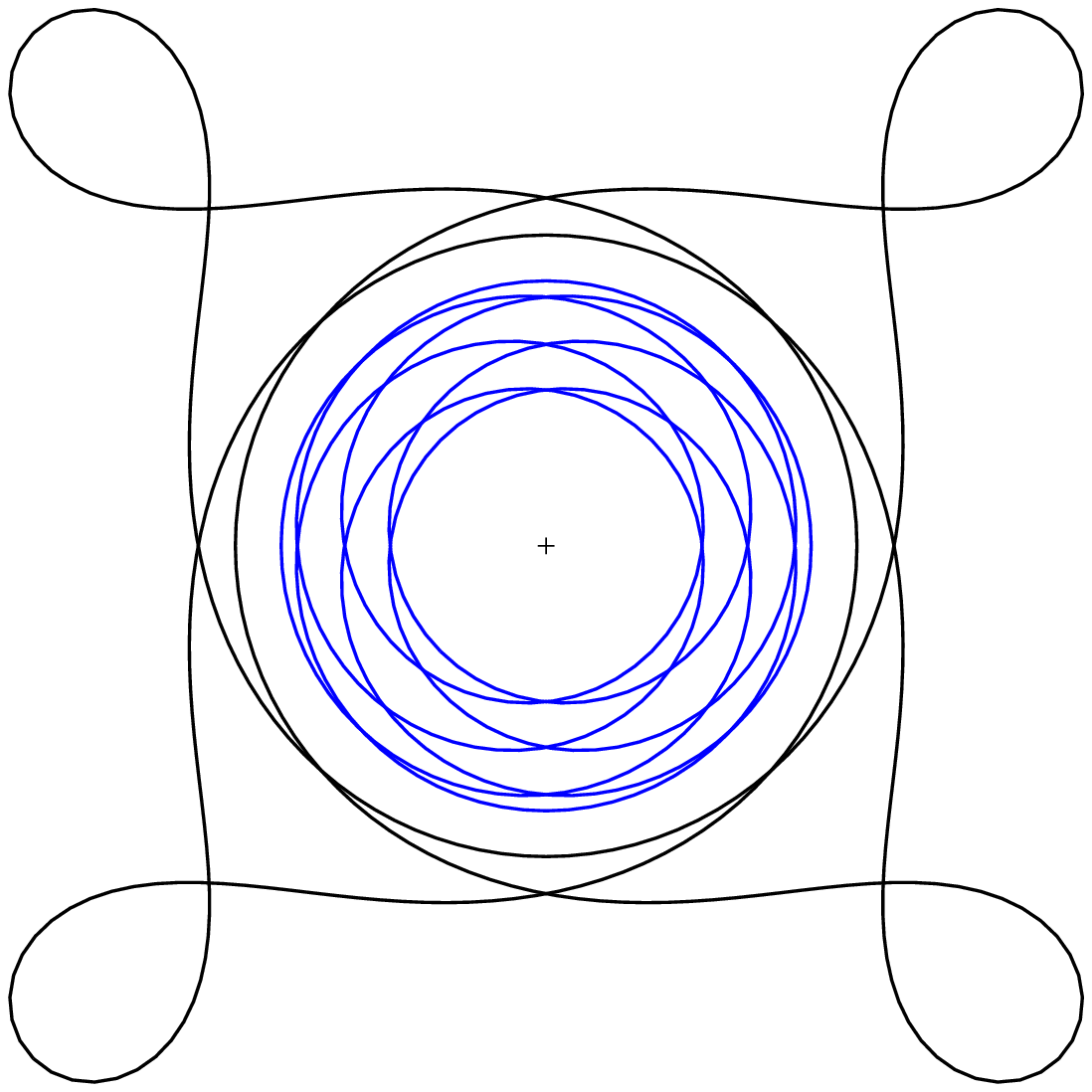,scale=0.4}\\
\Figur{fgp1p4i}{$m/n=1/4$, $\epsilon=2.5$}}
\etm

\btm
\parbox[b]{5cm}{\epsfig{file=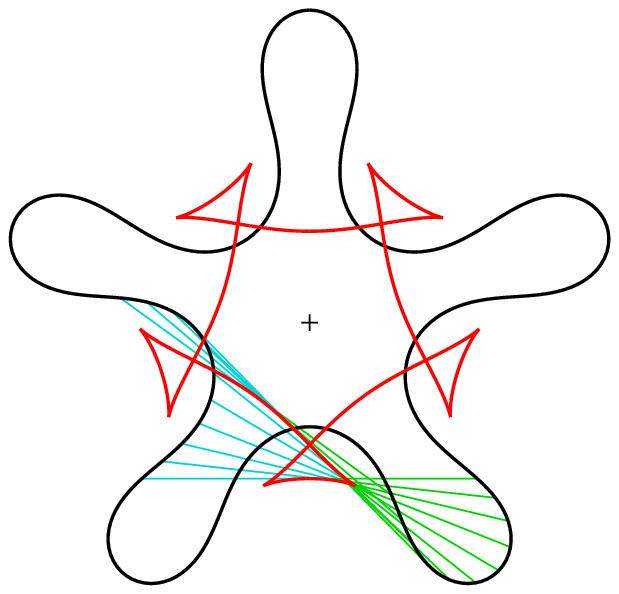,scale=0.5}\\
\Figurm{fgp1p5d}{$m/n=1/5$, $\epsilon=0.5$}}
\hspace{1cm}
\parbox[b]{5cm}{\epsfig{file=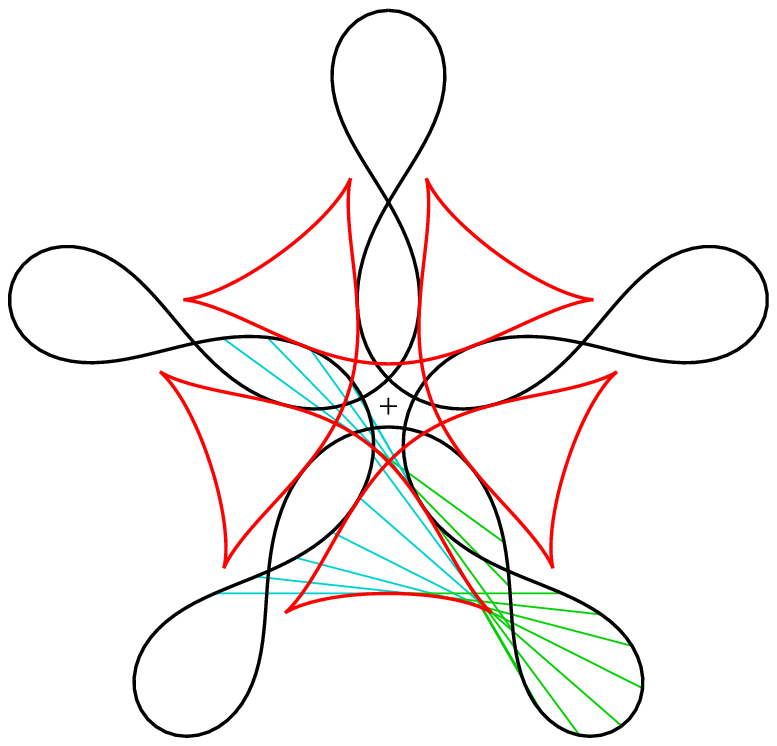,scale=0.5}\\
\Figurm{fgp1p5e}{$m/n=1/5$, $\epsilon=0.9$}}
\etm

\btm
\parbox[b]{5cm}{\epsfig{file=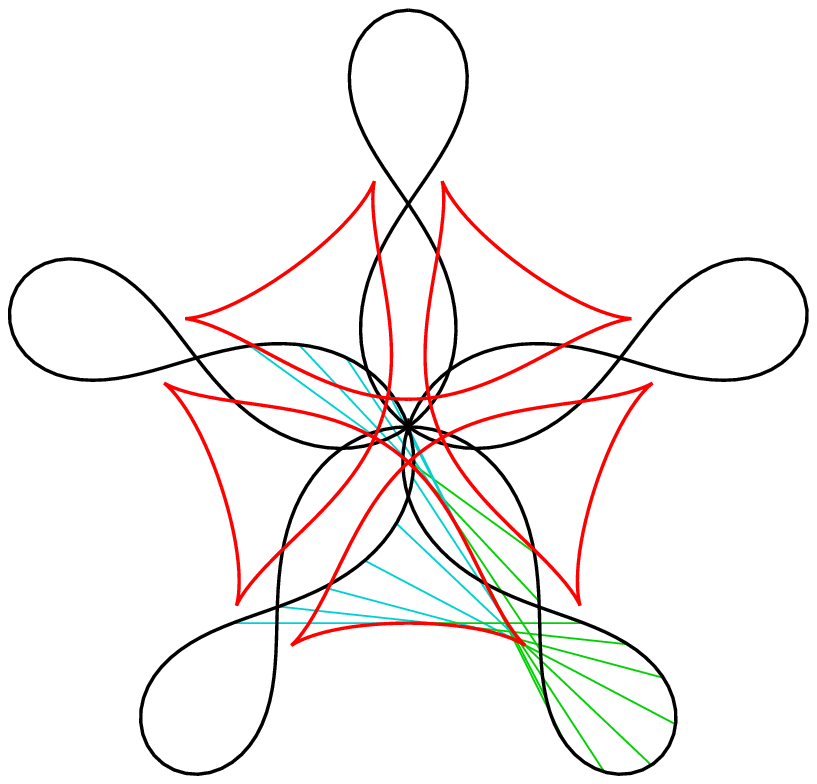,scale=0.5}\\
\Figurm{fgp1p5f}{$m/n=1/5$, $\epsilon=1.0$}}
\hspace{1cm}
\parbox[b]{5cm}{\epsfig{file=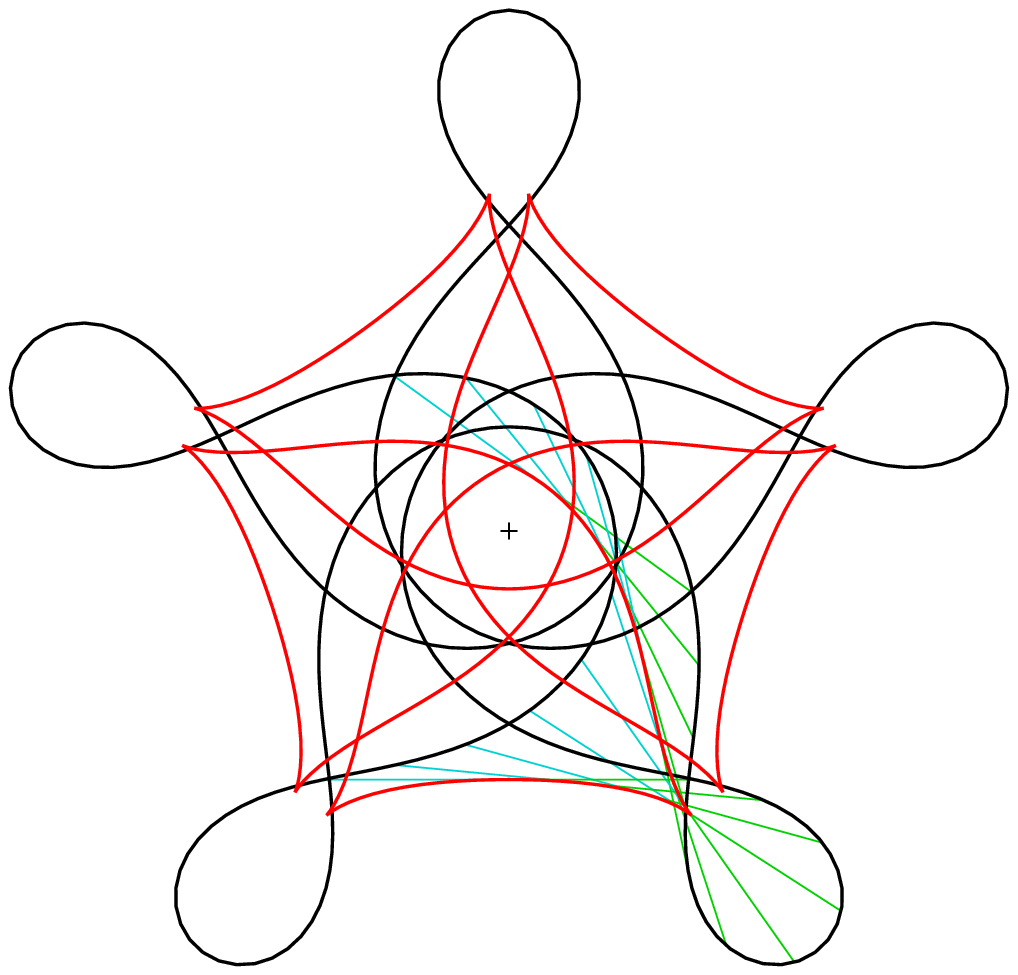,scale=0.5}\\
\Figurm{fgp1p5g}{$m/n=1/5$, $\epsilon=1.5$}}
\etm

\btm
\parbox[b]{5cm}{\epsfig{file=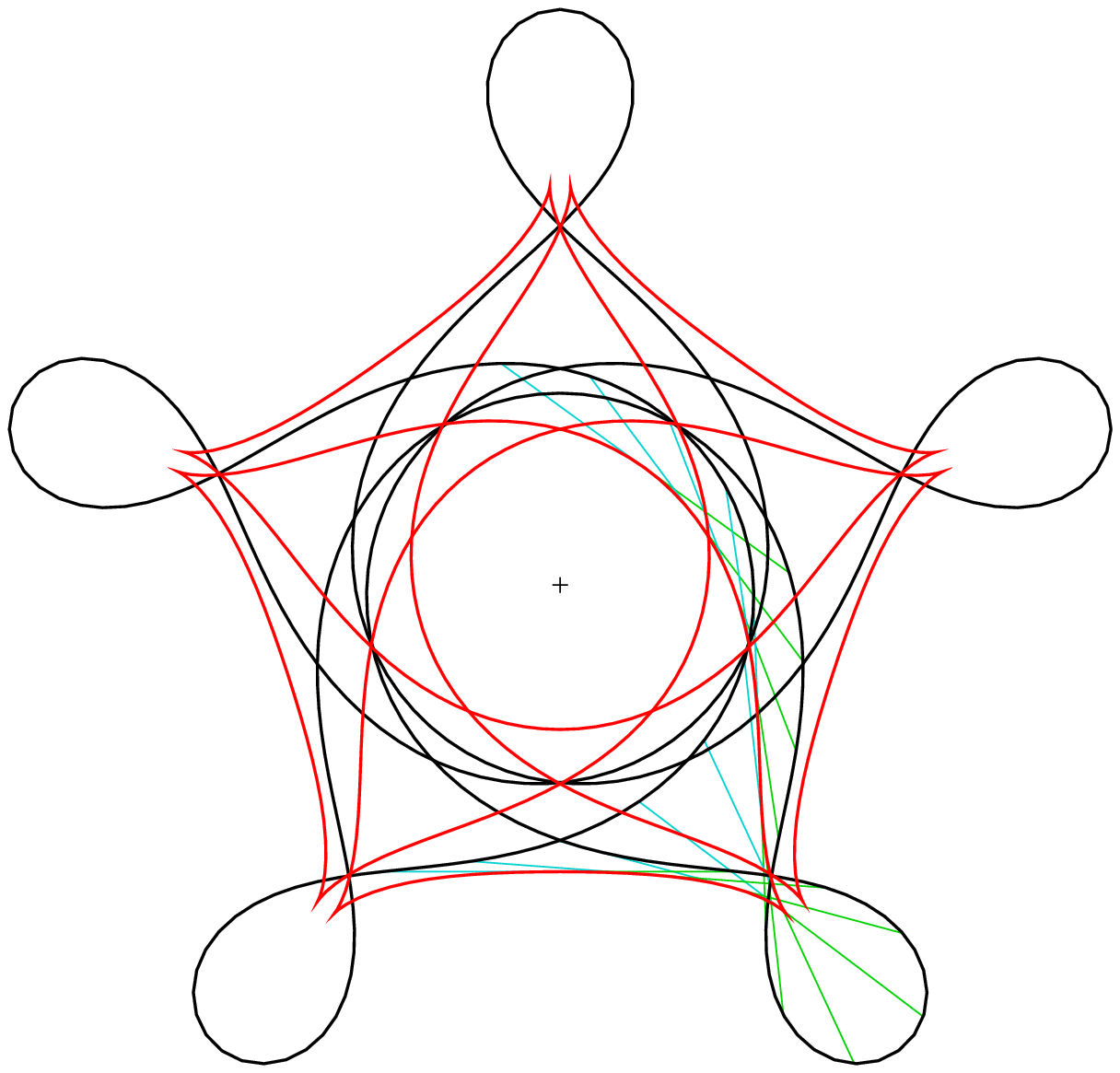,scale=0.4}\\
\Figurm{fgp1p5h}{$m/n=1/5$, $\epsilon=2.0$}}
\hspace{1cm}
\parbox[b]{5cm}{\epsfig{file=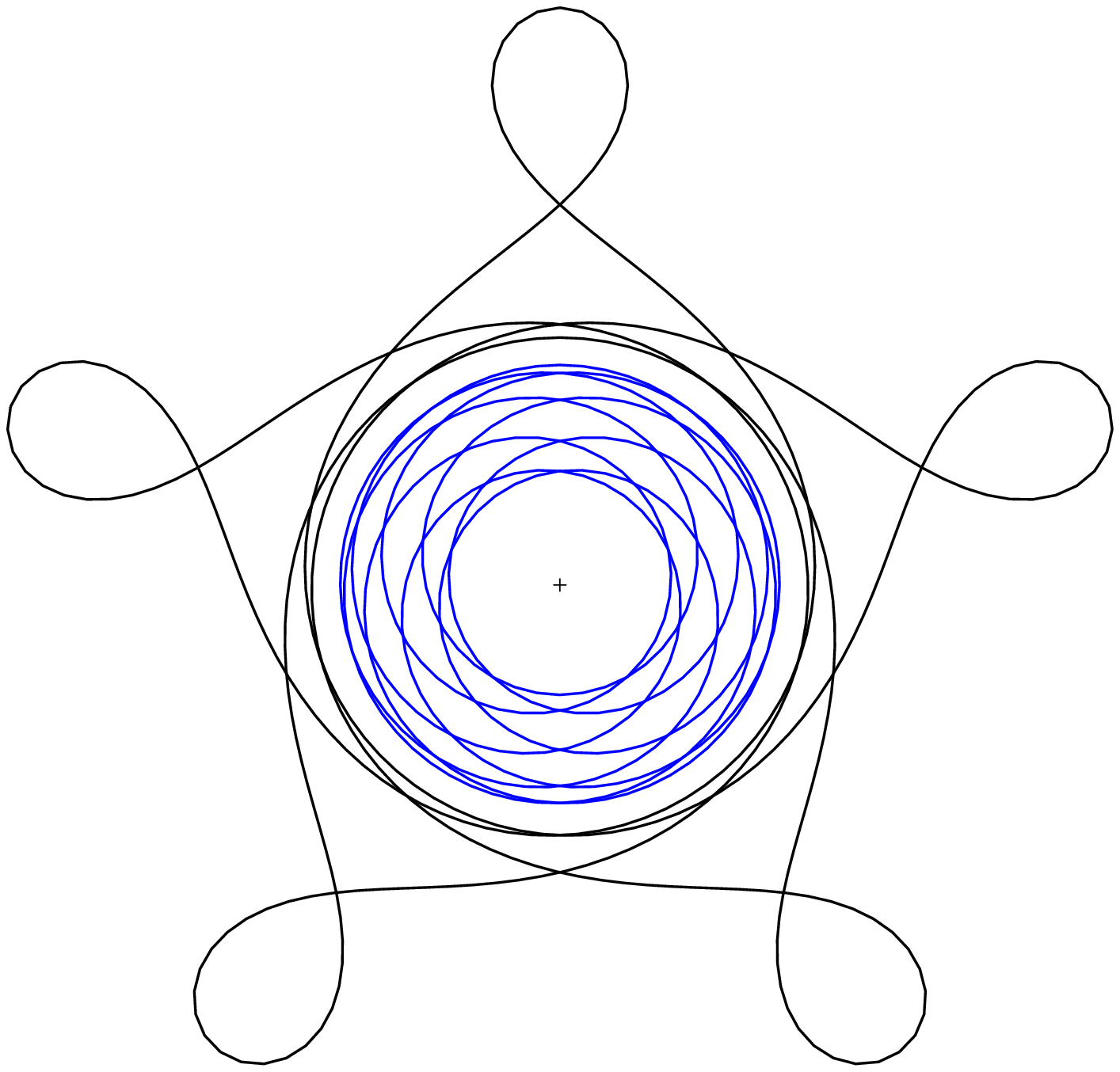,scale=0.3}\\
\Figur{fgp1p5i}{$m/n=1/5$, $\epsilon=2.5$}}
\etm

\btm
\epsfig{file=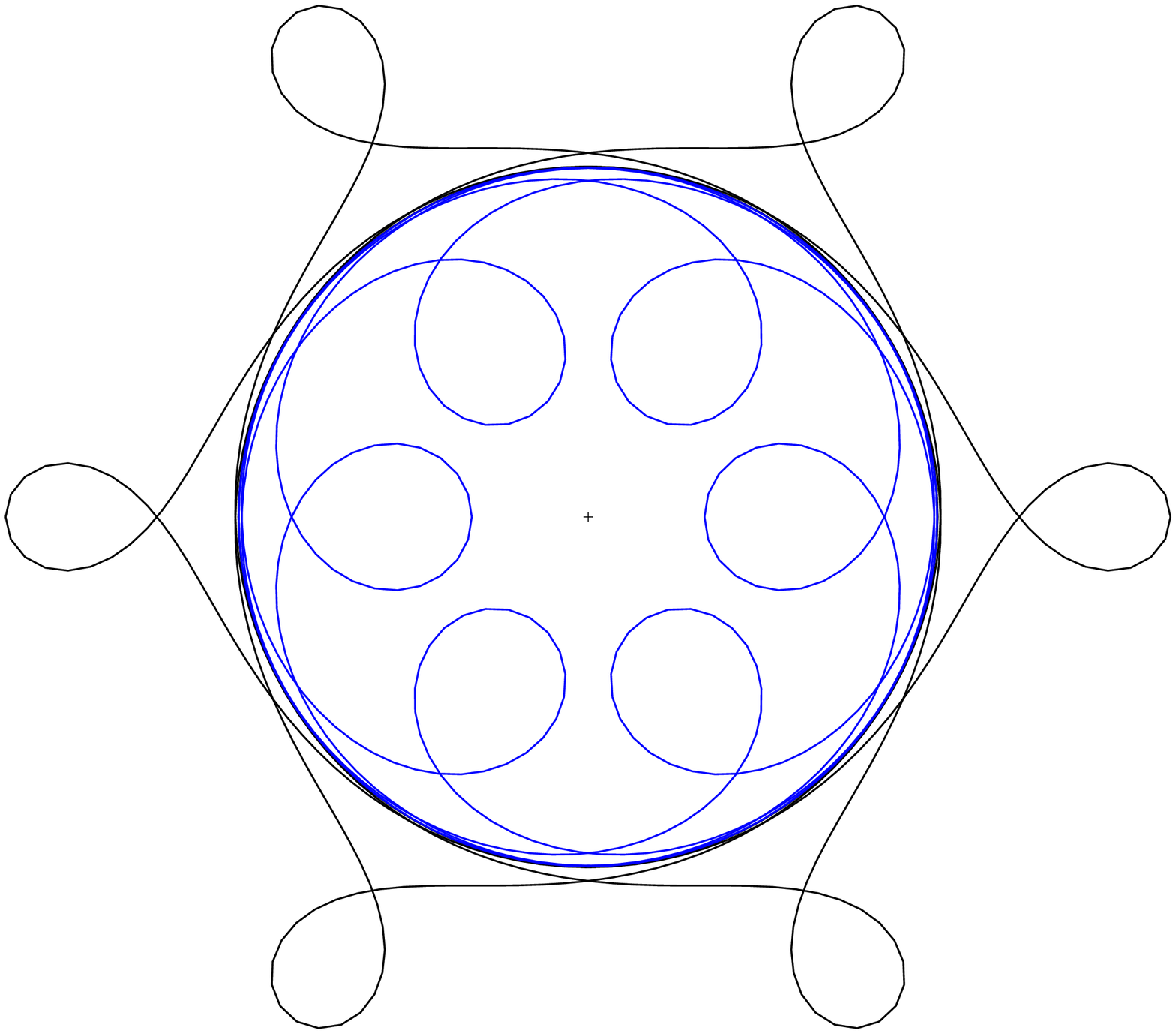,scale=0.3}
\hspace{1cm}
\parbox[b]{3.5cm}{\Figur{fgp1p6d}{$m/n=1/6$, $\epsilon=4.0$}}
\etm

Examples for $m/n=2/5$, $3/5$, $4/5$ are shown in figs. \ref{fgp2p5a} to
\ref{fgp4p5b}

\btm
\parbox[b]{5cm}{\epsfig{file=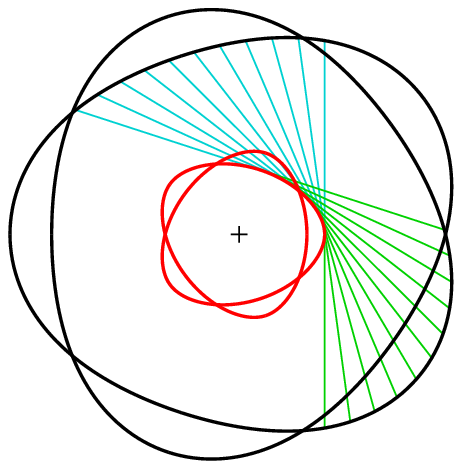,scale=0.5,angle=-90}\\
\Figurm{fgp2p5a}{$m/n=2/5$, $\epsilon=0.1$}}
\hspace{1cm}
\parbox[b]{5cm}{\epsfig{file=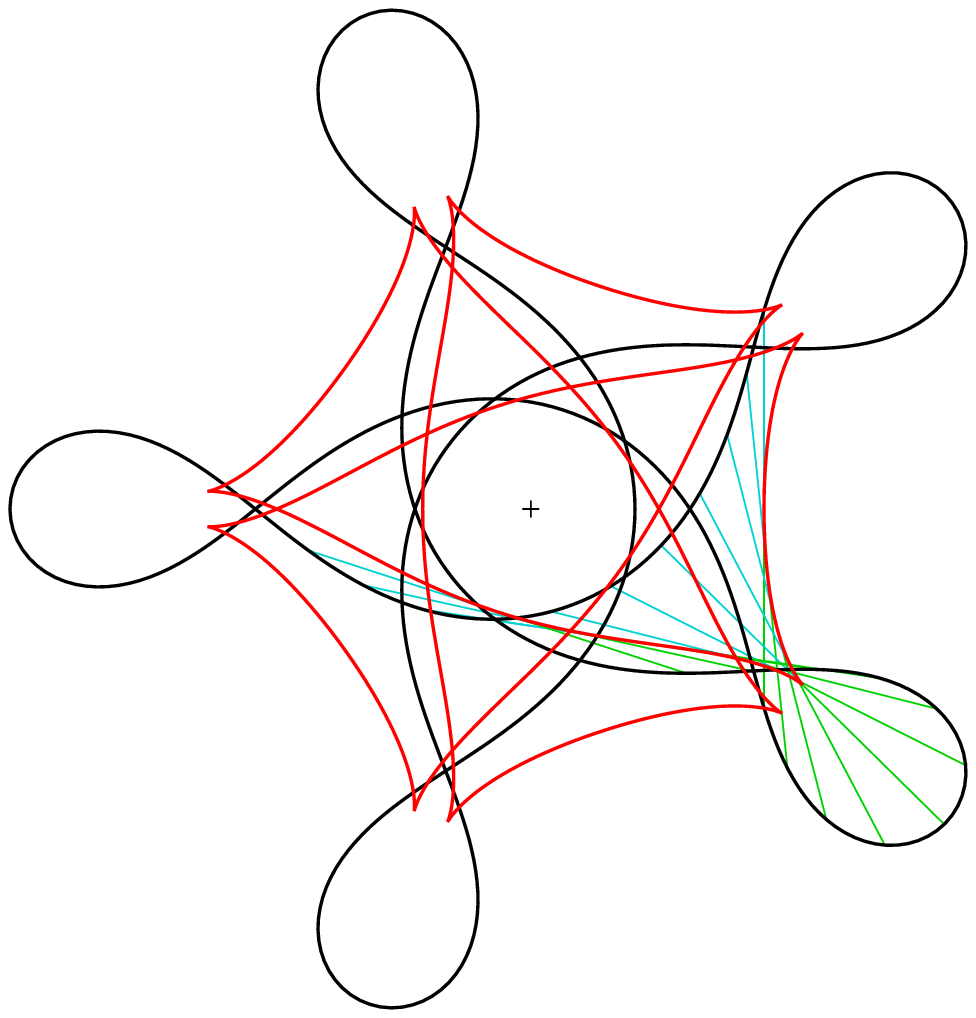,scale=0.4,angle=-90}\\
\Figur{fgp2p5b}{$m/n=2/5$, $\epsilon=1.5$}}
\etm

\btm
\parbox[b]{6cm}{\epsfig{file=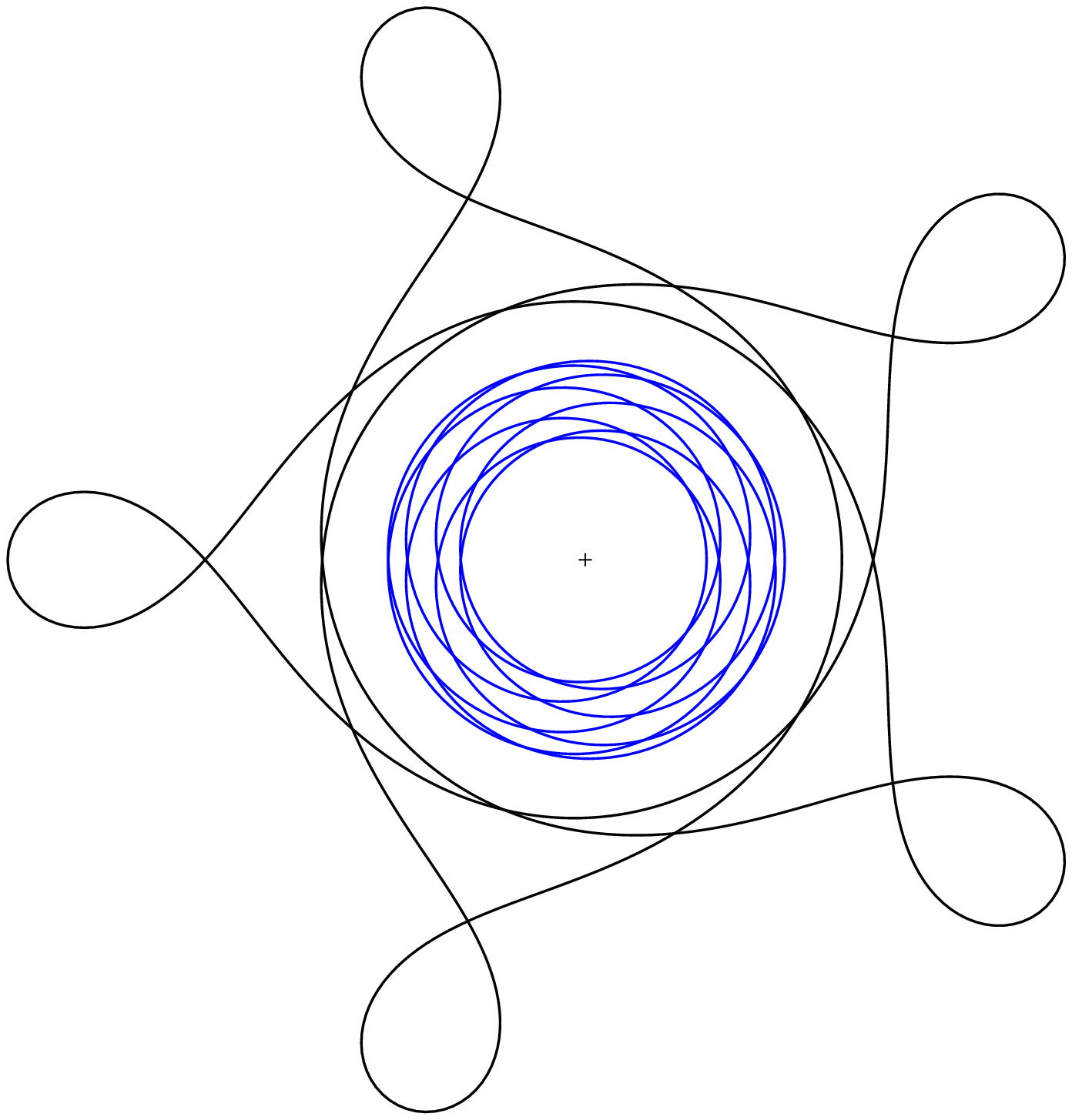,scale=0.3,angle=-90}\\
\Figur{fgp2p5c}{$m/n=2/5$, $\epsilon=2.6$}}
\hspace{0.8cm}
\parbox[b]{5cm}{\epsfig{file=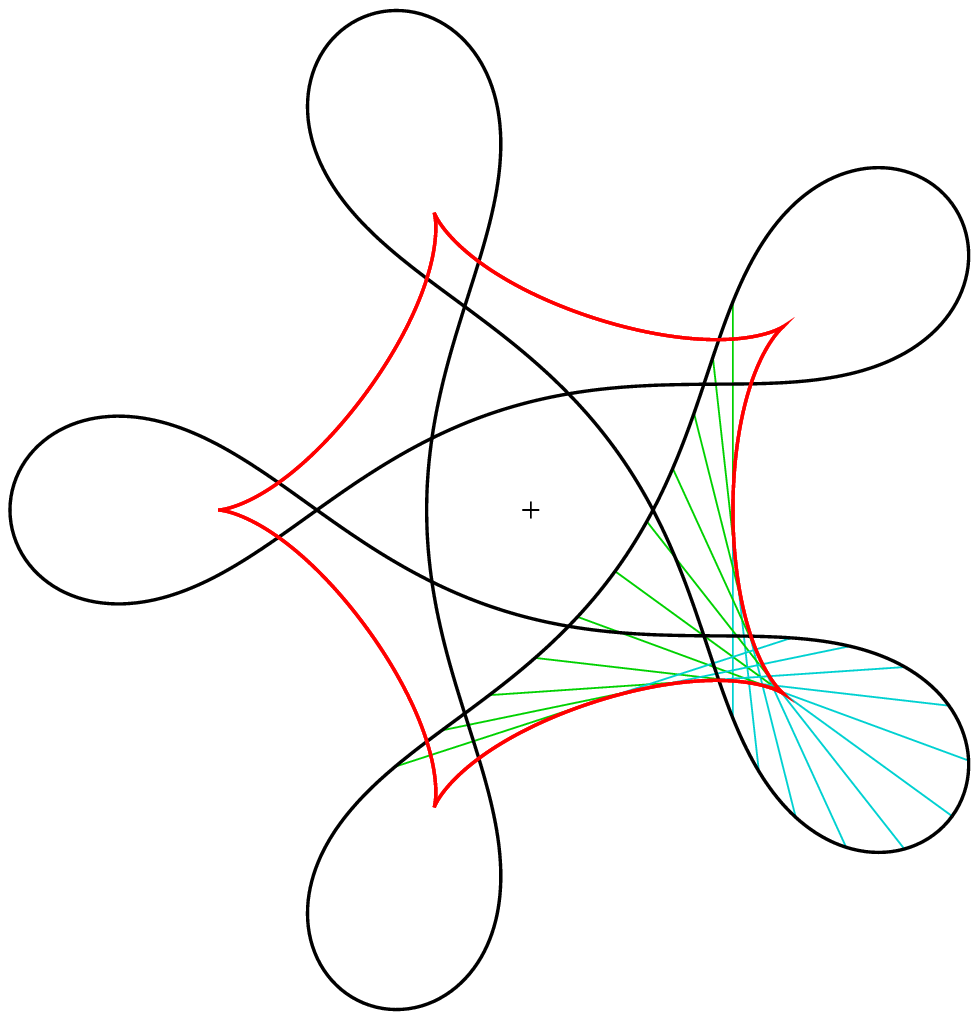,scale=0.4,angle=-90}\\
\Figur{fgp3p5a}{$m/n=3/5$, $\epsilon=1.5$}}
\etm

\btm
\parbox[b]{6cm}{\epsfig{file=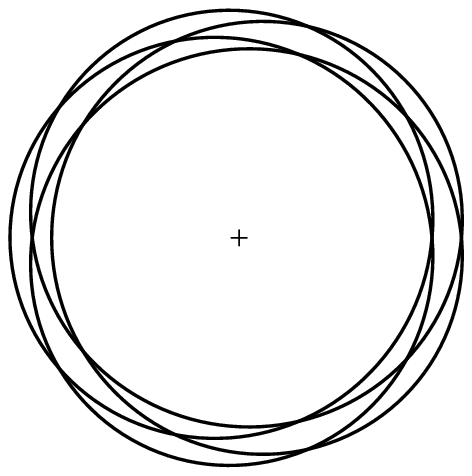,scale=0.5,angle=-90}\\
\Figur{fgp4p5a}{$m/n=4/5$, $\epsilon=0.1$}}
\hspace{1cm}
\parbox[b]{5cm}{\epsfig{file=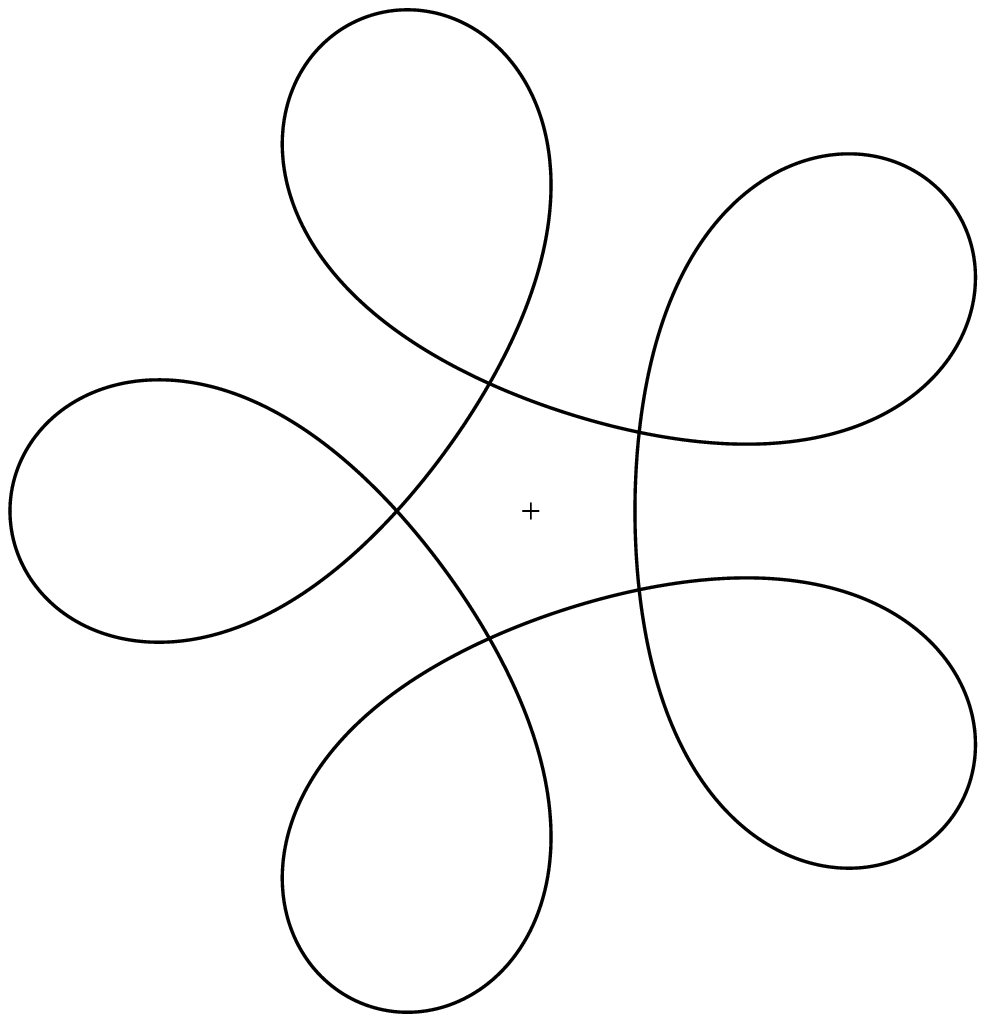,scale=0.4,angle=-90}\\
\Figur{fgp4p5b}{$m/n=4/5$, $\epsilon=1.5$}}
\etm

Note that the curve in fig. \ref{fgp4p5b} obeys the convexity condition
(\ref{convex}) ($\heps^2=-6.032657$).
There are no non-trivial solutions for the chord for $m/n=4/5$.
Some examples for negative $m/n$ are:
\medskip

\btm
\epsfig{file=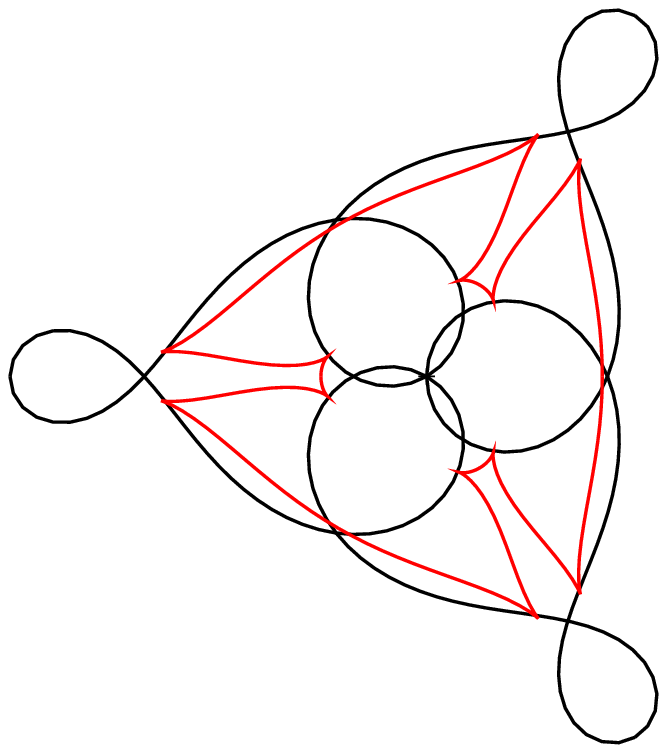,scale=0.5,angle=0}
\hspace{1cm}
\parbox[b]{4cm}{\Figur{fgm1p3b}{$m/n=-1/3$, $\epsilon=1.0$}}
\etm

\btm
\parbox[b]{5cm}{\epsfig{file=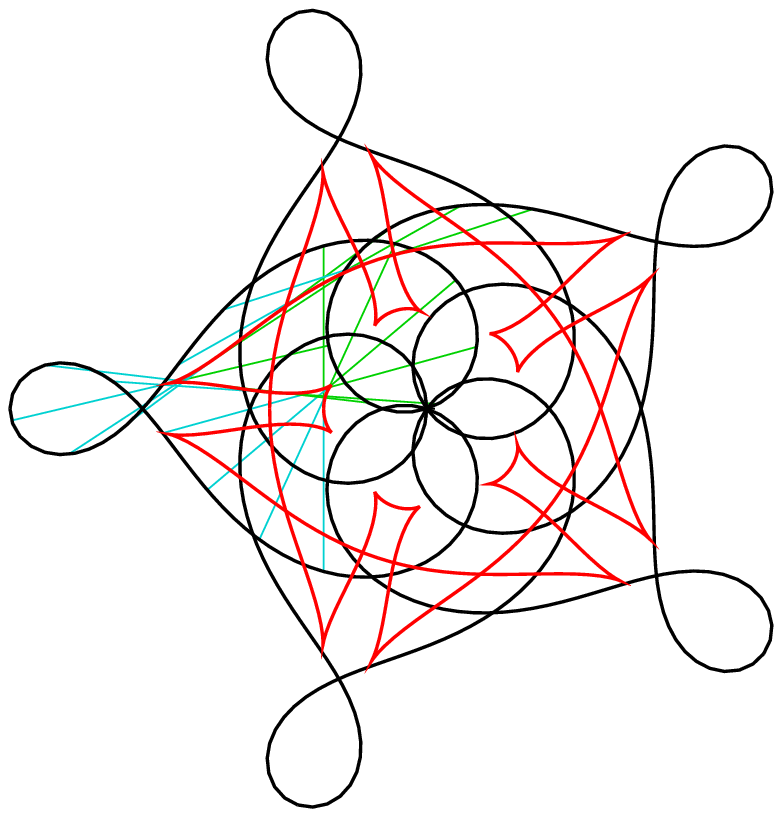,scale=0.5,angle=180}\\
\Figur{fgm2p5b}{$m/n=-2/5$, $\epsilon=1.0$}}
\hspace{1cm}
\parbox[b]{5cm}{\epsfig{file=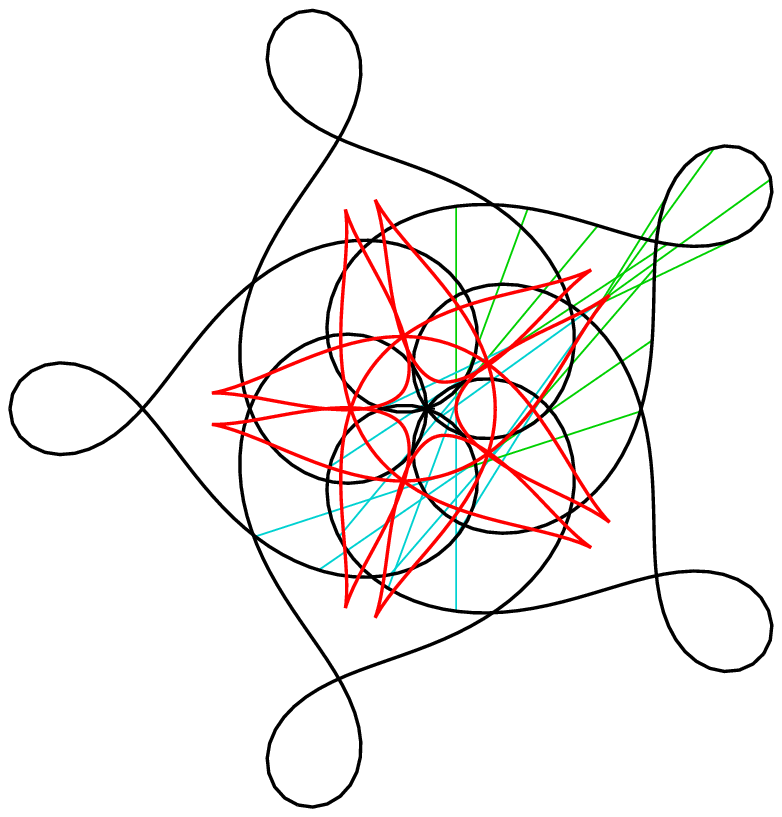,scale=0.5,angle=180}\\
\Figur{fgm2p5a}{$m/n=-2/5$, $\epsilon=1.0$}}
\etm

\btm
\parbox[b]{5cm}{\epsfig{file=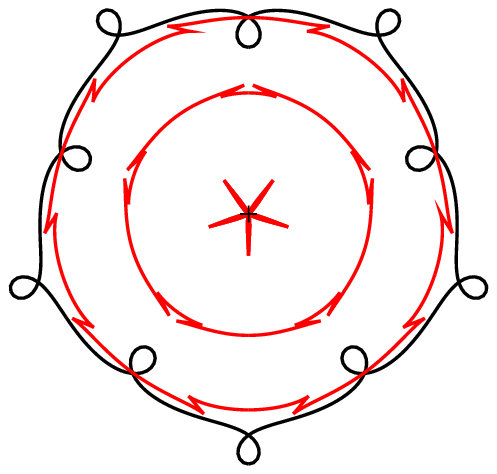,scale=0.5,angle=180}\\
\Figur{fgm1p5c}{$m/n=-1/5$, $\epsilon=0.2$}}
\hspace{1cm}
\parbox[b]{5cm}{\epsfig{file=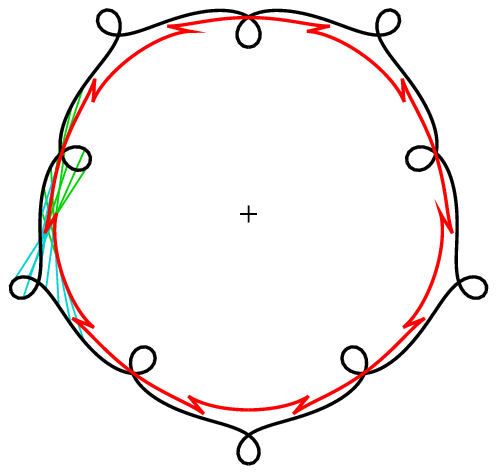,scale=0.5,angle=180}\\
\Figur{fgm1p5b}{$m/n=-1/5$, $\epsilon=0.2$}}
\etm

\btm
\parbox[b]{5cm}{\epsfig{file=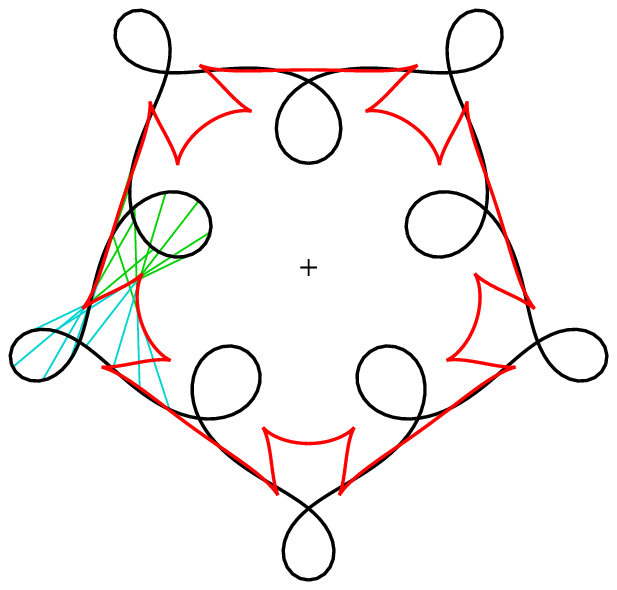,scale=0.5,angle=180}\\
\Figur{fgm1p5a}{$m/n=-1/5$, $\epsilon=0.5$}}
\hspace{1cm}
\parbox[b]{5cm}{\epsfig{file=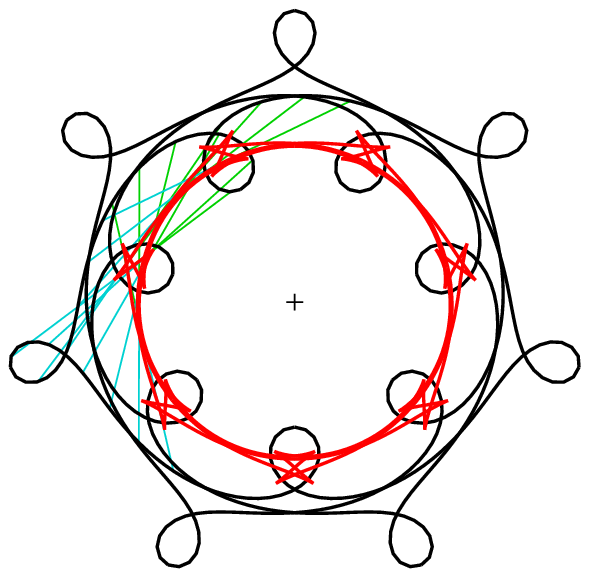,scale=0.5}\\
\Figur{fgm3p7a}{$m/n=-3/7$, $\epsilon=0.4$}}
\etm

\btm
\parbox[b]{5cm}{\epsfig{file=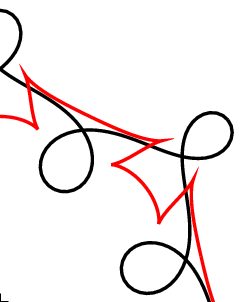,scale=0.5}\\
\Figur{fgm1p7b}{$m/n=-1/7$, $\epsilon=0.4$}}
\hspace{1cm}
\parbox[b]{5cm}{\epsfig{file=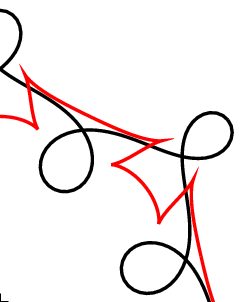,scale=0.5}\\
\Figurm{fgm1p7a}{$m/n=-1/7$, $\epsilon=0.4$}}
\etm

\btm
\parbox[b]{5cm}{\epsfig{file=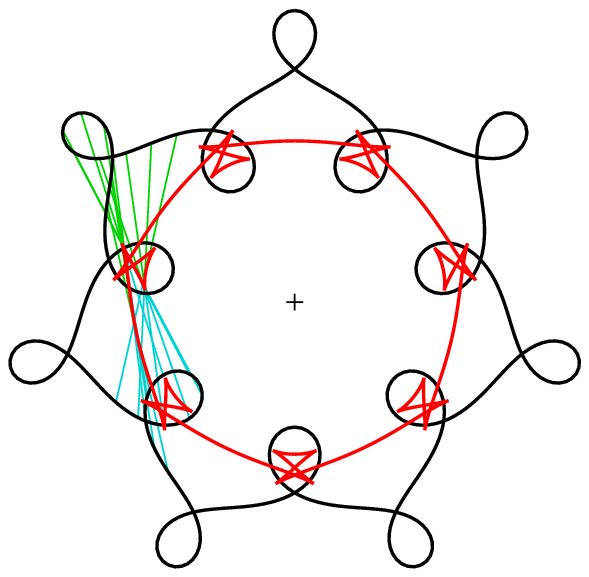,scale=0.5}\\
\Figurm{fgm1p7c}{$m/n=-1/7$, $\epsilon=0.4$}}
\hspace{1cm}
\parbox[b]{5cm}{\epsfig{file=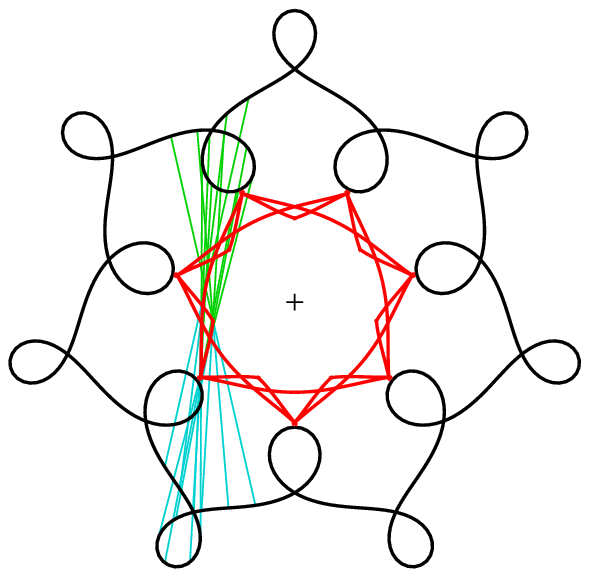,scale=0.5}\\
\Figurm{fgm1p7d}{$m/n=-1/7$, $\epsilon=0.4$}}
\etm

\btm
\parbox[b]{5cm}{\epsfig{file=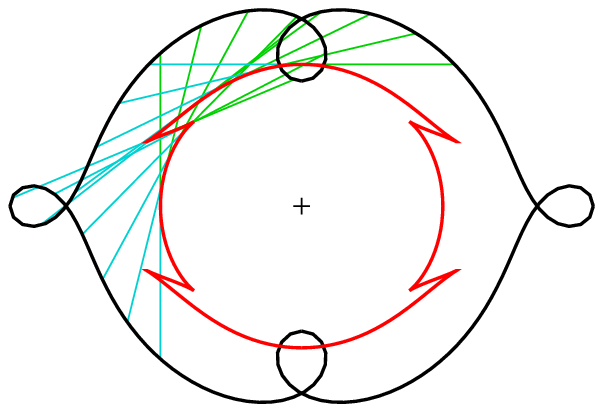,scale=0.5}\\
\Figur{fgm1p2a}{$m/n=-1/2$, $\epsilon=0.4$}}
\hspace{1cm}
\parbox[b]{5cm}{\epsfig{file=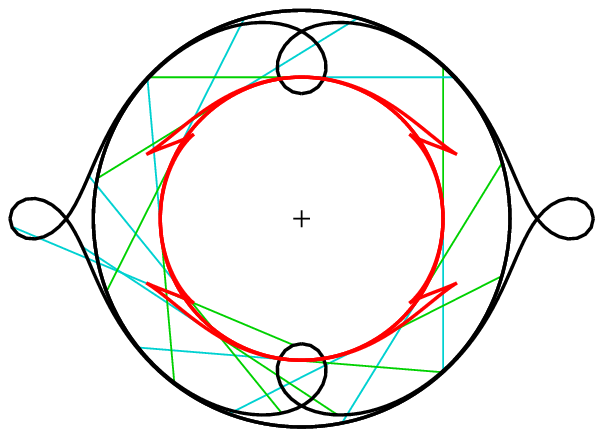,scale=0.5}\\
\Figur{fgm3p2a}{$m/n=-3/2$, $\epsilon=0.4$}}
\etm

For $m/n=0/1$ one obtains the {\it Eight}, which was already shown in figs.
\ref{fg0p1a} to \ref{fg0p1d}. Here some of the envelopes $\gamma$ are shown.
\medskip

\btm
\parbox[b]{5cm}{\epsfig{file=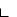,scale=0.5}\\
\Figurm{fg0p1f}{$m/n=0/1$, $\epsilon=0.5$}}
\hspace{1cm}
\parbox[b]{5cm}{\epsfig{file=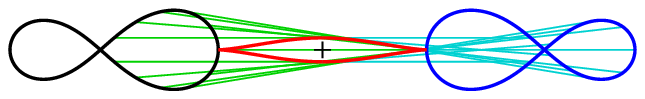,scale=0.5}\\
\Figurm{fg0p1g}{$m/n=0/1$, $\epsilon=0.5$}}
\etm

\btm
\parbox[b]{5cm}{\epsfig{file=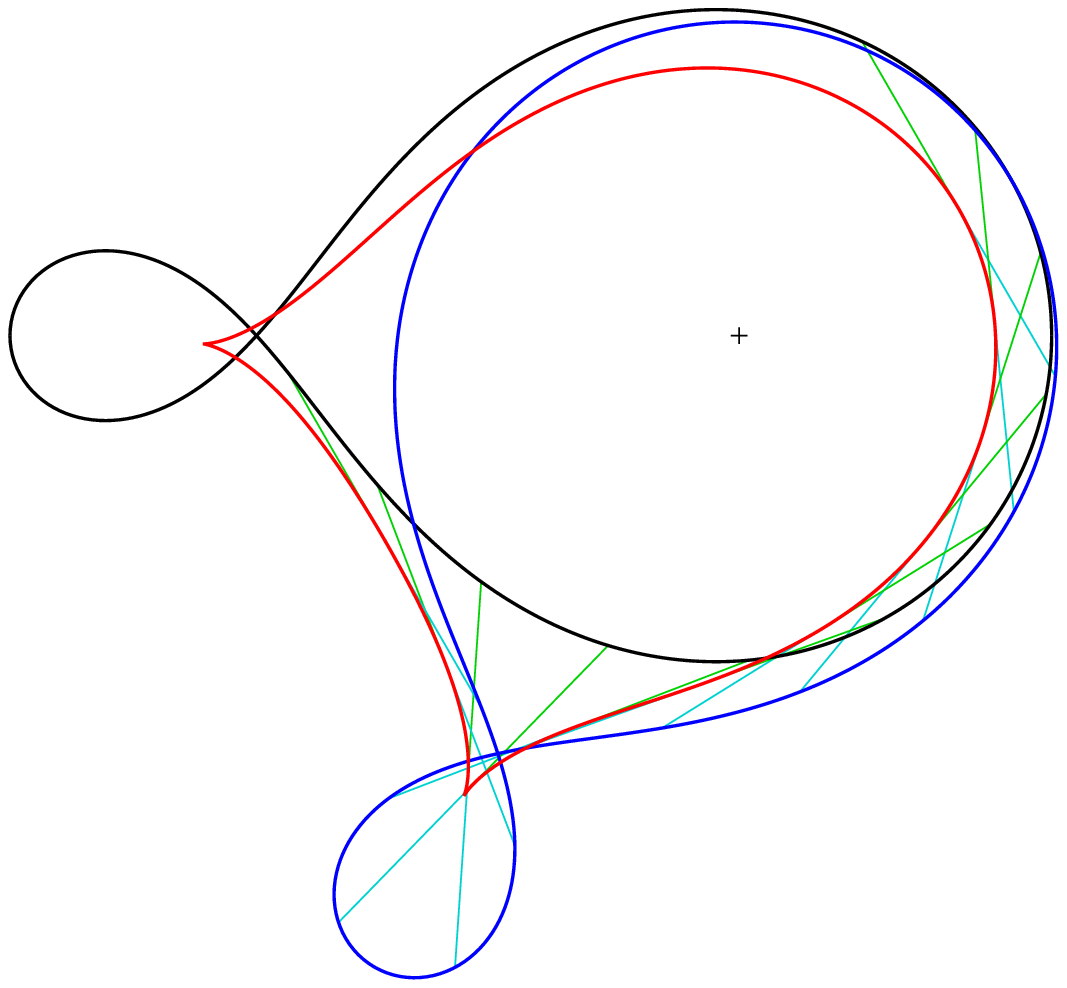,scale=0.4}\\
\Figur{fg0p1h}{$m/n=0/1$, $\epsilon=2.5$}}
\hspace{1cm}
\parbox[b]{5cm}{\epsfig{file=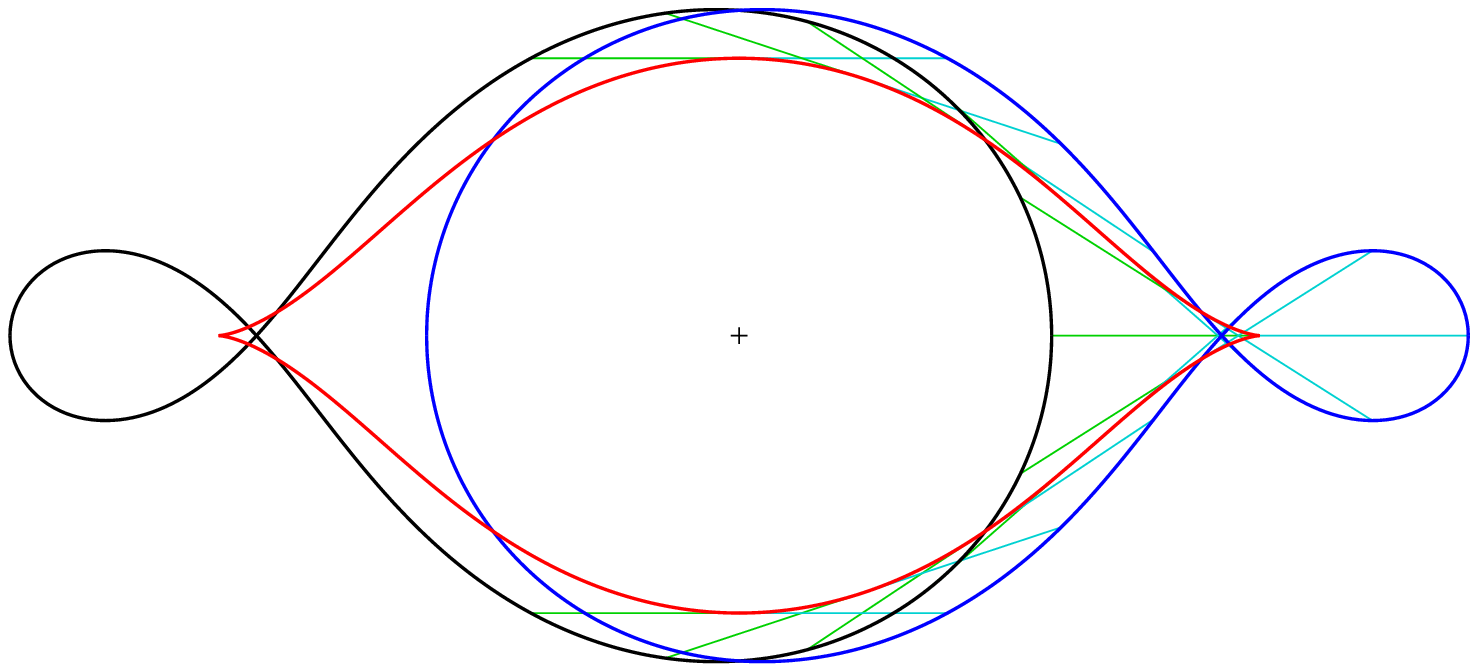,scale=0.4}\\
\Figur{fg0p1i}{$m/n=0/1$, $\epsilon=2.5$}}
\etm

\btm
\parbox[b]{5cm}{\epsfig{file=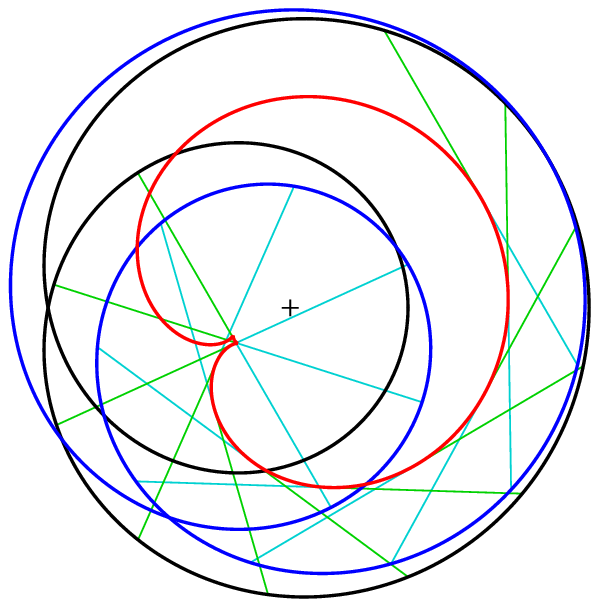,scale=0.4}\\
\Figur{fg0p1j}{$m/n=0/1$, $\epsilon=2.5$}}
\hspace{1cm}
\parbox[b]{5cm}{\epsfig{file=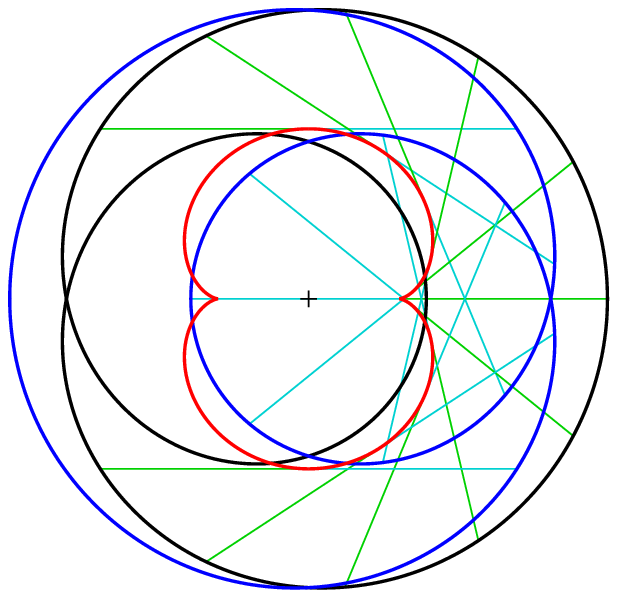,scale=0.4}\\
\Figur{fg0p1k}{$m/n=0/1$, $\epsilon=2.5$}}
\etm

\btm
\parbox[b]{5cm}{\epsfig{file=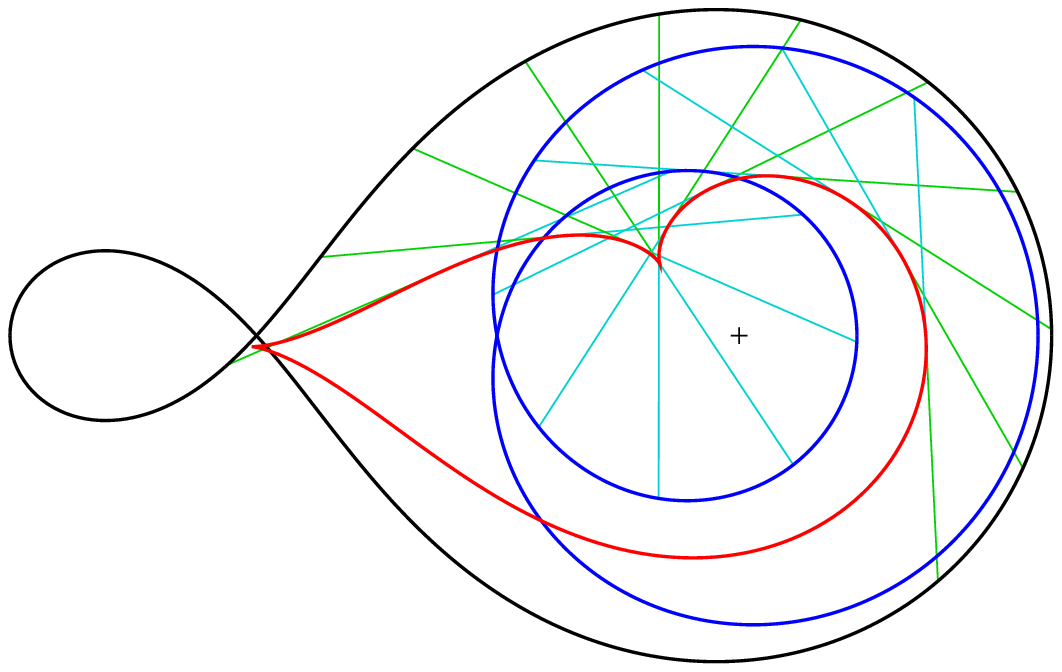,scale=0.4}\\
\Figur{fg0p1l}{$m/n=0/1$, $\epsilon=2.5$}}
\hspace{1cm}
\parbox[b]{5cm}{\epsfig{file=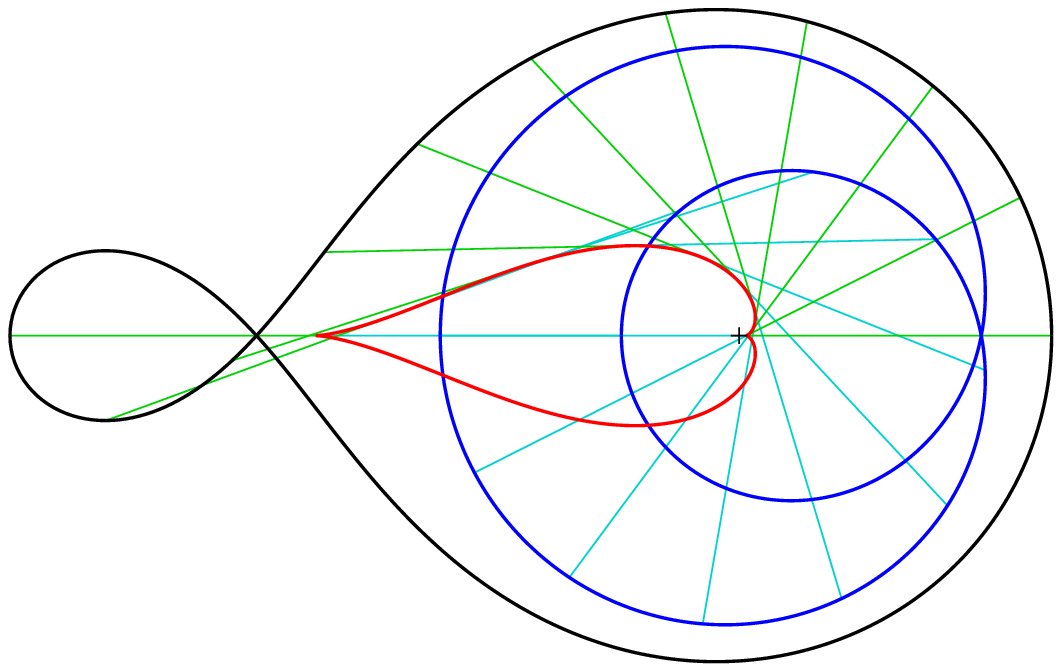,scale=0.4}\\
\Figur{fg0p1m}{$m/n=0/1$, $\epsilon=2.5$}}
\etm

Another example is shown in fig. \ref{fg0p1e}.

\section{The Limit Case $\mu=0$\label{calD}}

The discriminant $\cal D$ vanishes in
the limit $\mu=0$ and the elliptic functions degenerate to 
trigonometric and exponential functions.
In this limit two of the radii $r_i$ coincide.

\subsection{The case $\epsilon>2$}

In this region one obtains with
\be
r_4 = r_0(1+\epsilon), \quad r_3 = r_0(1-\epsilon), \quad r_2=r_1=-r_0
\ee
the expressions for $e_i$
\be
e_1=e_2 = \frac{4-\epsilon^2}{12\epsilon^2 r_0^2}<0, \quad
e_3 = \frac{\epsilon^2-4}{6\epsilon^2 r_0^2}>0,
\ee
and the functions
\bea
\wp(u) &=& \frac{2\lambda^2}3 + \lambda^2 \cot^2(\lambda u), \label{wp1}\\
\zeta(u) &=& \frac{\lambda^2 u}3 + \lambda \cot(\lambda u), \label{zet1} \\
\sigma(u) &=& \frac 1{\lambda} \ex{\lambda^2 u^2/6} \sin(\lambda u),
\label{sig1} \\
\lambda &=& \frac{\sqrt{\epsilon^2-4}}{2\epsilon r_0}.
\eea
In this degenerate case one has
\be
\omega = \frac{\pi}{2\lambda} = \frac{\pi\epsilon r_0}{\sqrt{\epsilon^2-4}},
\quad \omega'=\infty.
\ee
From
\be
\wp(v) = -\frac{(\epsilon+2)(\epsilon+10)}{12\epsilon^2 r_0^2} \label{pvlin}
\ee
one concludes
\be
\cot(\lambda v) = -\ie\sqrt{\frac{\epsilon+2}{\epsilon-2}}, \quad
\cos(2\lambda v) = \frac{\epsilon}2, \quad
\sin(2\lambda v) = \ie \frac{\sqrt{\epsilon^2-4}}2 \label{cotv}
\ee
and
\be
\psi_{\rm c} = 2\pi \left( 2-\frac{\epsilon}{\sqrt{\epsilon^2-4}}\right).
\ee
Evaluation of eq. (\ref{zu}) yields in a first step
\bea
z &=& \frac{P_r}{2\sigma^2(2v)} \frac{\sigma(u-3v)}{\sigma(u+v)}
\ex{2\zeta(2v)u} \\
&=& -\frac{r_0 \sin(\lambda(u-3v))}{\sin(\lambda(u+v))}
\ex{2u\lambda\cot(2\lambda v)}.
\eea
Multiplication of numerator and denominator by $\sin(\lambda(u-v))$ and use of
$\sin a$ $\sin b = \frac 12 (\cos(a-b) - \cos(a+b))$ yields
\be
z = -r_0\frac{\cos(2\lambda v)-\cos(\lambda(2u-4v))}
{\cos(2\lambda v)-\cos(2\lambda u)} \ex{2u\lambda\cot(2\lambda v)}.
\ee
From eqs. (\ref{cotv}) one obtains
\be
\cos(4\lambda v) = \frac{\epsilon^2}2-1, \quad
\sin(4\lambda v) = \ie \frac{\epsilon\sqrt{\epsilon^2-4}}2.
\ee
This yields
\be
z = r_0\frac{(\epsilon^2-2)\cos(2\lambda u)
+\ie\epsilon\sqrt{\epsilon^2-4}\sin(2\lambda u)-\epsilon}
{\epsilon-2\cos(2\lambda u)} \ex{-\ie u/r_0}.
\ee

Starting from eqs. (\ref{chdu}) and (\ref{ell0}) and using the expressions
(\ref{wp1}) to (\ref{sig1}) one obtains
the condition for the line segment of constant distance and its length
\bea
\tan(2\lambda\du) &=& 2\lambda r_0 \tan(\frac{\du}{r_0}-\frac{\chi-\hat\chi}2),
\\
\ell^2 &=& \frac{r_0^2}{1+\frac{\epsilon^2-4}{\epsilon^2} \cot^2(2\lambda\du)}.
\eea
Examples of the curves $\Gamma$ are shown in figs. \ref{fgp1p3r} to
\ref{fgm2p1a}.
\medskip

\btm
\parbox[b]{5.5cm}{\epsfig{file=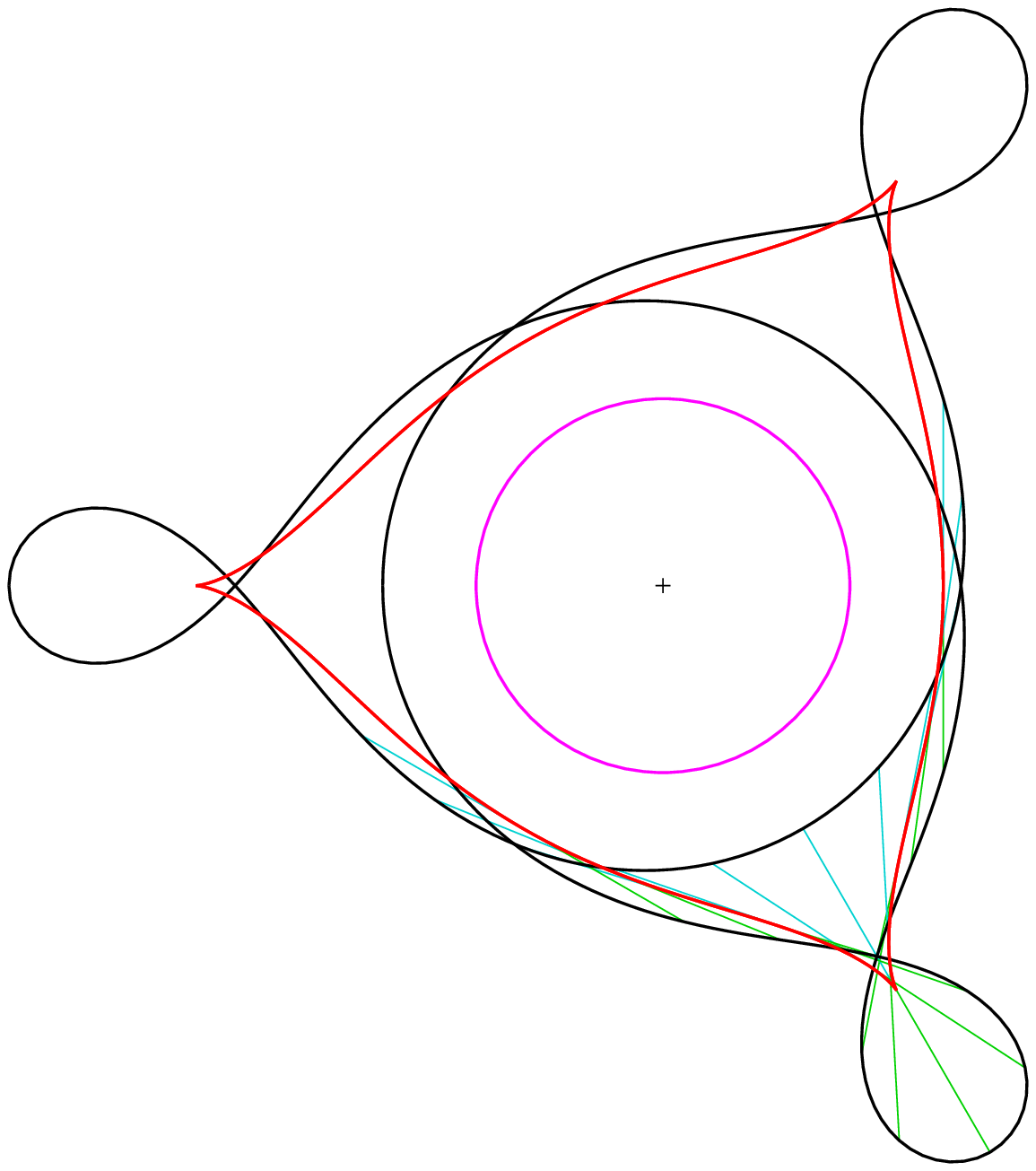,scale=0.4}\\
\Figurm{fgp1p3r}{$m/n=1/3$, $\epsilon=2.5$}}
\hfill
\parbox[b]{5.5cm}{\epsfig{file=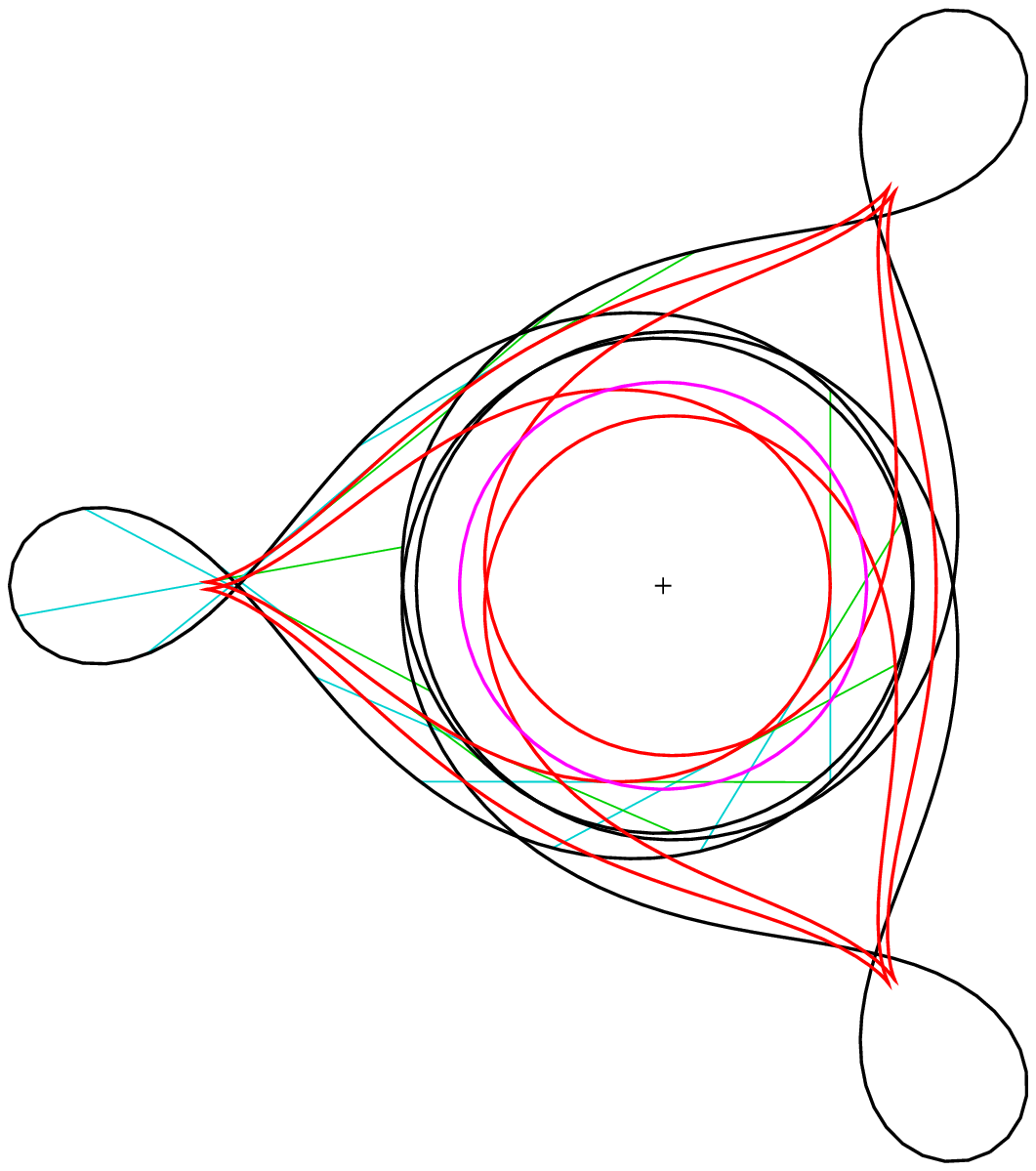,scale=0.4}\\
\Figurm{fgm1p3c}{$m/n=-1/3$, $\epsilon=2.213594$}}
\etm

\btm
\parbox[b]{5.5cm}{\epsfig{file=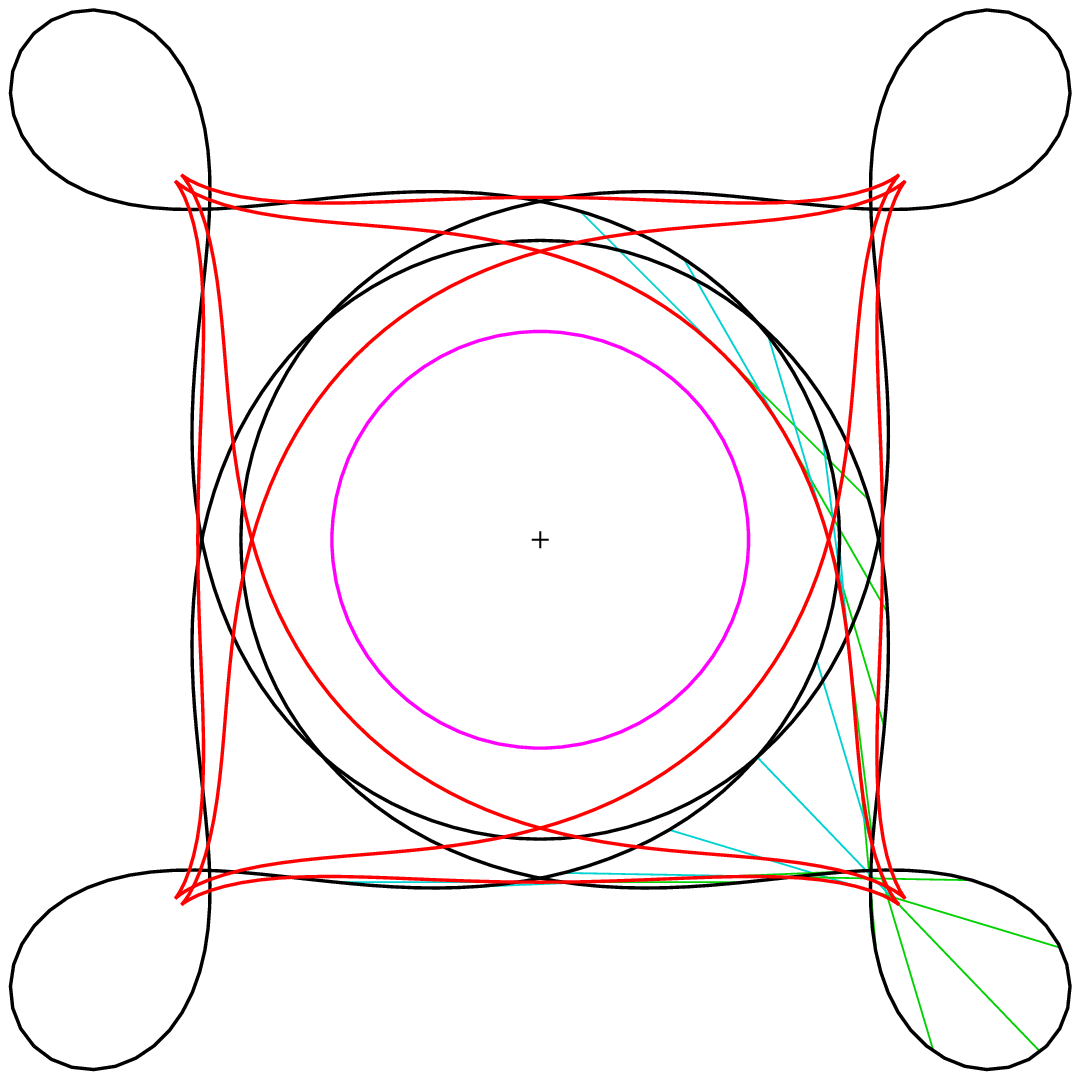,scale=0.4}\\
\Figur{fgp1p4j}{$m/n=1/4$, $\epsilon=2.437087$}}
\hfill
\parbox[b]{5.5cm}{\epsfig{file=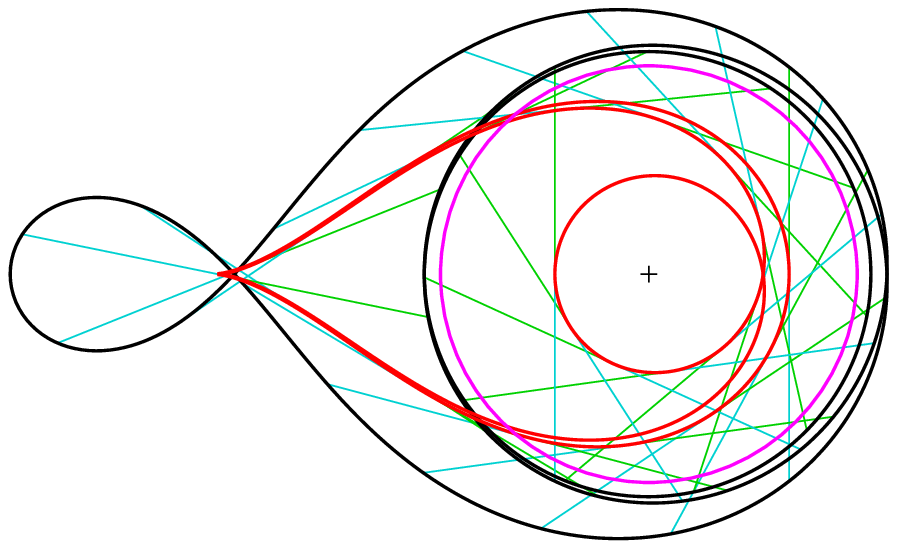,scale=0.5}\\
\Figurm{fgm2p1a}{$m/n=-2/1$, $\epsilon=2.06559$}}
\etm

\subsection{The case $\epsilon<2$}

In the case of $\epsilon<2$ one has
\be
r_4 = r_0(1+\epsilon), \quad r_3=r_2=-r_0, \quad r_1=r_0(1-\epsilon).
\ee
Then one obtains
\be
e_1=\frac{\epsilon^2-4}{6\epsilon^2 r_0^2}<0, \quad
e_2=e_3 = \frac{4-\epsilon^2}{12\epsilon^2 r_0^2}>0,
\ee
and the functions
\bea
\wp(u) &=& -\frac{2\lambda^2}3 + \lambda^2 \coth^2(\lambda u), \\
\zeta(u) &=& -\frac{\lambda^2 u}3 + \lambda \coth(\lambda u), \\
\sigma(u) &=& \frac 1{\lambda} \ex{-\lambda^2 u^2/6} \sinh(\lambda u), \\
\lambda &=& \frac{\sqrt{4-\epsilon^2}}{2\epsilon r_0}.
\eea
In this degenerate case the half-periods approach
\be
\omega_3=\infty, \quad
\omega_1=\frac{\pi\ie}{2\lambda} = \frac{\pi\ie\epsilon
r_0}{\sqrt{4-\epsilon^2}}.
\ee
Again one obtains (\ref{pvlin}) and concludes
\be
\coth(\lambda v) = -\ie\sqrt{\frac{2+\epsilon}{2-\epsilon}}, \quad
\cosh(2\lambda v) = \frac{\epsilon}2, \quad
\sinh(2\lambda v) = \ie \frac{\sqrt{4-\epsilon^2}}2. \label{cothv}
\ee
Eq. (\ref{zu}) is evaluated in a first step
\bea
z &=& \frac{P_r}{2\sigma^2(2v)} \frac{\sigma(u-3v)}{\sigma(u+v)}
\ex{2\zeta(2v)u} \\
&=& -r_0\frac{\sinh(\lambda(u-3v))}{\sinh(\lambda(u+v))}
\ex{2u\lambda\coth(2\lambda v)}.
\eea
Multiplication of numerator and denominator by $\sinh(\lambda(u-v))$ and use
of $\sinh a \sinh b = \frac 12 (\cosh(a+b) - \cosh(a-b))$ yields
\be
z = r_0\frac{\cosh(\lambda(2u-4v))-\cosh(2\lambda v)}
{\cosh(2\lambda u)-\cosh(2\lambda v)} \ex{2u\lambda\coth(2\lambda v)}.
\label{zem1}
\ee
From eqs. (\ref{cothv}) one obtains
\be
\cosh(4\lambda v) = \frac{\epsilon^2}2-1, \quad
\sinh(4\lambda v) = \ie \frac{\epsilon\sqrt{\epsilon^2-4}}2.
\ee
This yields
\be
z = r_0\frac{(2-\epsilon^2)\cosh(2\lambda u)
+\ie\epsilon\sqrt{4-\epsilon^2}\sinh(2\lambda u)+s\epsilon}
{2\cosh(2\lambda u)-s\epsilon} \ex{-\ie u/r_0} \label{zem}
\ee
with $s=+1$. Whereas for $\epsilon>2$ the radius of the curve oscillates between
$|r_3|$ and $r_4$, there are two curves for $\epsilon<2$, one with largest
distance $r_4=r_0(1+\epsilon)$, the other with minimal distance
$|r_1|=r_0|\epsilon-1|$ from the
origin. This second curve is obtained by adding $\omega_1$ to $v$, or
equivalently $\pi\ie/2$ to $\lambda v$. Thus $\cosh(4\lambda v)$,
$\sinh(4\lambda v)$, $\coth(2\lambda v)$ are unchanged, but
$\cosh(2\lambda v)$ changes sign. Starting from eq. (\ref{zem1}) one finds that
for the second curve one has to choose $s=-1$ in eq. (\ref{zem}). Both these
curves approach asymptotically the circle with radius $r_0$.

The condition for the line segment of constant length and its length read
\bea
\tanh(2\lambda\du) &=& 2\lambda r_0\tan(\frac{\du}{r_0}-\frac{\chi-\hat\chi}2),
\\
0 \le \frac{2\ell}{r_0} &=& 
\frac{2\epsilon}{\sqrt{(4-\epsilon^2)\coth^2(2\lambda\du)+\epsilon^2}}
\le \epsilon,
\eea
and that for the line segment connecting both curves
\bea
\tanh(2\lambda\du) &=& \frac 1{2\lambda r_0}
\cot(\frac{\du}{r_0}-\frac{\chi-\hat\chi}2), \\
\epsilon \le \frac{2\ell}{r_0} &=&
\frac{2\epsilon}{\sqrt{(4-\epsilon^2)\tanh^2(2\lambda\du)+\epsilon^2}}
\le 2.
\eea
Curves for $\epsilon=1.6$ are shown in figs. \ref{fgp1zo} to \ref{fgp1zn}.
Figure \ref{fgp1zn} is shown in the movie\cite{movie} for a different solution
$\du$.
\medskip

\btm
\parbox[b]{6cm}{\epsfig{file=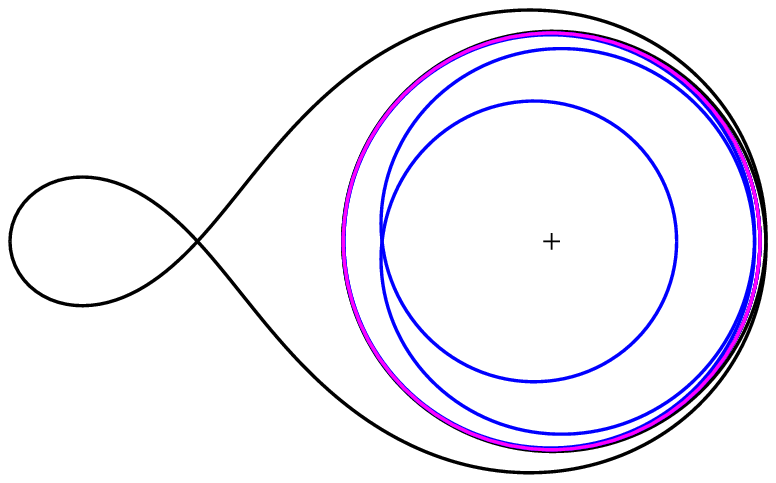,scale=0.5}\\
\Figur{fgp1zo}{$m/n=1/0$, $\epsilon=1.6$}}
\hfill
\parbox[b]{6cm}{\epsfig{file=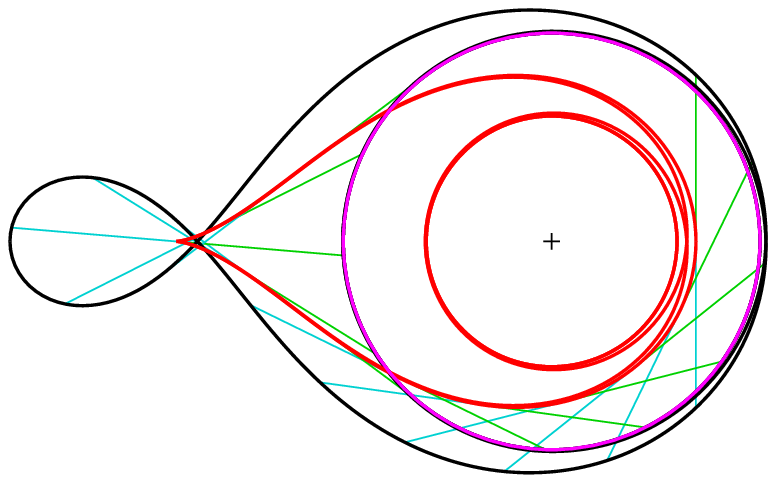,scale=0.5}\\
\Figurm{fgp1zm}{$m/n=1/0$, $\epsilon=1.6$}}
\etm

\btm
\parbox[b]{6cm}{\epsfig{file=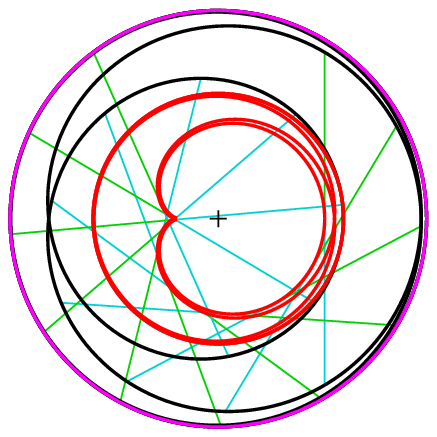,scale=0.5}\\
\Figurm{fgp1zl}{$m/n=1/0$, $\epsilon=1.6$}}
\hfill
\parbox[b]{6cm}{\epsfig{file=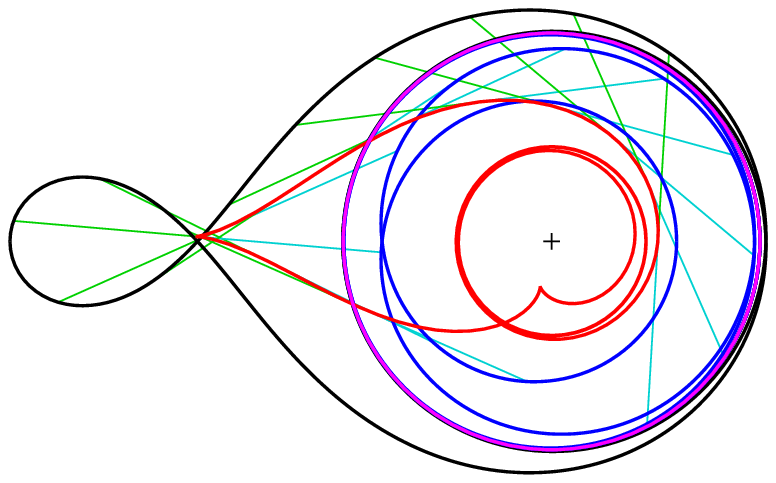,scale=0.5}\\
\Figurm{fgp1zn}{$m/n=1/0$, $\epsilon=1.6$}}
\etm

\subsection{The limit $\epsilon=2$}

In the limit $\epsilon=2$ one has
\be
r_4=3r_0, \quad r_3=r_2=r_1=-r_0.
\ee
This yields
\be
e_1=e_2=e_3=0.
\ee
The Weierstrass-functions are degenerate to
\be
\wp(u) = \frac 1{u^2}, \quad
\zeta(u) = \frac 1u, \quad
\sigma(u) = u
\ee
and there is no periodicity left, $\omega=\omega'=\infty$.
With $\wp(v) = -1/r_0^2$ one obtains
\be
v = \frac{\ie}{r_0}
\ee
and
\be
z=-\frac{u-3\ie r_0}{u/r_0+\ie} \ex{-\ie u/r_0}
=\frac{3r_0+\ie u}{1-\ie u/r_0} \ex{-\ie u/r_0}.
\ee
Again this curve approaches asymptotically the unit circle.
The condition for the line segment of constant length reduces to
\be
\ex{\ie(-2\du/r_0+\chi-\hat\chi)} = \frac{1-\ie\du/r_0}{1+\ie\du/r_0},
\ee
equivalently this may be written
\be
\frac{\du}{r_0} = \tan\left(\frac{\du}{r_0}-\frac{\chi-\hat\chi}2\right)
\ee
and $\ell$ is given by
\be
\ell^2 = \frac{r_0^2\du^2}{r_0^2+\du^2}.
\ee

\btm
\parbox[b]{5cm}{\epsfig{file=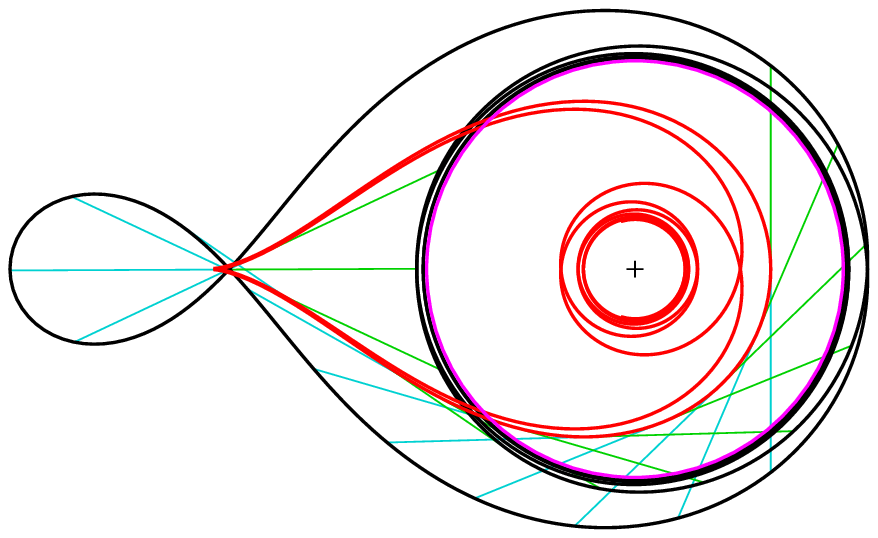,scale=0.5}\\
\Figurm{fgp1zf}{$m/n=1/0$, $\epsilon=2$}}
\hfill
\parbox[b]{6.5cm}{\epsfig{file=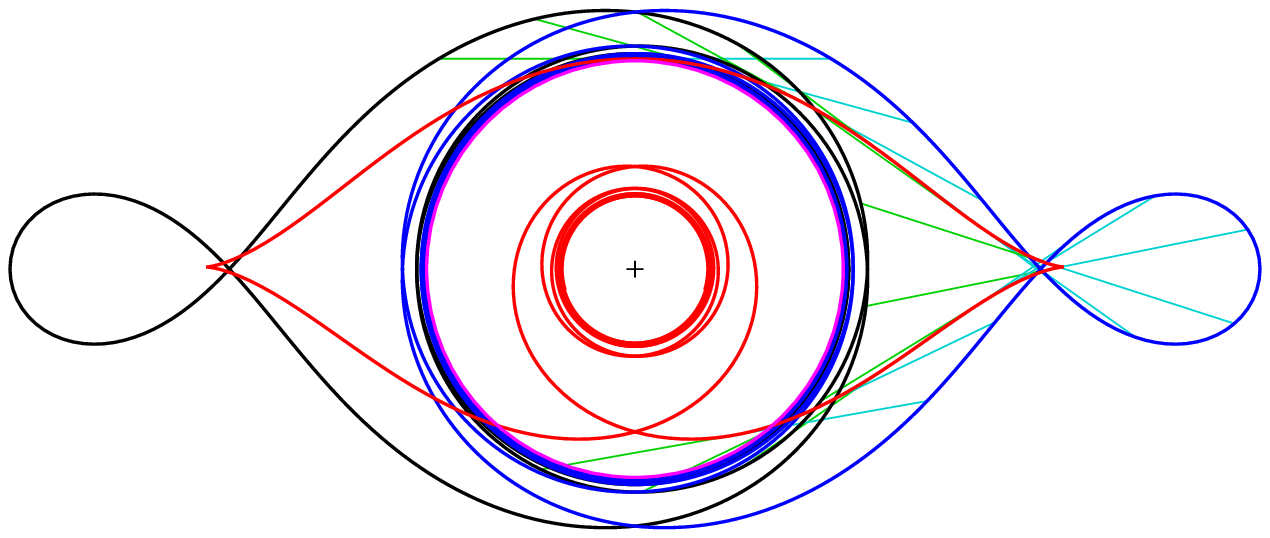,scale=0.5}\\
\Figur{fgp1zd}{$m/n=1/0$, $\epsilon=2$}}
\etm

\subsection{Elementary construction\label{elem}}

The circle
\be
\hat z= r_0 \ex{-\ie u/r_0}
\ee
is a second solution for $\epsilon>2$. For $\epsilon\le 2$ this circle is
approached asymptotically for $u\rightarrow\pm\infty$. In all three cases the
length of the line segment between this circle and the
non-circular curves given above is
\be
2\ell = \epsilon r_0.
\ee

This property allows an interesting construction of these curves.
Consider a straight line (indicated in the figures by cyan and green) of
length $2\ell=\epsilon r_0$. One end of it moves along the (magenta) circle of
radius $r_0$ in figs. \ref{fgp1p3n} to \ref{fgp1zi}. The middle point of this
line moves always in direction of the line itself, as if it were carried by a
wheel aligned in this direction, creating the red envelope. Then the other
end-point moves along the curve $\Gamma$.

If $\epsilon<2$, then there are two such curves, one outside and
one inside the circle. Both approach asymptotically the circle for
$u\rightarrow\pm\infty$. If $\epsilon\ge2$, then there is only one curve, which
for $\epsilon=2$ still approaches asymptotically the circle with
$r-r_0\propto 1/u^2$, whereas for $\epsilon<2$ the difference $r-r_0$ decays
exponentially with $u$. For $\epsilon>2$ there is only one curve, which
oscillates between radii $r_4$ and $r_3$, which allows for periodically closed
curves.

Examples for $\epsilon>2$ are shown in figs. \ref{fgp1p3n} to \ref{fgm2p1b}
below.
\medskip

\btm
\parbox[b]{5.5cm}{\epsfig{file=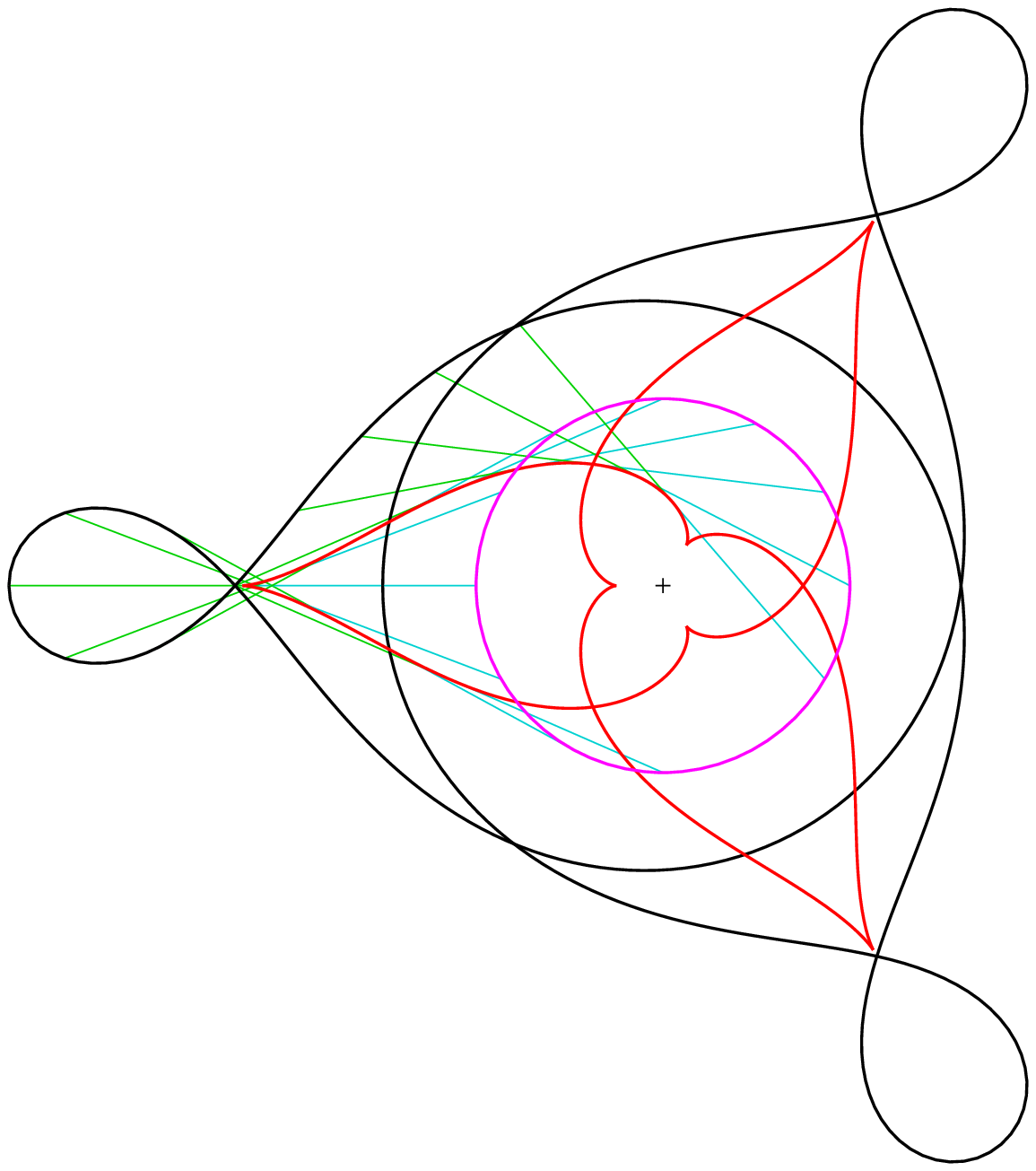,scale=0.4}\\
\Figurm{fgp1p3n}{$m/n=1/3$, $\epsilon=2.5$}}
\hfill
\parbox[b]{5.5cm}{\epsfig{file=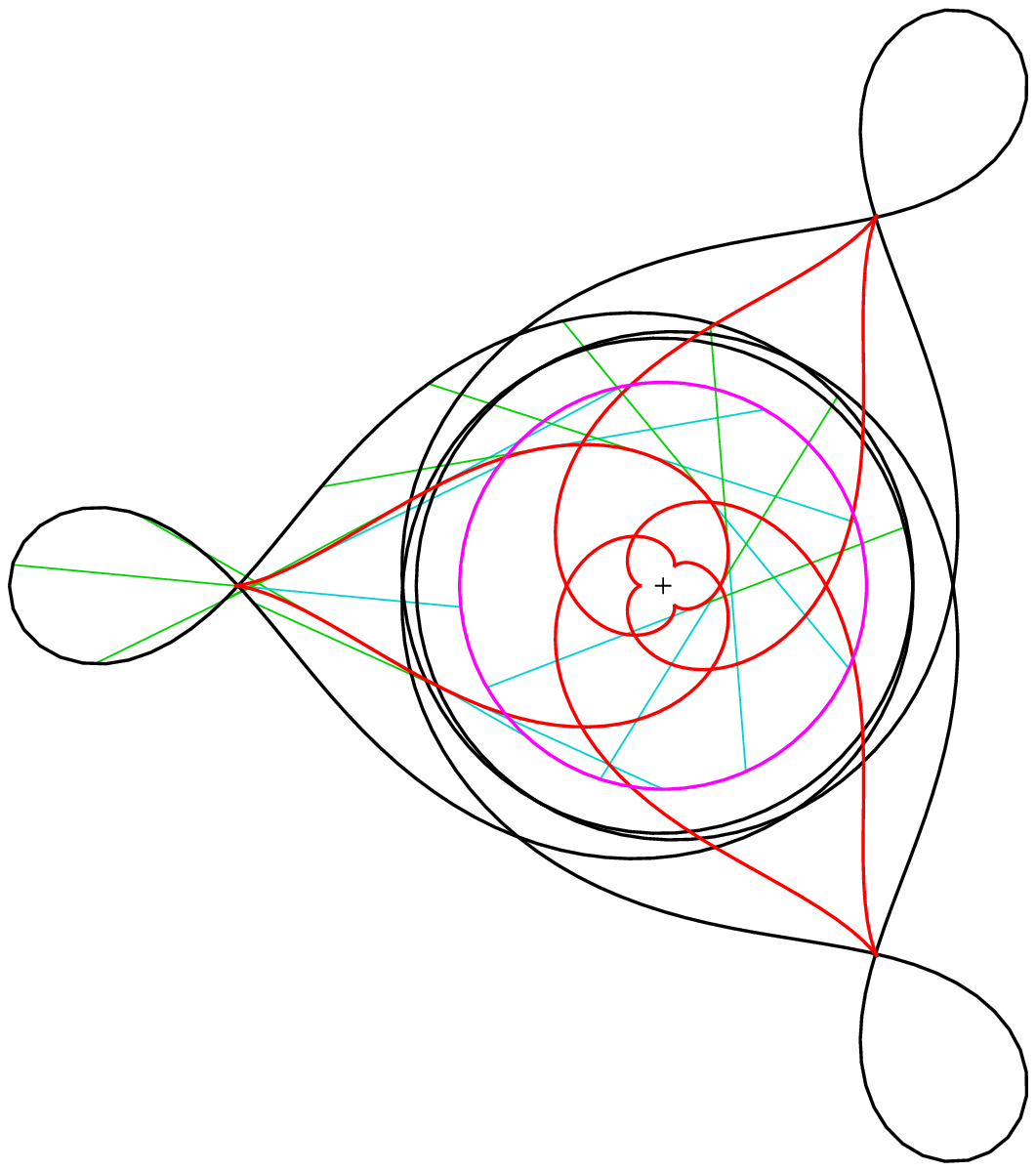,scale=0.4}\\
\Figurm{fgm1p3a}{$m/n=-1/3$, $\epsilon=2.213594$}}
\etm

\btm
\parbox[b]{5.5cm}{\epsfig{file=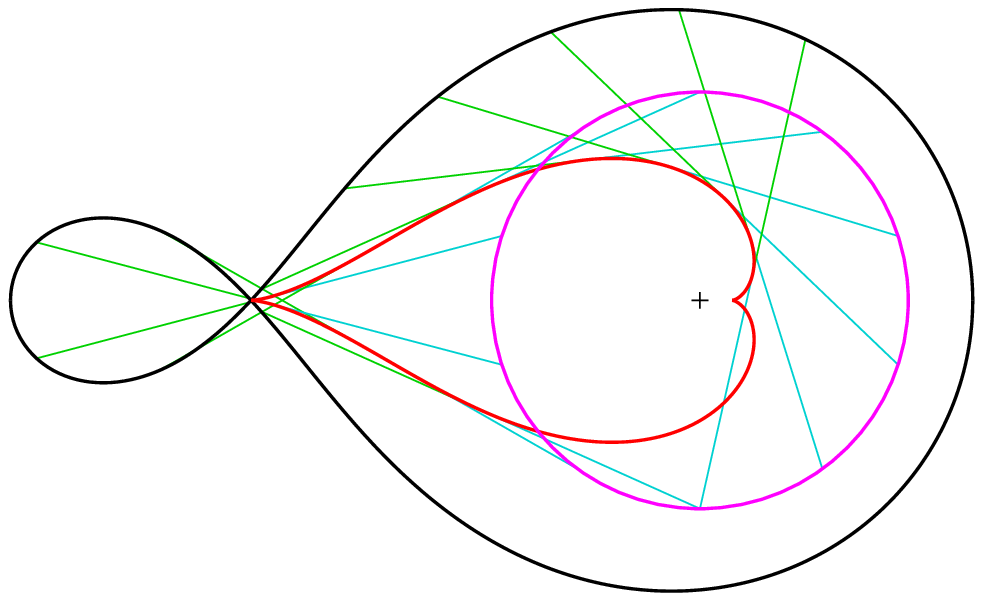,scale=0.4}\\
\Figur{fg0p1e}{$m/n=0/1$, $\epsilon=2.309401$}}
\hfill
\parbox[b]{5.5cm}{\epsfig{file=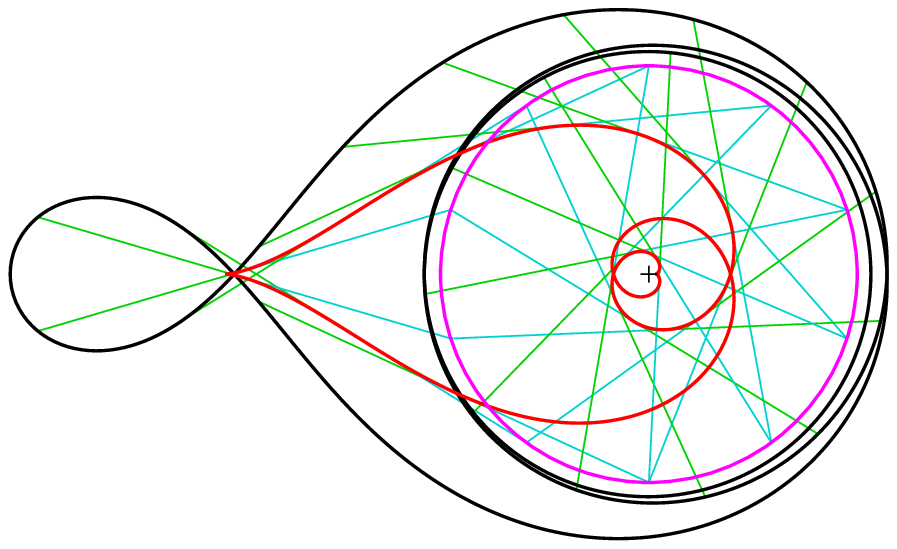,scale=0.4}\\
\Figurm{fgm2p1b}{$m/n=-2/1$, $\epsilon=2.06559$}}
\etm
Examples for $\epsilon<2$ are given in figs. \ref{fgp1zp} and \ref{fgp1zq}
showing the outer and inner curve.
\medskip

\btm
\parbox[b]{6cm}{\epsfig{file=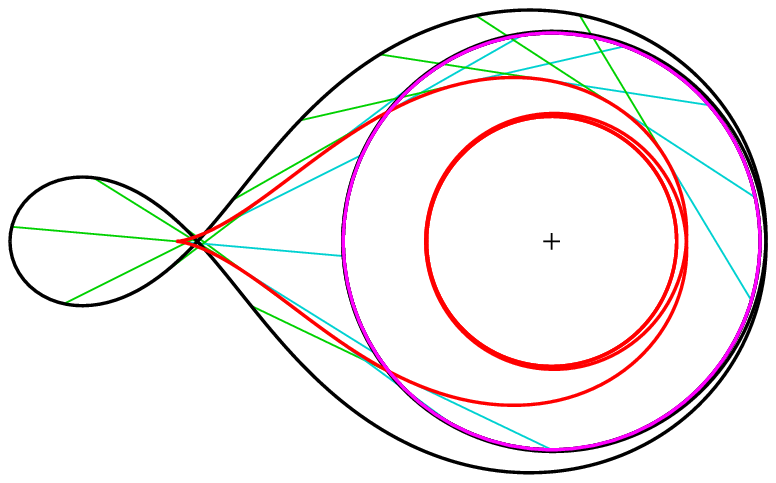,scale=0.5}\\
\Figurm{fgp1zp}{$m/n=1/0$, $\epsilon=1.6$}}
\hfill
\parbox[b]{6cm}{\epsfig{file=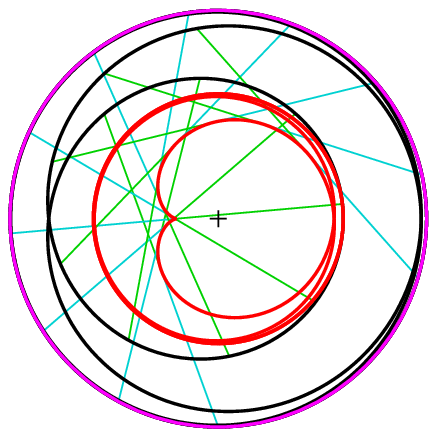,scale=0.5}\\
\Figurm{fgp1zq}{$m/n=1/0$, $\epsilon=1.6$}}
\etm
Finally the construction for $\epsilon=2$ is shown.
\medskip

\btm
\parbox[b]{5cm}{\epsfig{file=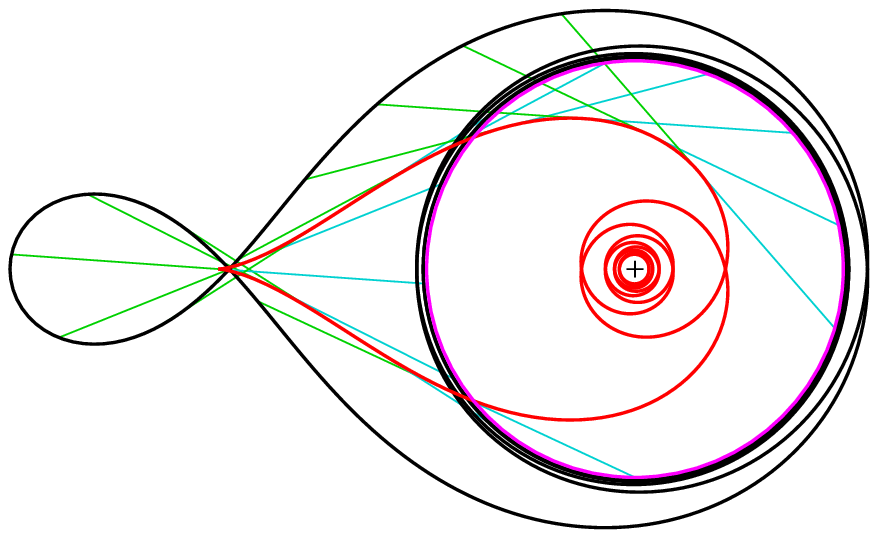,scale=0.5}}
\hspace{1cm}
\parbox[b]{5cm}{\Figurm{fgp1zi}{$m/n=1/0$, $\epsilon=2$}}
\etm

\section{Linear Case\label{linear}}

In this section the limit is considered, in which $r_0$
tends to infinity, but the differences between the radii $r_i$ is kept finite.
In subtracting this average radius by taking the limit
\be
\lim_{r_0\rightarrow \infty} (z(u) -r_0)
\ee
one obtains curves, which repeat under translation in $y$-direction. Therefore
this is called the linear case.
One can distinguish three cases.

\subsection{Two branches \label{lin2}}

The choice
\be
r_4=r_0+d, \quad r_3=-r_0-\hd, \quad
r_2=-r_0+\hd, \quad r_1=r_0-d, \quad d>\hd>0 \label{lin2a}
\ee
yields
\bea
a &=& \frac 1{2r_0(d^2-\hd^2)}, \label{lin2b} \\
e_3 &=& \frac{\hd^2+d^2+6d\hd}{3(d^2-\hd^2)^2} - \frac 1{12r_0^2}, \\
e_2 &=& \frac{\hd^2+d^2-6d\hd}{3(d^2-\hd^2)^2} - \frac 1{12r_0^2}, \\
e_1 &=& \frac{-2(\hd^2+d^2)}{3(d^2-\hd^2)^2} + \frac 1{6r_0^2}, \\
\wp(2v) &=& e_1 - \frac 1{4r_0^2}.
\eea
$v$ lies infinitesimal close to $\omega_1/2$,
\bea
\wp(\frac{\omega_1}2) &=& \frac{\hd^2-5d^2}{3(d^2-\hd^2)^2} + O(1/r_0^2), \\
\wp'(\frac{\omega_1}2) &=& -\frac{4d\ie}{(d^2-\hd^2)^2} + O(1/r_0^2), \\
\wp(v) &=& \wp(\frac{\omega_1}2) - \frac d{r_0(d^2-\hd^2)} + O(1/r_0^2).
\eea
This yields
\be
v = \frac{\omega_1}2 + \dv, \quad
\dv= -\ie \frac{d^2-\hd^2}{4r_0}.
\ee
One starts from
\be
z=r_4 \frac{\sigma(u-3v)\sigma(v)}{\sigma(u+v)\sigma(-3v)}\, \ex{2u\zeta(2v)}.
\ee
Due to the factor $r_4$ in front which is of order $r_0$, one has to take into
account all terms of order $1/r_0$ in the fraction and in the exponential on
the right hand side of this equation.
With
\be
\sigma(u-3v) = \sigma(u-\frac 32\omega_1-3\dv)
= -\sigma(u+\frac 12\omega_1-3\dv)
\ex{2(\frac 12\omega_1-u+3\dv)\zeta(\omega_1)}
\ee
and
\be
\sigma(u+\frac 12\omega_1-3\dv) = \sigma(u+v-4\dv)
= \sigma(u+v) \ex{-4\dv\zeta(u+v)}
\ee
one obtains
\be
z=r_4 \ex{2u(\zeta(2v)-\zeta(\omega_1))
+4\dv(\zeta(\frac 12\omega_1)-\zeta(u+\frac 12\omega_1))}.
\ee
Further
\be
\zeta(2v)-\zeta(\omega_1) = \zeta(\omega_1+2\dv) - \zeta(\omega_1) =
-2\dv\wp(\omega_1) = -2\dv e_1
\ee
and
\be
\zeta(u-v)-\zeta(u+v)+2\zeta(v) = \wp'(v)\, \cE 2u{\emptyset}{v,-v}
=\frac{\wp'(v)}{\wp(u)-\wp(v)}
\ee
yields
\bea
z &=& r_4 \ex{4\dv(-e_1 u+\zeta(\frac 12\omega_1)-\zeta(u+\frac 12\omega_1))} \\
&=& r_0+d-2r_0\dv
\Big(\zeta(u-v)+\zeta(u+v)-\frac{\wp'(v)}{\wp(u)-\wp(v)}+2 e_1 u\Big).
\label{z2c}
\eea
Adding a constant $y_0$ to $y$, which corresponds to the angle $\chi$ with
$r_0\chi=y_0$ the final result is
\bea
x-r_0 &=& \pm d
\mp \frac{2d(d^2-\hd^2)}{(d^2-\hd^2)^2\wp(u)-\frac 13(\hd^2-5d^2)}, \label{xlin}
\\
y &=& y_0 + \frac{d^2-\hd^2}2 \left(\zeta(u-\frac 12\omega_1)
+\zeta(u+\frac 12\omega_1) -\frac{4(d^2+\hd^2)}{3(d^2-\hd^2)^2} u\right),
\eea
where the upper signs apply.

Instead of starting from $r_4$ one can start from $r_1$. Then the signs
for the expression for $x-r_0$ have to be changed, as indicated by the lower
signs in equation (\ref{xlin}).

By inserting the extreme values of $\wp(u)$ one finds that $x-r_0$ oscillates
between $d$ and $\hd$ for one curve and between $-d$ and $-\hd$ for the other
curve. This yields the two branches. Shifting the arc parameter by $2\omega_3$
shifts the curve by $y_{\rm
per}$ in the $y$-direction and leaves $x$ invariant,
\bea
x(u+2\omega_3) = x(u), &&
y(u+2\omega_3) = y(u) + y_{\rm per}, \\
y_{\rm per} &=& -\frac{4(d^2+\hd^2)}{3(d^2-\hd^2)}\omega_3
+2(d^2-\hd^2)\zeta(\omega_3).
\eea

In order to obtain the condition for the line segment of constant length one
starts out
again from eq. (\ref{chdu}) and expands both sides up to order $O(1/r_0)$
\bea
\ex{\ie(\chi-\hat\chi)} &=& 1+\ie\frac{y_0-\hat y_0}{r_0}, \\
\zeta(2v)+\zeta(2\hat v) &=& \zeta(\omega_1+2\dv) +
\zeta(\omega_1+2n\omega_1+2\dv) \nn
&=& 2(n+1)\zeta(\omega_1)-4 e_1\dv, \\
(-)^{1-n}\frac{\sigma(2\du+v+\hat v)}{\sigma(2\du-v-\hat v)} &=&
\ex{2(1+n)(2\du-2\dv)\zeta(\omega_1)}\nn
&& \times
\frac{\sigma(2\du+(n+1)\omega_1+2\dv)}{\sigma(2\du+(n+1)\omega_1-2\dv)}, \\
\frac{\sigma(2\du+(n+1)\omega_1+2\dv)}{\sigma(2\du+(n+1)\omega_1-2\dv)}
&=& \ex{4\dv \zeta(2\du+(n+1)\omega_1)}.
\eea
Then the terms in leading order cancel and the terms of order $O(1/r_0)$ yield
the condition for $\du$
\be
y_0-\hat y_0 = \frac{4(\hd^2+d^2)}{3(d^2-\hd^2)}\du
-(d^2-\hd^2)(\zeta(2\du+(1+n)\omega_1)-(1+n)\zeta(\omega_1))
\ee
which for one and the same branch (even $n$) yields
\be
y_0-\hat y_0 = \frac{4(\hd^2+d^2)}{3(d^2-\hd^2)}\du
-(d^2-\hd^2)(\zeta(2\du+\omega_1)-\zeta(\omega_1))
\ee
and for a line segment between the two different branches (odd $n$)
\be
y_0-\hat y_0 = \frac{4(d^2+\hd^2)}{3(d^2-\hd^2)}\du
-(d^2-\hd^2)\zeta(2\du).
\ee
From eq. (\ref{bound1}) one learns that the length of the line segment of
constant length
for the same curve shifted by some $y_0-\hat y_0$ cannot exceed $d-\hd$.
Examples are shown in figs. \ref{flpg} to \ref{flpf}.

\bt
\parbox[b]{2.5cm}{\epsfig{file=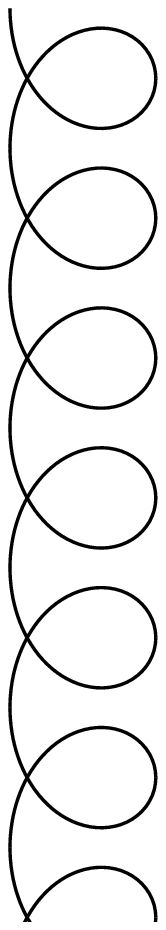,scale=0.5}\\
\Figur{flpg}{$\hd/d=0.3$}}
\hspace{5mm}
\parbox[b]{2.5cm}{\epsfig{file=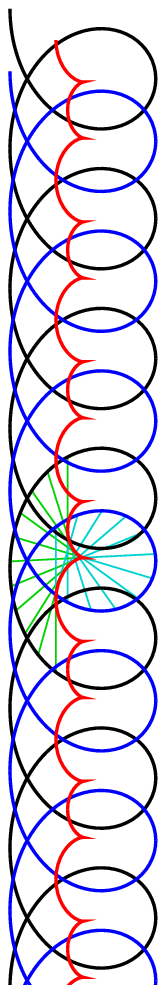,scale=0.5}\\
\Figurm{flpe}{$\hd/d=0.3$}}
\hspace{5mm}
\parbox[b]{2.5cm}{\epsfig{file=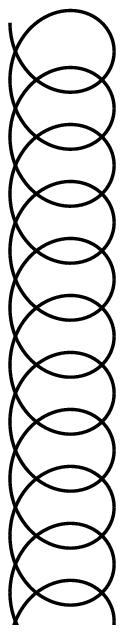,scale=0.5}\\
\Figur{flpf}{$\hd/d=0.5$}}
\et

These line segments are {\it not} the trivial ones parallel to the $y$-axis of
length $y_{\rm per}$ and integer multiples.
The length of a line segment of constant length between two different curves has
a
lower bound $d+\hd$, but no upper bound. Indeed one obtains an infinity of
solutions.
A pair of curves is shown in figs. \ref{flpa} to \ref{flpd} with three different
lengths of the line segment.

\bt
\parbox[b]{2.5cm}{\epsfig{file=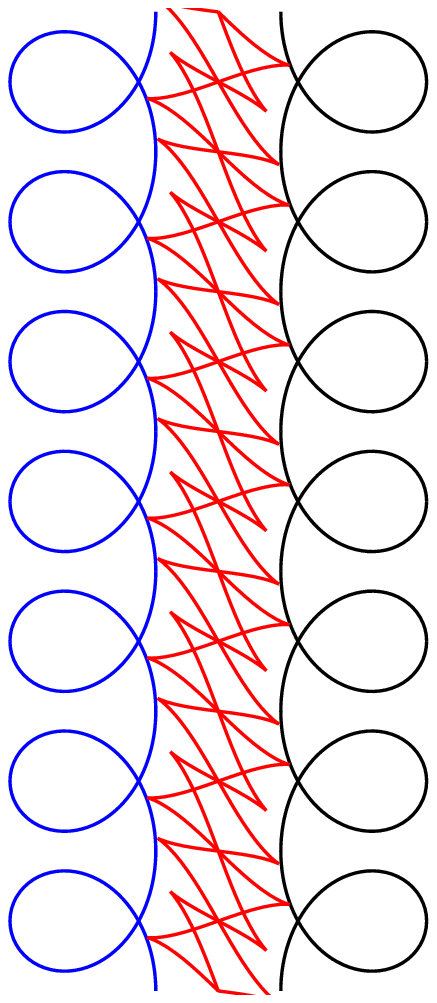,scale=0.5}\\
\Figur{flpa}{$\hd/d=0.3$}}
\hfill
\parbox[b]{2.5cm}{\epsfig{file=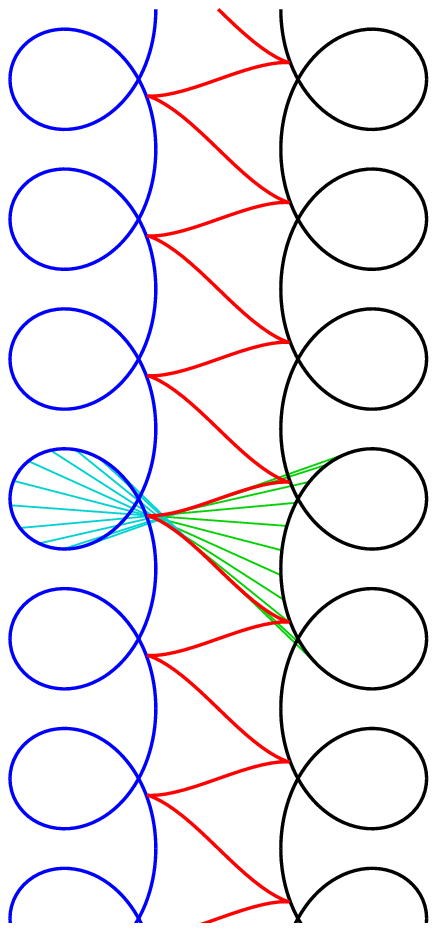,scale=0.5}\\
\Figurm{flpb}{$\hd/d=0.3$}}
\hfill
\parbox[b]{2.5cm}{\epsfig{file=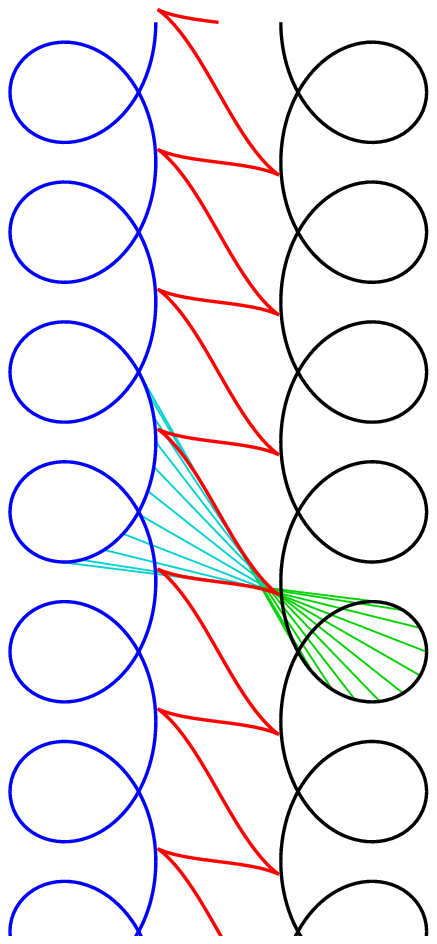,scale=0.5}\\
\Figurm{flpc}{$\hd/d=0.3$}}
\hfill
\parbox[b]{2.5cm}{\epsfig{file=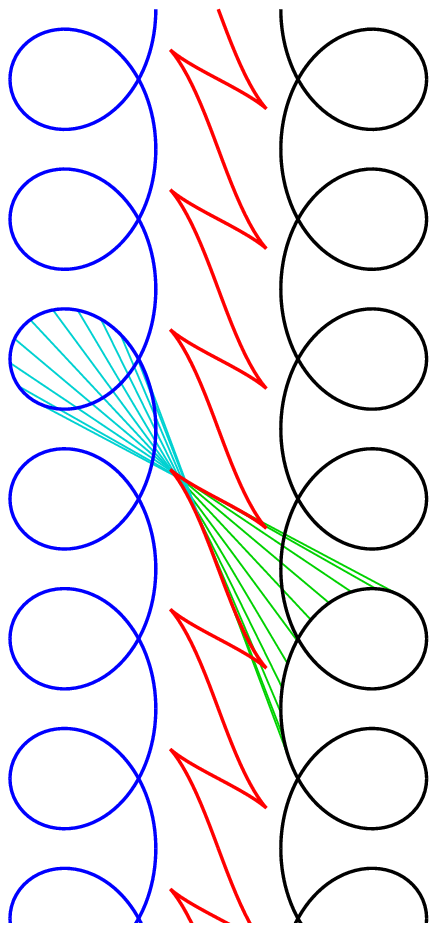,scale=0.5}\\
\Figurm{flpd}{$\hd/d=0.3$}}
\et

The curves shown here are also trajectories of massive charges in a
perpendicular magnetic field. In this linear limit one obtains from eqs.
(\ref{Br}, \ref{kappa}, \ref{lin2a}, \ref{lin2b}) that $B$ increases
linearly
\be
B = -\frac{4mw}{e(d^2-\hd^2)} (x-r_0). \label{linB}
\ee
Note that on both trajectories in figs. \ref{flpa} to \ref{flpd} the charges
move in the same direction. Such trajectories as well as those described in the
following subsections have been obtained by Evers et al \cite{Evers}. An
example of a pair of drifting orbits is shown in fig. 1 of their paper.

\subsection{The limit $\hd=0$}

In the limit $\hd=0$ one obtains
\bea
e_2=e_3 &=& \frac 1{3d^2}, \\
e_1 &=& - \frac 2{3d^2}, \\
\wp(u) &=& -\frac 2{3d^2} + \frac 1{d^2} \coth^2(\frac ud), \\
\zeta(u) &=& -\frac u{3d^2} + \frac 1d \coth(\frac ud), \\
x-r_0 &=& \frac{\pm d}{\cosh(2u/d)}, \\
y &=& y_0-u +d\tanh(2u/d), \\
\omega_1&=& \frac{d\pi\ie}2, \quad \omega_3=\infty.
\eea
The curve is no longer periodic. The conditions for line segments of constant
length are
\be
y_0-\hat y_0 -2\du = \left\{\begin{array}{cc}
-d \tanh(2\du/d), & \mbox{same sign of } x-r_0, \\
-d \coth(2\du/d), & \mbox{different sign of } x-r_0. \end{array}\right.
\ee
Thus there is no chord of constant length for one and the same curve.
The length of the line segments is given by
\be
2\ell = |y_0-\hat y_0 +2\du|.
\ee
There are always two solutions for line segments between pairs of curves.
\medskip

\btm
\parbox[b]{2.6cm}{\epsfig{file=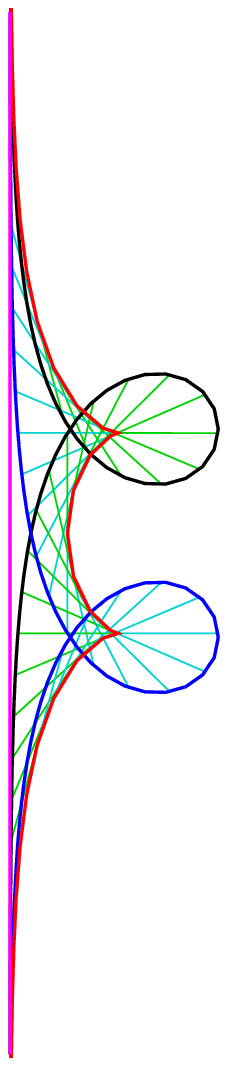,scale=0.5}\\
\Figurm{flza}{$\hd=0$}}
\hspace{0.3cm}
\parbox[b]{2.6cm}{\epsfig{file=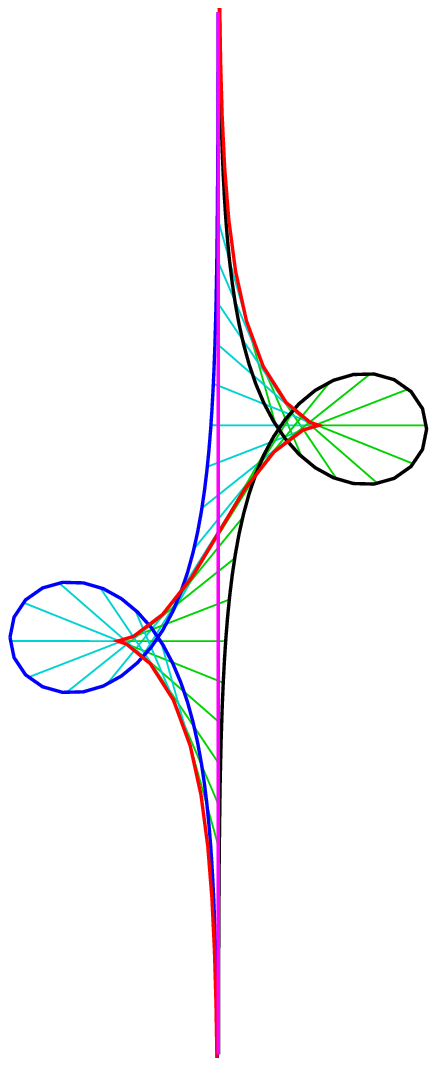,scale=0.5}\\
\Figurm{flzb}{$\hd=0$}}
\hspace{0.3cm}
\parbox[b]{2.6cm}{\epsfig{file=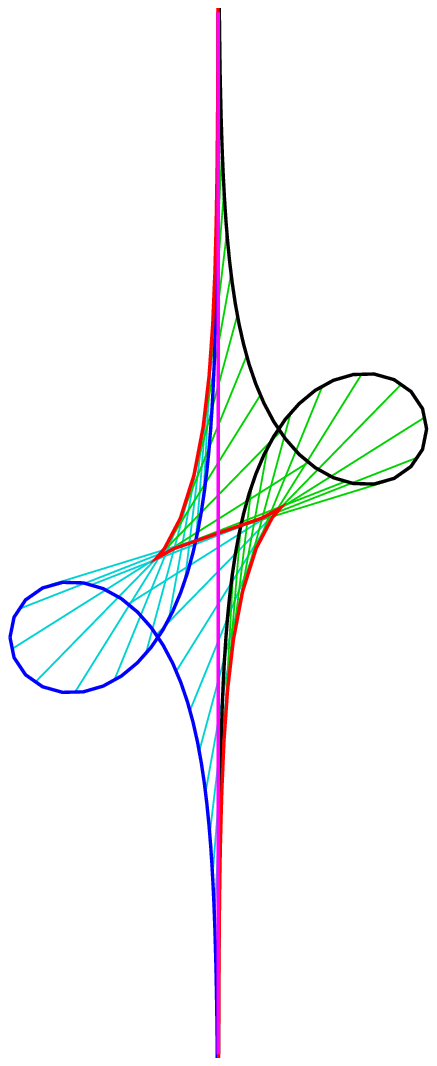,scale=0.5}\\
\Figurm{flze}{$\hd=0$}}
\hspace{0.3cm}
\parbox[b]{2.6cm}{\epsfig{file=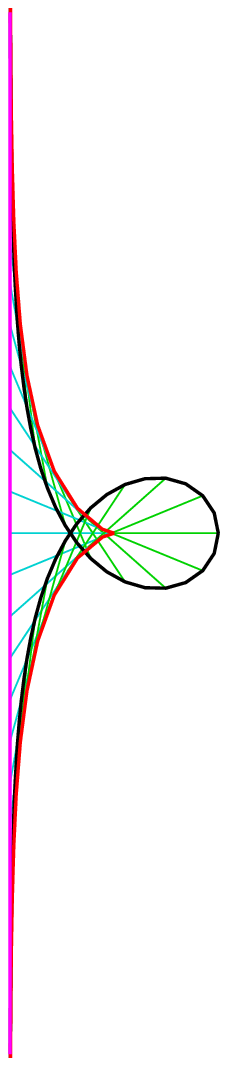,scale=0.5}\\
\Figurm{flzc}{$\hd=0$}}
\etm

A similar elementary construction as in subsect. \ref{elem} is possible.
Now one end of the straight line of length $2\ell$ moves along the straight line
$x-r_0=0$ as shown in fig. \ref{flzc}. Auerbach\cite{Auerbach} has composed his examples for $\rho=1/2$ from several copies of the central part of the curve and the corresponding piece of the straight line of fig. \ref{flzc}.

Evers et al. \cite{Evers} show in the upper fig. 2 the variation of the
classical trajectory of an electron as the gradient of $B$ is varied, which
corresponds here to a variation of $\hd^2/d^2$. The drifting orbit transforms
into snake states, which will be considered in the next subsection. The limit
case
is the state $\hd=0$ shown here.

\subsection{One branch}

In the case of one branch (two $r_i$ are real, two $r_i$ are complex) the choice
is
\be
r_4=r_0+d, \quad r_3=r_0-d, \quad
r_2=-r_0+\ie s, \quad r_1=-r_0-\ie s. \label{linr}
\ee
The calculation is rather similar to that in subsection \ref{lin2}. Indices 1
and 3 have to be exchanged and $\hd$ becomes $\ie s$.
This yields
\bea
a &=& \frac 1{2r_0(d^2+s^2)}, \label{lina} \\
e_1 &=& \frac{d^2-s^2+6\ie ds}{3(d^2+s^2)^2} - \frac 1{12r_0^2}, \\
e_2 &=& \frac{d^2-s^2-6\ie ds}{3(d^2+s^2)^2} - \frac 1{12r_0^2}, \\
e_3 &=& \frac{2(s^2-d^2)}{3(d^2+s^2)^2} + \frac 1{6r_0^2}, \\
\wp(2v) &=& e_3 - \frac 1{4r_0^2}.
\eea
Now $v$ lies infinitesimal close to $\omega'/2$. One determines
\bea
\wp(\frac{\omega'}2) &=& \frac{-s^2-5d^2}{3(d^2+s^2)^2} + O(1/r_0^2), \\
\wp'(\frac{\omega'}2) &=& -\frac{4d\ie}{(d^2+s^2)^2} + O(1/r_0^2), \\
\wp(v) &=& \wp(\frac{\omega'}2) - \frac d{r_0(d^2+s^2)} + O(1/r_0^2).
\eea
From this one concludes
\be
v = \frac{\omega'}2 + \dv, \quad
\dv= -\ie \frac{d^2+s^2}{4r_0}.
\ee
In the following calculation one has to replace $\omega_1$ by $\omega'$ and
$e_1$ by $e_3$ and one obtains in analogy to (\ref{z2c})
\be
z=r_0+d-2r_0\dv
\Big(\zeta(u-v)+\zeta(u+v)-\frac{\wp'(v)}{\wp(u)-\wp(v)}+2 e_3 u\Big).
\ee
\newcommand{\longfigure}
{\btm
\parbox[b]{2.5cm}{\epsfig{file=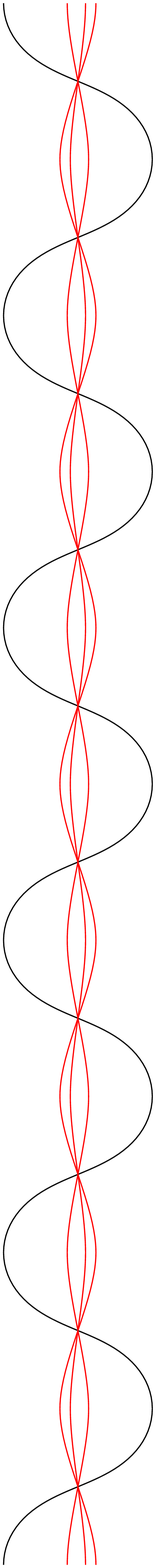,scale=0.333}\\
\Figur{flma}{$s/d=1.5$}}
\hspace{3mm}
\parbox[b]{2.5cm}{\epsfig{file=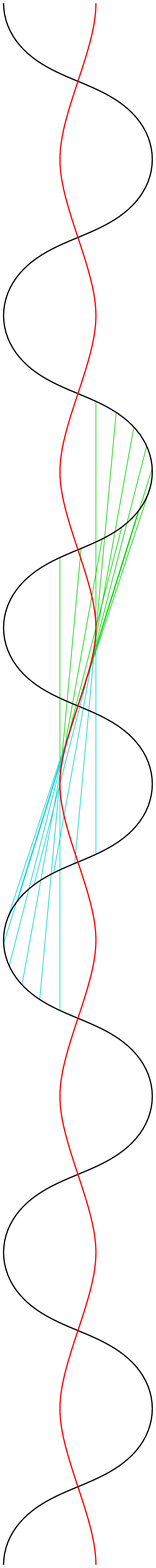,scale=0.333}\\
\Figurm{flmb}{$s/d=1.5$}}
\hspace{3mm}
\parbox[b]{2.5cm}{\epsfig{file=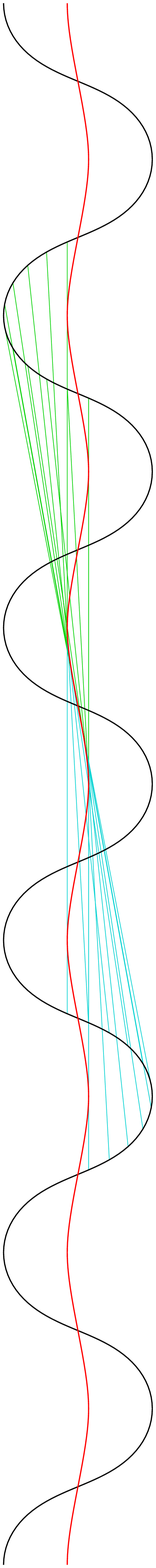,scale=0.333}\\
\Figurm{flmc}{$s/d=1.5$}}
\hspace{3mm}
\parbox[b]{2.5cm}{\epsfig{file=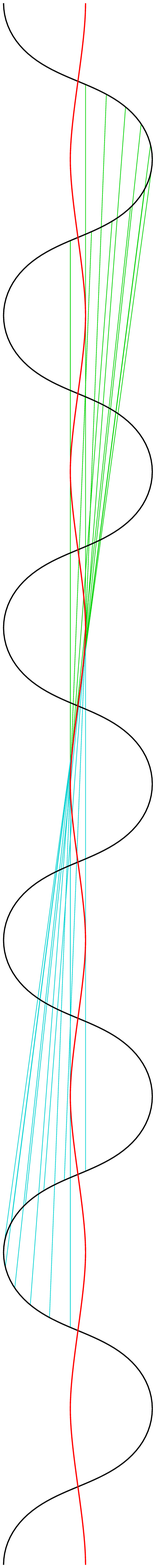,scale=0.333}\\
\Figurm{flmg}{$s/d=1.5$}}
\etm}
\longfigure

Thus the final result is
\bea
x-r_0 &=& d - \frac{2d(d^2+s^2)}{(d^2+s^2)^2\wp(u)+\frac 13(s^2+5d^2)}, \\
y &=& y_0 + \frac{d^2+s^2}2 \left(\zeta(u-\frac 12\omega')
+\zeta(u+\frac 12\omega')-\frac{4(d^2-s^2)}{3(d^2+s^2)^2} u\right).
\eea
$x-r_0$ oscillates between $d$ and $-d$. Moreover one
has
\bea
x(u+n\omega_3) - r_0 &=& (-)^n (x(u)-r_0), \\
y(u+n\omega_3) &=& y(u) + \frac n2 y_{\rm per}, \\
y_{\rm per} &=& -\frac{4(d^2-s^2)}{3(d^2+s^2)}\omega_3
+2(d^2+s^2)\zeta(\omega_3).
\eea
The condition for the line segment of constant length can be obtained similarly
as for the case of two curves. One obtains
\be
y_0-\hat y_0 = -\frac{4(d^2-s^2)}{3(d^2+s^2)}\du
+(d^2+s^2)(\zeta(2\du+\omega_3)-\zeta(\omega_3)),
\ee
where one replaces
$\zeta(2\du+\omega')-\zeta(\omega')=\zeta(2\du+\omega_3)-\zeta(\omega_3)$,
since $\omega'$ and $\omega_3$ differ by a period.

The examples of figs. \ref{flma} to \ref{flmg} show curves along which a
bicyclist could drive. One and the same curve $\Gamma$ for the front tire
corresponds to different traces of the rear tire for bicycles of different
lengths.

In figs. \ref{flmd} to \ref{flmf} the traces are more artistic, since
periodically the bicyclist has to move back and forth with the rear tire, but
always forward with the front tire.
\medskip

\btm
\parbox[b]{3.5cm}{\epsfig{file=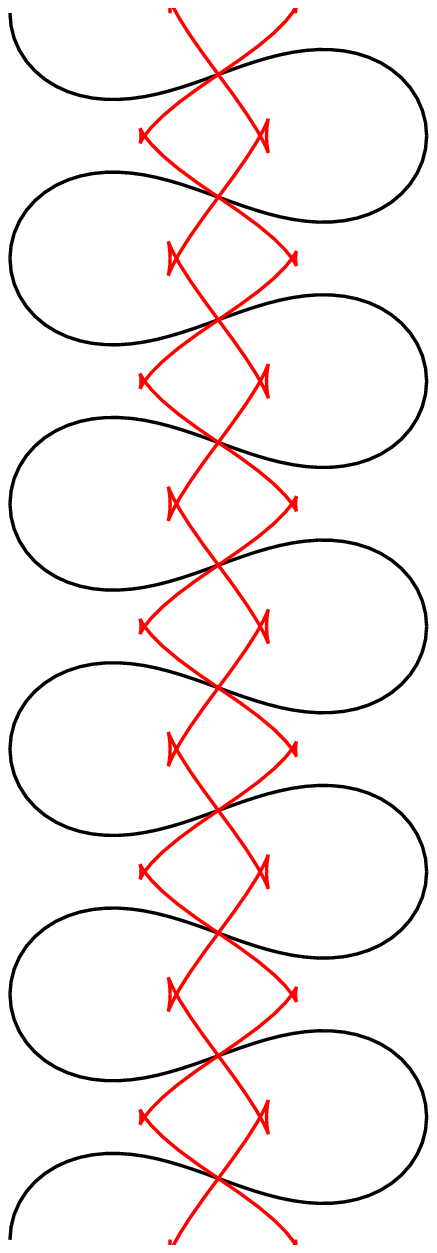,scale=0.5}\\
\Figur{flmd}{$s/d=0.7$}}
\parbox[b]{3.5cm}{\epsfig{file=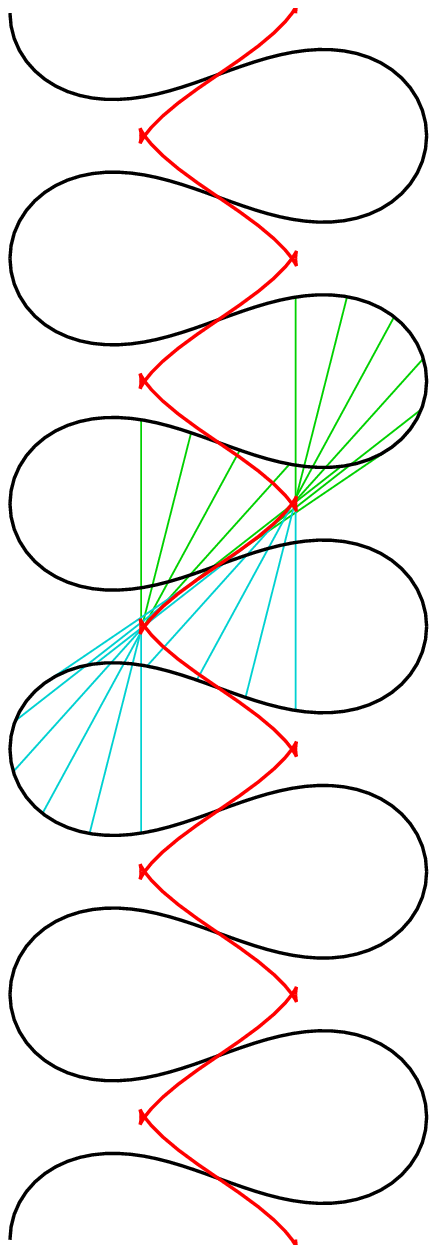,scale=0.5}\\
\Figur{flme}{$s/d=0.7$}}
\parbox[b]{3.5cm}{\epsfig{file=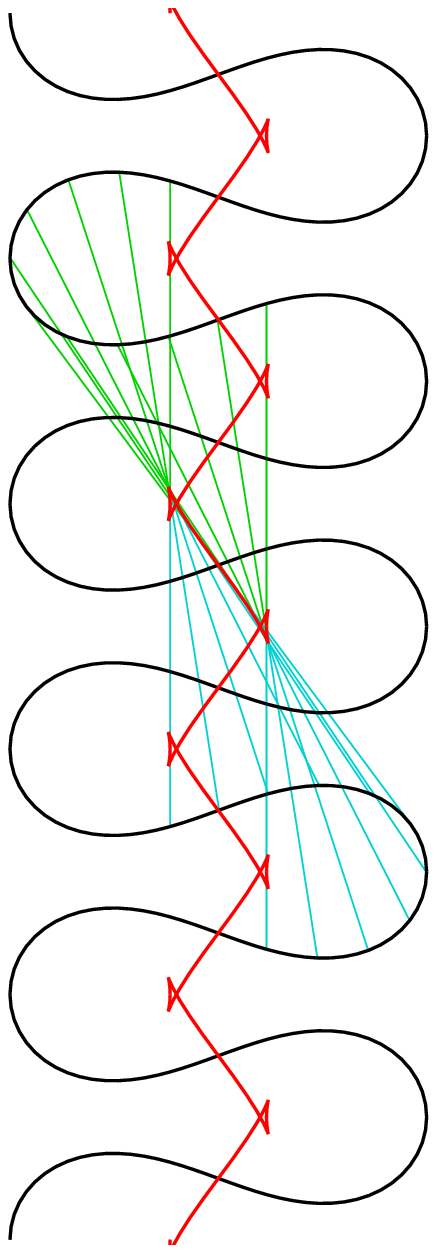,scale=0.5}\\
\Figur{flmf}{$s/d=0.7$}}
\etm

If $s/d$ decreases further the loops overlap stronger until at $s/d=0.458786$
the charge moves in an eight shown in fig. \ref{flml}. For this value of $s/d$
one has
$y_{\rm per}=0$. $y_{\rm per}$ becomes negative for smaller values of $s/d$.
The expression (\ref{linB}) for the magnetic field reads now
\be
B = -\frac{4mw}{e(d^2+s^2)} (x-r_0).
\ee
These eights have still the property that between two of them one has line
segments of
constant length. When $s/d$ becomes even smaller then the charges move into the
opposite direction. In fig. 1 of ref. \cite{Evers} two snake state trajectories
are shown moving in opposite directions. In fig. 2 of this paper the eight can
be seen where the electron is reflected in the magnetic bottle neck.
\medskip

\btm
\parbox[b]{3cm}{\epsfig{file=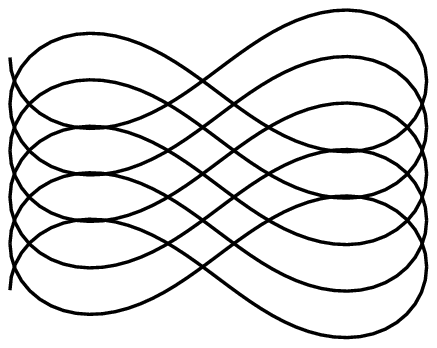,scale=0.5}\\
\Figur{flmj}{$s/d=0.5$}}
\parbox[b]{3cm}{\epsfig{file=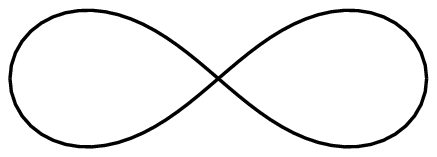,scale=0.5}
\vspace{4mm}

\epsfig{file=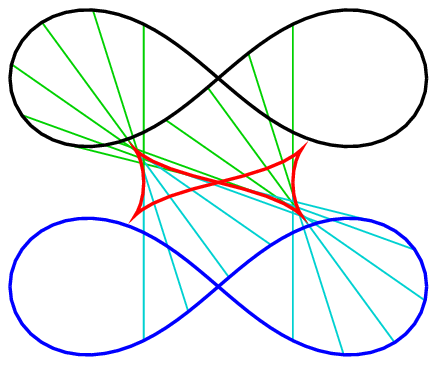,scale=0.5}
\Figurm{flml}{$s/d=\\0.458786$}}
\parbox[b]{3cm}{\epsfig{file=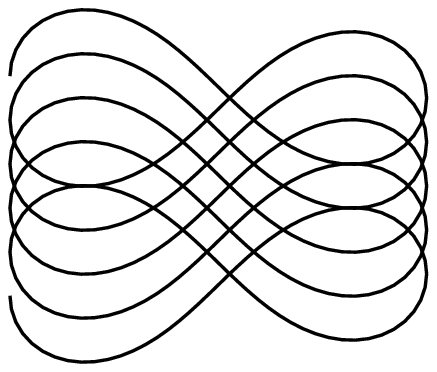,scale=0.5}\\
\Figur{flmm}{$s/d=0.4219$}}
\parbox[b]{3cm}{\epsfig{file=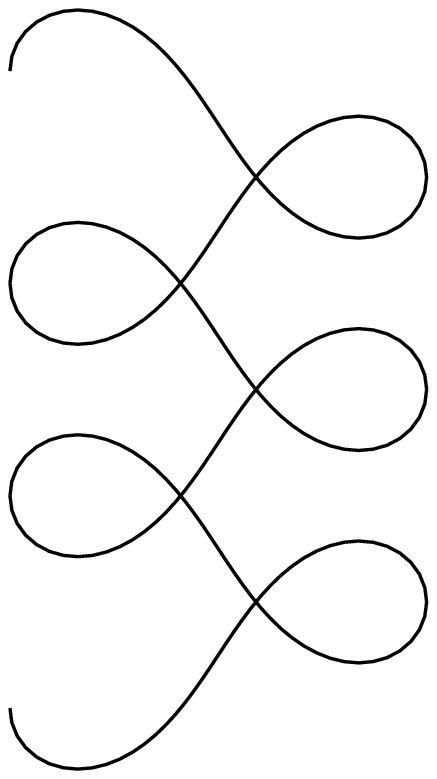,scale=0.5}\\
\Figur{flmn}{$s/d=0.3$}}
\etm

\section{The Carousel by Montejano, Oliveros and Bracho}

Oliveros, Montejano and Bracho\cite{Oliveros,Bracho} have considered special
cases of these curves. They consider a carousel, which is a dynamical
equilateral (not equiangular) polygon in which the midpoint of each edge
travels parallel to it. Then the trace of the vertices constitute the curves
$\Gamma$ whereas the midpoints outline the envelopes $\gamma$. They perform
explicitly
calculations for pentagons. In general they allow for five different curves for
the vertices.

Denoting the curves by indices $j$ one obtains solutions for the carousel from
the curves introduced here by defining
\be
z_j(u) := \ex{2j\ie\delta\chi} z(u),
\ee
and requiring
\be
|z_j(u+\du)-z_{j-1}(u-\du)| = 2\ell, \quad
z_5(u+10\du) = z_0(u).
\ee
The first condition is fulfilled with $\ex{2\ie\delta\chi}=$ r.h.s. of
(\ref{chdu}).
The second condition which guarantees that the pentagon is closed yields
\be
z_5(u+10\du) = \ex{10\ie\delta\chi} z_0(u+10\du) = z_0(u).
\ee
Thus $10\du$ has to be an integer multiple of $2\omega_3$,
\be
\du = \frac k5 \omega_3
\ee
with integer $k$. Then the condition reduces to
\be
\ex{\ie(10\delta\chi+k\psi_{\rm per})} = 1. \label{cond5}
\ee

\bt
\parbox[b]{5.5cm}{\epsfig{file=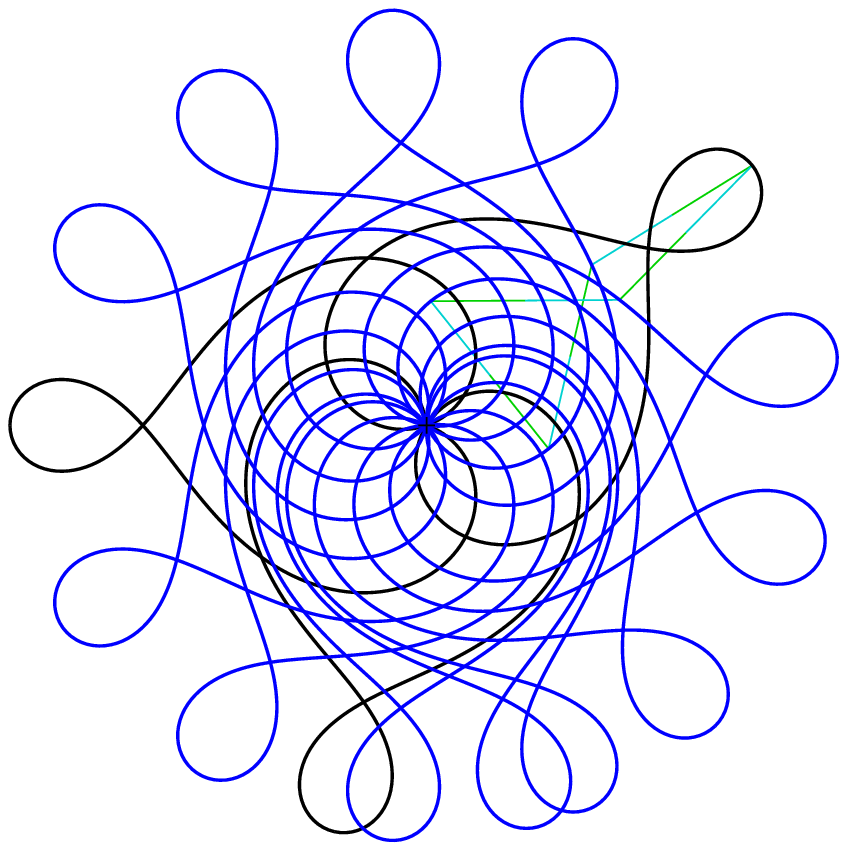,scale=0.5}
\Figurm{fgmpspez1}{$\epsilon=1$, $\heps^2=-0.025835$}}
\hspace{0.5cm}
\parbox[b]{5.5cm}{\epsfig{file=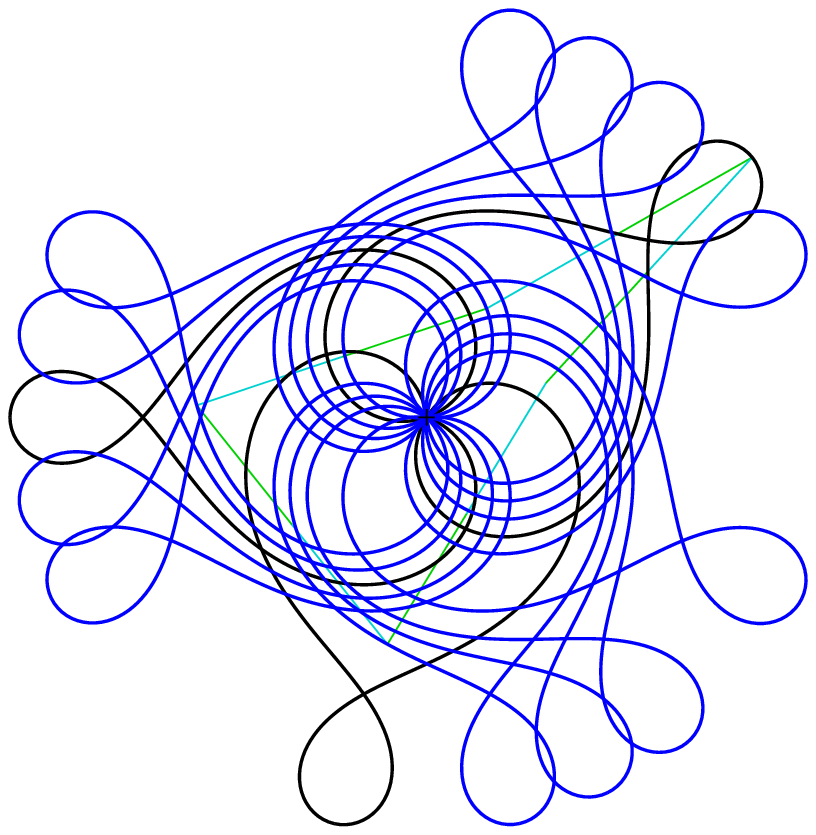,scale=0.5}
\Figurm{fgmpspez2}{$\epsilon=1$, $\heps^2=-0.025835$}}
\et

An example is fig. 6 of ref. \cite{Oliveros}. Since the curves go through the
origin,
one has $\epsilon=1$. One realizes that the curves seem to be close to those for
$m/n=-1/3$, $\epsilon=1$ in fig. \ref{fgm1p3b} which yields
$\heps^2=-0.034862$. One finds that condition (\ref{cond5}) is fulfilled for
$\heps^2=-0.025835$ and apparently for two different $\kappa$ and two different
lengths $2\ell$ of the line segments. This is shown in figs. \ref{fgmpspez1} and
\ref{fgmpspez2}. The second one seems to reproduce fig. 6 of ref.
\cite{Oliveros}.
\medskip

\nicht{
\bt
\parbox[b]{4cm}{\epsfig{file=fgmpspez1.eps,scale=0.5}
\Figurm{fgmpspez1}{$\epsilon=1$, $\heps^2=-0.025835$}}
\hspace{1cm}
\parbox[b]{4cm}{\epsfig{file=fgmpspez2.eps,scale=0.5}
\Figurm{fgmpspez2}{$\epsilon=1$, $\heps^2=-0.025835$}}
\et}

A solution for five 'Eights' is obtained for $m/n=0/1$, $\epsilon=0.689454$ in
figs. \ref{fg0p5cr} and \ref{fg0p5dr}. The ratio of the arc length of the outer
and the inner loop of the eights is 2/3. Fig. 8 in \cite{Oliveros} shows the
same eights (to the extend one can judge this with the eyes), but the pentagon
shown there is misleading. It should be as in fig. \ref{fg0p5dr}.
\medskip

\btm
\parbox[b]{3.6cm}{\epsfig{file=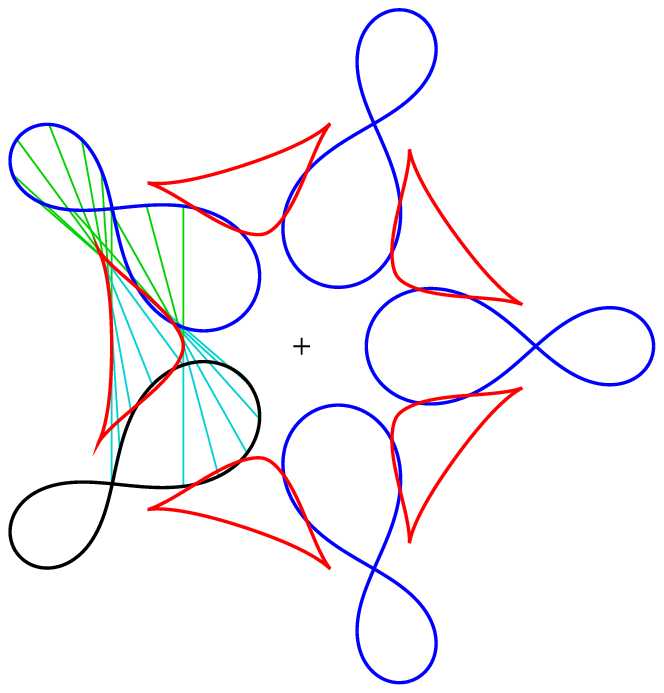,scale=0.5}
\Figur{fg0p5cr}{$m/n=0/1$, $\epsilon=0.689454$}}
\hspace{1.4cm}
\parbox[b]{3.6cm}{\epsfig{file=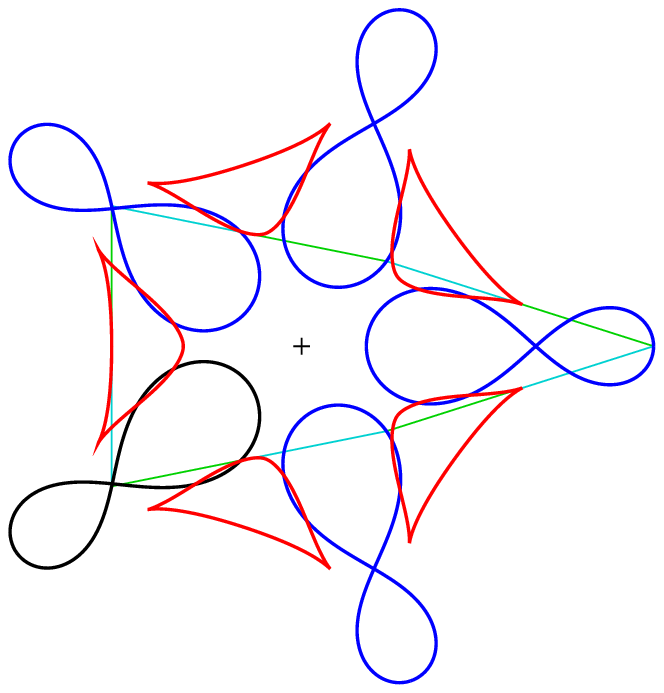,scale=0.5}
\Figurm{fg0p5dr}{$m/n=0/1$, $\epsilon=0.689454$}}
\etm

The corresponding carousel for the linear case is shown in fig. \ref{flmh} and
\ref{flmi}. This solution looks like that of fig. 2 of \cite{Bracho}.
\medskip

\btm
\parbox[b]{5cm}{\epsfig{file=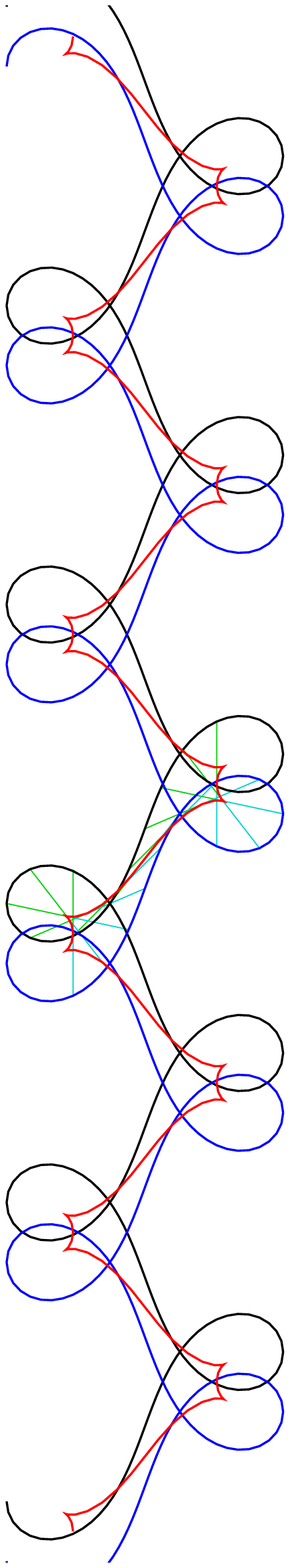,scale=0.5}
\Figur{flmh}{$s^2/d^2=0.0315653$}}
\hspace{1cm}
\parbox[b]{5cm}{\epsfig{file=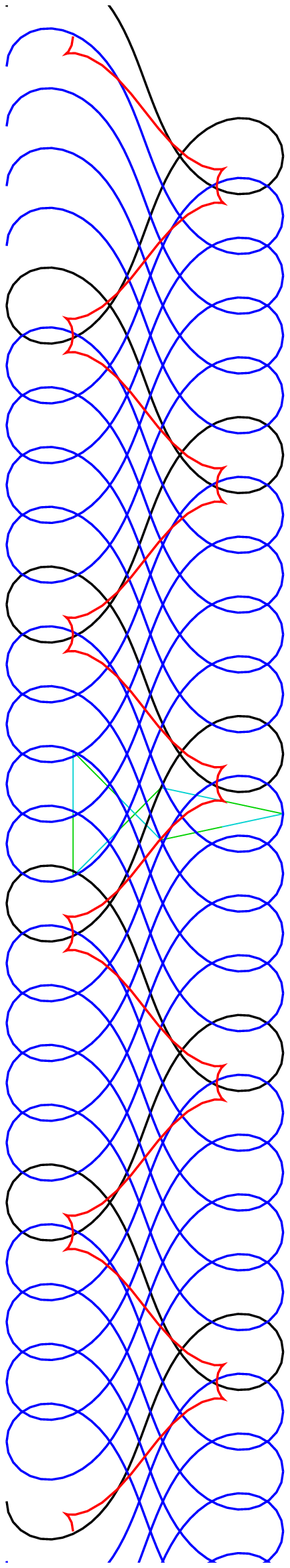,scale=0.5}
\Figurm{flmi}{$s^2/d^2=0.0315653$}}
\etm

Of interest is the case where all five figures fall onto each other and the
curve is closed. Then
\be
\psi_{\rm per} = \frac{2\pi}n, \quad
\du=\frac{k'n\omega_3}5
\ee
with integer $k'$. One with $n=7$-fold symmetry is obtained for
$\epsilon=0.459104$ in figs. \ref{fgp1p7a}, \ref{fgp1p7c} and should be
compared with fig. 9 of \cite{Oliveros} and fig. 3 of \cite{Bracho}, the other
one with $n=12$-fold symmetry for $\epsilon=0.5727853$ in figs. \ref{fgp1p12a},
\ref{fgp1p12b} and is to be compared with fig. 1 of \cite{Bracho}.

\btm
\parbox[b]{4cm}{\epsfig{file=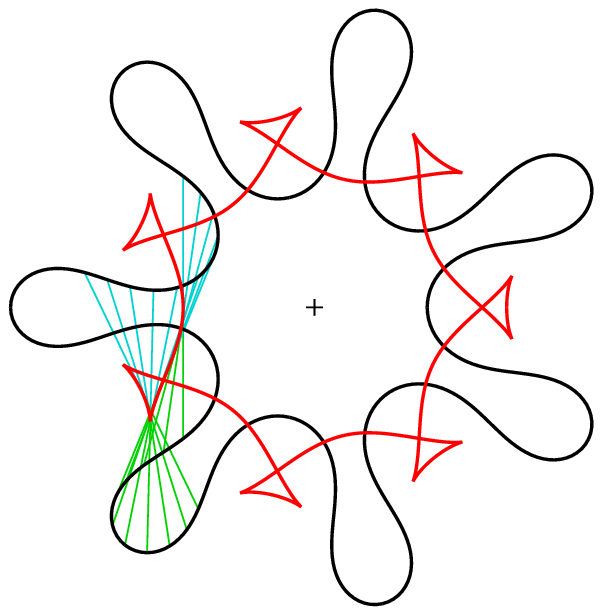,scale=0.5}
\Figur{fgp1p7a}{$m/n=1/7$, $\epsilon=0.459104$}}
\hspace{1cm}
\parbox[b]{4cm}{\epsfig{file=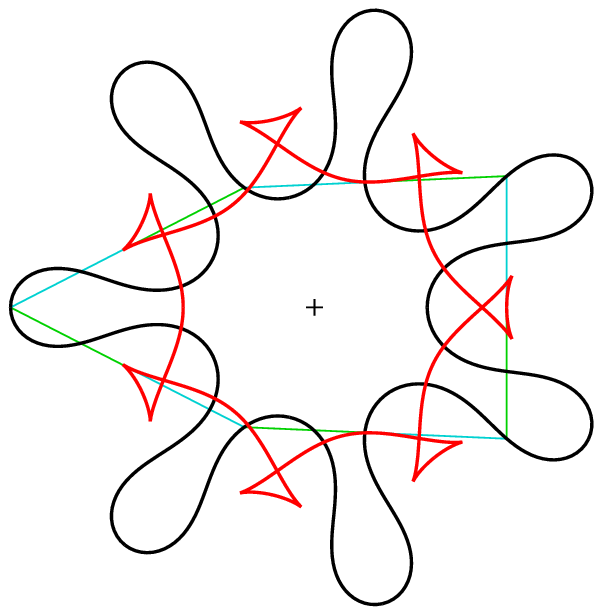,scale=0.5}
\Figurm{fgp1p7c}{$m/n=1/7$, $\epsilon=0.459104$}}
\etm

\btm
\parbox[b]{4cm}{\epsfig{file=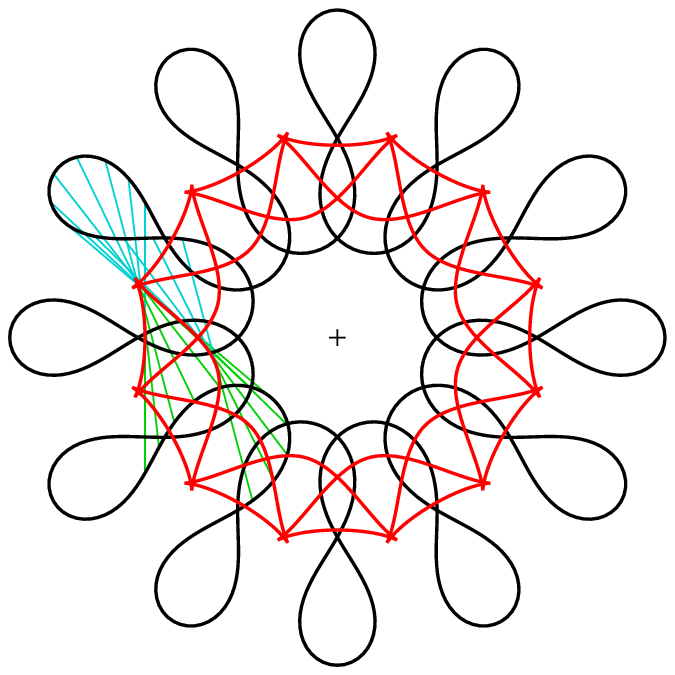,scale=0.5}
\Figur{fgp1p12a}{$m/n=1/12$, $\epsilon=0.5727853$}}
\hspace{1cm}
\parbox[b]{4cm}{\epsfig{file=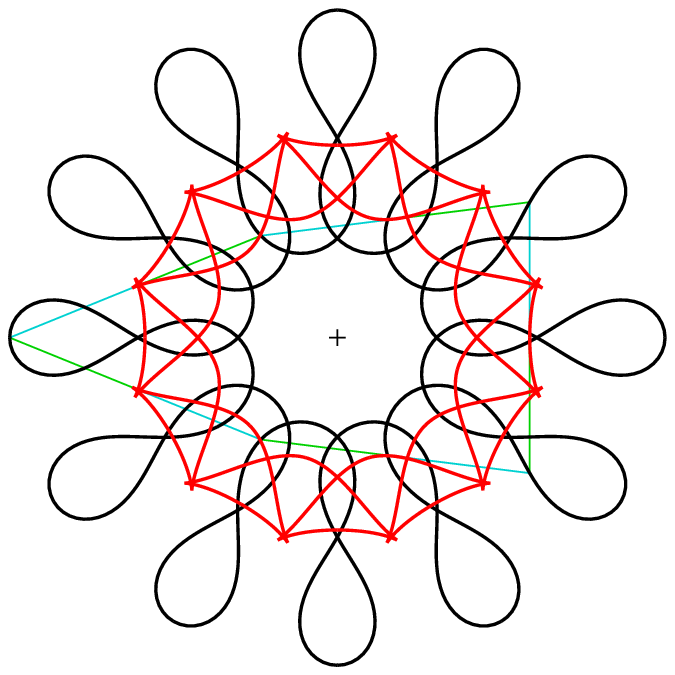,scale=0.5}
\Figurm{fgp1p12b}{$m/n=1/12$, $\epsilon=0.5727853$}}
\etm

\paragraph{Acknowlegment} I am indebted to Serge Tabachnikov for useful
correspondence and to Peter W\"olfle for useful discussions.

\begin{appendix}

\section{Weierstrass Functions\label{Wei}}

\subsection{Definitions and relations of Weierstrass functions\label{formulae}}

The Weierstrass function $\wp$ is defined by
\be
\wp^{\prime 2}(z) = 4\wp^3(z)-g_2\wp(z)-g_3
= 4(\wp(z)-e_1)(\wp(z)-e_2)(\wp(z)-e_3) \label{diff}
\ee
with the requirement that one of the singularities is at $z=0$ and
\be
e_1+e_2+e_3=0.
\ee
Commonly the two integrals are defined
\bea
\zeta(a) &=& \frac 1a - \int_0^a (\wp(z)-\frac 1{z^2}) \de z, \label{zetaint} \\
\sigma(a) &=& a \exp\left(\int_0^a (\zeta(z)-\frac 1z) \de z\right),
\label{sigmaint}
\eea
The function $\wp$ is an even function of its argument, $\zeta$ and $\sigma$ are
odd functions. The Laurent and Taylor expansions start with
\bea
\wp(a) &=& \frac 1{a^2} + \frac{g_2}{20} a^2 + \frac{g_3}{28} a^4 + ... \\
\zeta(a) &=& \frac 1a - \frac{g_2}{60} a^3  - \frac{g_3}{140} a^5 - ... \\
\sigma(a) &=& a - \frac{g_2}{240} a^5 - \frac{g_3}{840} a^7 - ...
\eea
There exist addition theorems
\bea
\wp(a+b) &=& -\wp(a) -\wp(b) + \frac{(\wp'(a)-\wp'(b))^2}{4(\wp(a)-\wp(b))^2},
\label{wpadd} \\
\zeta(a+b) &=& \zeta(a)+\zeta(b) +\frac 12 \frac{\wp'(a)-\wp'(b)}
{\wp(a)-\wp(b)}, \label{zetaadd} \\
\sigma(a+b)\sigma(a-b) &=& -\sigma^2(a)\sigma^2(b)(\wp(a)-\wp(b)).
\label{sigmaadd}
\eea
If $\omega$ is a half-period, that is $\omega$ itself is not a period of $\wp$,
but $2\omega$ is, then the following relations hold for integer $n$
\bea
\wp(a+2n\omega) &=& \wp(a), \label{wpper} \\
\zeta(a+2n\omega) &=& \zeta(a) + 2n\zeta(\omega), \label{zetaper} \\
\sigma(a+2n\omega) &=& (-)^n\sigma(a) \ex{2n(a+n\omega)\zeta(\omega)}.
\label{sigmaper}
\eea
The increase of the argument of $\sigma(a)$, if the real part of $a$ is
increased by $2\omega_3$, is obtained from eq. (\ref{sigmaper})
\be
\arg\sigma(a+2\omega_3) - \arg\sigma(a)= -\pi(2m+1) + 2\Im a \,\zeta(\omega_3)
\label{argsigma}
\ee
with integer $m$, where $-\pi(2m+1)$ comes from the minus-sign in
(\ref{sigmaper}) and $m$ has to be determined. If $a$ is real, then $\sigma$
changes sign, whenever $a$ is an integer multiple of $2\omega_3$. If $a$ has a
small positive (negative) imaginary part, then $a$ passes the zero clockwise
(counter-clockwise) in the complex plane, yielding a decrease (an increase) of
the argument by
$\pi$.
Further jumps by $2\pi$ occur, whenever the imaginary part of $a$ crosses an
integer multiple of $\Im\omega_1$. Thus $m$ in eq. (\ref{argsigma}) is given by
\be
2m \Im \omega_1 < \Im a < 2(m+1) \Im \omega_1. \label{argsigma1}
\ee
From (\ref{zetaadd}) one obtains
\be
\zeta(a+b)+\zeta(a-b) = 2\zeta(a)+\frac{\wp'(a)}{\wp(a)-\wp(b)}.
\label{zetaadd2}
\ee

Legendre's relation reads
\be
\eta\omega'-\eta'\omega=\frac{\pi\ie}2 \quad \mbox{ for } \quad
\Im\frac{\omega'}{\omega}>0, \label{Legendre}
\ee
where $2\omega$ and $2\omega'$ span one elementary cell.

\subsection{Double-periodic functions\label{cE}}

Two basic theorems on double-periodic functions are

\paragraph{Theorem I}
Two elliptic functions which have the same period, the same poles, and the same
principal parts at each pole differ by a constant.
\paragraph{Theorem II}
The quotient of two elliptic functions whose periods, poles, and zeroes (and
multiplicities of poles and zeroes) are the same, is a constant.\\
Due to this second theorem the elliptic function
\be
\cE{n-m} u {u_1,u_2 ... u_m}{v_1,v_2 ... v_n},
\ee
is defined as the periodic function with zeroes at $u_1$, $u_2$, ... $u_m$
and poles at $v_1$, $v_2$, ... $v_n$ in an elementary cell. Zeroes and poles at
the origin are counted separately. If $n>m$, then the function has
an $n-m$-fold zero at $u=0$. If $n<m$, then the function has instead a pole of
order $m-n$ at $u=0$. The function $\cal E$ is normalized so that
\be
{\cal E}_{n-m}(u) = u^{n-m} +  O(u^{n-m+1}).
\ee
Moreover one has to require that the sums $\sum_i u_i$ and $\sum_i v_i$ differ
only by a period $2\omega$
\be
\sum_{i=1}^m u_i = \sum_{i=1}^n v_i+2\sum_j k_j\omega_j
\ee
with integer $k_j$. The functions $\cal E$ can explicitly be expressed in terms
of the functions $\sigma$
\bea
&& \cE{n-m} u {u_1,u_2 ... u_m}{v_1,v_2 ... v_n} \nn
&=& \sigma^{n-m}(u) \prod_{i=1}^m \frac{\sigma(u-u_i)}{\sigma(-u_i)}
\prod_{i=1}^n \frac{\sigma(-v_i)}{\sigma(u-v_i)}
\ex{2u\sum_j k_j \zeta(\omega_j)}.
\eea
Using eqs. (\ref{sigmaper}) and (\ref{Legendre}) one finds that this function is
double-periodic.
$g_2$ and $g_3$ are not explicitly indicated. In these eqs. it is always
assumed that they are the same in all functions.

Multiplication yields
\bea
&& \cE{n-m} u{u_1,u_2 ... u_m}{v_1,v_2 ... v_n}
\cE{n'-m'} u{u'_1,u'_2 ... u'_{m'}}{v'_1,v'_2 ... v'_{n'}} \\
&=& \cE{n+n'-m-m'} u{u_1,u_2 ... u_m,u'_1,u'_2 ... u'_{m'}}
{v_1,v_2 ... v_n,v'_1,v'_2 ... v'_{n'}}.
\eea
If locations of zeroes and poles coincide they have to be cancelled
\be
\cE{n-m} u{a_1,a_2,...,a_k,u_1,u_2 ... u_m}{a_1,a_2,...,a_k,v_1,v_2 ... v_n}
=\cE{n-m} u{u_1,u_2 ... u_m}{v_1,v_2 ... v_n}.
\ee

Examples are
\bea
\wp(u)-\wp(v) &=& \cE{-2}u{v,-v}{\emptyset}, \\
\frac{\wp(u)-\wp(v_3)}{\wp(u)-\wp(v)} &=& \cE 0u{v_3,-v_3}{v,-v}, \\
\wp'(u) &=& -2\cE{-3}u{\omega_1,\omega_2,\omega_3}{\emptyset}
=-\frac{\sigma(2u)}{\sigma^4(u)}. \label{wpssigma}
\eea
and
\bea
\left|\begin{array}{ccc} 1 & \wp(u) & \wp'(u) \\ 1 & \wp(v) & \wp'(v) \\
1 & \wp(w) & \wp'(w) \end{array}\right|
&=& -2\cE{-3}u{v,w,-v-w}{\emptyset} \cE{-2}v{w,-w}{\emptyset} \nn
&=& \frac{-2\sigma(u-v)\sigma(u-w)\sigma(v-w)\sigma(u+v+w)}
{\sigma^3(u)\sigma^3(v)\sigma^3(w)},
\eea
from which one concludes that the determinant vanishes, if $u+v+w=0$,
\be
\left|\begin{array}{ccc} 1 & \wp(u) & \wp'(u) \\ 1 & \wp(v) & \wp'(v) \\
1 & \wp(u+v) & -\wp'(u+v) \end{array}\right| = 0. \label{det}
\ee
In the limit $u\rightarrow v$ one obtains
\be
\left|\begin{array}{ccc} 1 & \wp(v) & \wp'(v) \\ 0 & \wp'(v) & \wp\dpr(v) \\
1 & \wp(2v) & -\wp'(2v) \end{array}\right| = 0. \label{det2}
\ee

\subsection{Representations of Weierstrass functions\label{Weier}}

The Weierstrass function $\zeta(z)$ with elementary periods $2\omega$ and
$2\omega'$ has simple poles with residues 1 at
\be
z=2m\omega+2n\omega'
\ee
for integer $m$ and $n$. It reads
\bea
\zeta(z) &=& \frac{\eta}{\omega}z
+ \frac{\pi}{2\omega} \left( \cot(\frac u2)
+\sum_{n=1}^{\infty} \big(\cot(\frac u2+\frac{n\pi\omega'}{\omega})
+\cot(\frac u2-\frac{n\pi\omega'}{\omega})\big) \right) \nn
&=& \frac{\eta}{\omega}z
+\frac{\pi}{2\omega} \sin u \left(\frac 1{1-\cos u}
+4\sum_{n=1}^{\infty} \frac{q^n}{1-2q^n\cos u+q^{2n}}\right) \label{repz}
\eea
with
\bea
u &=& \frac{\pi z}{\omega}, \\
q &=& \exp(2\ie\pi\omega'/\omega), \\
\eta &=& \zeta(\omega) = \omega \left(\frac{\pi}{2\omega}\right)^2
\left(\frac 13 +\sum_{n=1}^{\infty}
\frac 2{\sin^2(\frac{n\pi\omega'}{\omega})}\right)\nn
&=& \omega \left(\frac{\pi}{2\omega}\right)^2 \left(\frac 13
-8\sum_{n=1}^{\infty} \frac{q^n}{(1-q^n)^2}\right).
\eea
The derivative of the $\zeta$-function yields the Weierstrass $\wp$-function
\bea
\wp(z) &=& -\frac{\de\zeta(z)}{\de z} \nn
&=& -\frac{\eta}{\omega} +\left(\frac{\pi}{2\omega}\right)^2
\left( \frac 1{\sin^2(\frac u2)} +\sum_{n=1}^{\infty}
\big(\frac 1{\sin^2(\frac u2+\frac{n\pi\omega'}{\omega})}
+ \frac 1{\sin^2(\frac u2 -\frac{n\pi\omega'}{\omega})} \big) \right) \nn
&=& -\frac{\eta}{\omega} +2\left(\frac{\pi}{2\omega}\right)^2
\left(\frac 1{1-\cos u} -4\sum_{n=1}^{\infty}
\frac{q^n(\cos u -2q^n +q^{2n}\cos u)}{(1-2q^n\cos u+q^{2n})^2} \right).
\eea
The constant term to $\wp(z)$ fixes the constant term in the Laurent expansion
of $\wp(z)=\frac 1{z^2} + 0z^0 +...$ to vanish. The derivative of the
Weierstrass function reads
\bea
\wp'(z) &=& -4\left(\frac{\pi}{2\omega}\right)^3 \sin u
\left( \frac 1{(1-\cos u)^2} \right. \nn
&& \left. -4\sum_{n=1}^{\infty} 
\frac{q^n(1+2q^n\cos u -6q^{2n}+2q^{3n}\cos u +q^{4n})}
{(1-2q^n\cos u+q^{2n})^3} \right).
\eea
Finally the integral of $\zeta$ is obtained
\bea
\sigma(z) &=& z\exp\left(\int_0^z (\zeta(z')-\frac 1{z'})\de z'\right) \nn
&=& \frac{2\omega}{\pi} \sin\frac u2 \exp\left(\frac{\eta z^2}{2\omega}\right)
\prod_{n=1}^{\infty} \frac{1-2q^n\cos u +q^{2n}}{1-2q^n+q^{2n}}. \label{reps}
\eea
These representations of the Weierstrass functions have been used for the
calculation of the figures.

\section{Symmetric Polynomials\label{symm}}

Four symmetric polynomials $P_q$, $P_r$, $P_{\rm m}$ and $\hat P$ are
introduced. Symmetric means that they are invariant under permutation of the
$r_i$ and $q_i$, resp. Three of them are defined by
\bea
P_q &=& \frac 18 (q_1+q_2-q_3-q_4) (q_1-q_2+q_3-q_4) (q_1-q_2-q_3+q_4),
\label{Pq} \\
P_r &=& \frac 18 (r_1+r_2-r_3-r_4) (r_1-r_2+r_3-r_4) (r_1-r_2-r_3+r_4) \nn
&=& -(r_1+r_2) (r_1+r_3) (r_2+r_3) = (r_4+r_1) (r_4+r_2) (r_4+r_3), \label{Pr}
\\
P_{\rm m} &=& \sum_{i<j} q_i q_j
= \frac 12 (q_1+q_2+q_3+q_4)^2 - \frac 12 (q_1^2+q_2^2+q_3^2+q_4^2). \label{Pm}
\eea
The identity for $P_r$ is derived from $\sum_i r_i = 0$.
Expanding the last expression in (\ref{Pr}) as polynomial and subtracting
$r_4(r_4+r_1+r_2+r_3)$, which vanishes one obtains
\be
P_r = r_1r_2r_3 +r_1r_2r_4+r_1r_3r_4+r_2r_3r_4. \label{Pra}
\ee
Realizing that
\be
q_1+q_2-q_3-q_4 = r_1^2+r_2^2-r_3^2-(r_1+r_2+r_3)^2 = -2(r_3+r_1)(r_3+r_2)
\label{qq}
\ee
and similarly for permutations one finds
\be
P_r^2 = P_q. \label{Prq}
\ee

Consider
\bea
r_4 P_r &=& r_4(r_4+r_1)(r_4+r_2)(r_4+r_3) \nn
&=& r_4 \big( r_4^3 +r_4^2(r_1+r_2+r_3)
+r_4(r_1r_2+r_1r_3+r_2r_3) + r_1r_2r_3 \big) \nn
&=& \frac 12 q_4(q_4-q_1-q_2-q_3) + r_1r_2r_3r_4, \label{r4Pr}
\eea
where use is made of
\be
r_1r_2+r_1r_3+r_2r_3 = \frac 12 \big((r_1+r_2+r_3)^2 -r_1^2-r_2^2-r_3^2 \big)
=\frac 12 (q_4-q_1-q_2-q_3).
\ee
Using also permutations one obtains
\be
0 = (r_1+r_2+r_3+r_4) P_r = 4r_1r_2r_3r_4 - 4 \hat P
\ee
with
\bea
\hat P &=& -\frac 18 (q_1^2+q_2^2+q_3^2+q_4^2)
+ \frac 14 P_{\rm m} \nn
&=& \frac 18 (q_1+q_2+q_3+q_4)^2 - \frac 14 (q_1^2+q_2^2+q_3^2+q_4^2).
\label{hatP}
\eea
This yields
\be
r_1r_2r_3r_4 = \hat P \label{Prr}
\ee
in terms of the squares $q_i=r_i^2$.

\nicht{
\section{Various Quantities in Terms of $\epsilon$ and $\heps$}

\bea
r_4 &=& r_0(1+\epsilon), \\
r_3 &=& r_0(1-\epsilon), \\
r_2 &=& r_0(-1-\heps), \\
r_1 &=& r_0(-1+\heps).
\eea

\bea
P_{\rm m} &=&
r_0^4(6+2\epsilon^2+2\heps^2+\epsilon^4+4\epsilon^2\heps^2+\heps^4), \\
\hat P &=& r_0^4(1-\epsilon^2)(1-\heps^2), \\
P_r = \frac 1a &=& 2r_0^3 (\epsilon^2-\heps^2), \\
P_q = P_r^2 &=& 4r_0^6 (\epsilon^2-\heps^2)^2.
\eea

\bea
e_1 &=& \frac{r_0^4a^2(4\epsilon^2+4\heps^2-24\epsilon\heps
-\epsilon^4+2\epsilon^2\heps^2-\heps^4)}3, \\
e_2 &=& \frac{r_0^4a^2(4\epsilon^2+4\heps^2+24\epsilon\heps
-\epsilon^4+2\epsilon^2\heps^2-\heps^4)}3, \\
e_3 &=& -\frac{2r_0^4a^2(4\epsilon^2+4\heps^2
-\epsilon^4+2\epsilon^2\heps^2-\heps^4)}3.
\eea

\bea
p_1^{(4)} &=&
\frac{r_0^4a^2(-20\epsilon^2+4\heps^2-12\epsilon^3+12\epsilon\heps^2
-\epsilon^4+2\epsilon^2\heps^2-\heps^4)}3, \\
p_1^{(3)} &=&
\frac{r_0^4a^2(-20\epsilon^2+4\heps^2+12\epsilon^3-12\epsilon\heps^2
-\epsilon^4+2\epsilon^2\heps^2-\heps^4)}3, \\
p_1^{(2)} &=& \frac{r_0^4a^2(4\epsilon^2-20\heps^2+12\epsilon^2\heps-12\heps^3
-\epsilon^4+2\epsilon^2\heps^2-\heps^4)}3, \\
p_1^{(1)} &=& \frac{r_0^4a^2(4\epsilon^2-20\heps^2-12\epsilon^2\heps+12\heps^3
-\epsilon^4+2\epsilon^2\heps^2-\heps^4)}3, \\
p_2 &=&
-\frac{r_0^4a^2(8\epsilon^2+8\heps^2+\epsilon^4-2\epsilon^2\heps^2+\heps^4)}3.
\eea
} 

\end{appendix}


\begin{thebibliography}{99}

\bibitem{Scottish} R.D. Mauldin (ed.), {\it The Scottish Book}, Birkh\"auser
Boston 1981

\bibitem{WegnerIII} F. Wegner, {\it Floating Bodies of Equilibrium. Explicit
Solution}, e-Print archive physics/0603160

\bibitem{Auerbach} H. Auerbach, {\it Sur un probleme de M. Ulam
concernant l'equilibre des corps flottant}, Studia Math. 7 (1938) 121-142

\bibitem{Bracho} J. Bracho, L. Montejano, D. Oliveros, {\it Carousels,
Zindler curves and the floating body problem}, Per. Math. Hung. 49 (2004) 9-23

\bibitem{Oliveros} D. Oliveros and L. Montejano, {\it De volantines,
espir\'ographos y la flotaci\'on de los cuerpos}, Revista Ciencias 55-56 (1999)
46-53

\bibitem{Tabachnikov} S. Tabachnikov {\it Tire track geometry: variations on a
theme} Israel J. of Math. 151 (2006) 1-28 archive math.DG/0405445

\bibitem{WegnerI} F. Wegner, {\it Floating Bodies of Equilibrium I},
e-Print archive physics/0203061

\bibitem{Wegner} F. Wegner, {\it Floating Bodies of Equilibrium} Studies in
Appl. Math. 111 (2003) 167-183

\bibitem{Finn} D.L. Finn, {\it Which way did you say that bicycle went?} Math.
Mag. 77 (2004) 357-367

\bibitem{FinnI} D.L. Finn, {\it Which way did you say that bicycle went?}
http://www.rose-hulman.edu/~finn/research/bicycle/tracks.html

\bibitem{Doyle} A.C. Doyle, {\it The Adventure of the Priory School} in {\it The
Return of Sherlock Holmes}

\bibitem{WegnerII} F. Wegner, {\it Floating Bodies of Equilibrium II},
e-Print ar\-chi\-ve physics/0205059

\bibitem{movie} F. Wegner, {\it Three Problems - One Solution},
http://www.tphys.uni-heidelberg.de/$\sim$wegner/Fl2mvs/Movies.html;\\
{\it Drei Probleme - Eine L\"osung},
http://www.tphys.uni-heidelberg.de/$\sim$wegner/Fl2mvs/Filme.html

\bibitem{Evers} F. Evers, A.D. Mirlin, D.G. Polyakov, P. Woelfle, {\it
Semiclassical theory of transport in a random magnetic field},
cond-mat/9901070, Phys. Rev. B 60 (1999) 8951

\bibitem{Abramowitz} M. Abramowitz and I. A. Stegun, eds. {\it Handbook of
Mathematical Functions} Dover Publ., New York

\end{thebibliography}
\end{document}